\documentclass[11pt,centertags,reqno]{amsart}

\usepackage[foot]{amsaddr}
\usepackage{latexsym}
\usepackage[english]{babel}
\usepackage[T1]{fontenc}
\usepackage[numbers]{natbib}
\usepackage{amssymb}
\usepackage{fancyhdr}
\usepackage{url}
\usepackage{hyperref}
\usepackage{verbatim}
\usepackage{leftidx}
\usepackage{bm}
\usepackage{color,graphicx}
\usepackage{tikz}
\usepackage{bbm}

\usepackage{mathrsfs, mathtools}
\usepackage{stmaryrd}

\usepackage{marvosym}

\usepackage[margin=2.5cm]{geometry}
\usepackage{graphicx}
\usepackage{subcaption}  
\usepackage{float}
\usepackage{amsmath}

\usepackage[normalem]{ulem}
\definecolor{MGchange}{RGB}{165,70,0}

\numberwithin{equation}{section} \makeatletter
\renewcommand{\subsection}{\@startsection
{subsection}{2}{0mm}{\baselineskip}{-0.25cm}
{\normalfont\normalsize\bf}} \makeatother

\newtheorem{theorem}{Theorem}[section]
\newtheorem{lemma}[theorem]{Lemma}
\newtheorem{corollary}[theorem]{Corollary}
\newtheorem{definition}[theorem]{Definition}
\newtheorem{remark}[theorem]{Remark}
\newtheorem{proposition}[theorem]{Proposition}
\newtheorem{example}[theorem]{Example}
\newtheorem{assumption}[theorem]{Assumption}

\def \F {\mathcal F}

\def \R {\mathbb R}
\def \bF {\mathbb F}

\usepackage[textsize={tiny}]{todonotes}

\begin{document}

 \author[O.~Bonesini]{Ofelia Bonesini*}
\address{*Ofelia Bonesini, Department of Mathematics, London School of Economics and Political Science, London, United Kingdom}\email{o.bonesini@lse.ac.uk}

\author[G.~Callegaro]{Giorgia Callegaro}
\address{Giorgia Callegaro, Department of Mathematics, University of
Padova, Padova, Italy}\email{gcallega@math.unipd.it}

\author[M.~Grasselli]{Martino Grasselli}
\address{Martino Grasselli, De Vinci Higher Education, De Vinci Research Center, Paris, France, and Department of Mathematics,
University of Padova, Padova, Italy}\email{grassell@math.unipd.it}

\author[G.~Pag\`es]{Gilles Pag\`es}
\address{Gilles Pag\`es, LPSM, Sorbonne Universit\'e, Paris, France}\email{gilles.pages@upmc.fr}

\title[]{Efficient simulation of a new class of Volterra-type SDE\tiny s}

\date{\today}

\allowdisplaybreaks

\begin{abstract}
    We propose a new theoretical framework that exploits convolution kernels to transform a Volterra-type path-dependent (non-Markovian) stochastic process into a standard (Markovian) diffusion process. The transformation is reversible.
    
    We discuss existence and path-wise regularity of solutions for our class of stochastic differential equations. 
    In the fractional-kernel case, when $H\in(0,\frac12)$, where $H$ is the Hurst coefficient, we propose a numerical simulation scheme which exhibits a strong convergence rate of order $1/2$, improving upon the rate typically obtained by Euler schemes for stochastic Volterra equations with comparably rough trajectories. This improvement is made possible by the distinctive structure of the proposed class, characterized by a non-Markovian process whose coefficients are driven by an associated Markovian one.
    \end{abstract}

\maketitle



\section{Introduction}

Over the past 15 years Stochastic Volterra Equations (SVEs) have gathered more and more interest among the academic community. This is due to their non-Markovian nature together with the fact that their sample paths can be as regular as prescribed making them suitable to be applied in various contexts such as biology, physics and math finance, see \cite{di2016fractional}. Despite numerous attempts over these years their simulation still poses some challenges, in particular the rate of (strong and weak) convergence quickly deteriorates with the regularity of the paths. For a recent account of approximation and simulation methods for SVEs, we refer the interested reader to \cite{alfonsi2024approximation}.
Any alternative attempt to introduce a family of processes that is still non-Markovian and whose trajectories have low regularity but is easier to solve and simulate is obviously welcome.

Our paper fits into this perspective: we propose a new theoretical framework where we exploit the possibility to transform, via a convolution kernel, a Volterra path-dependent (non-Markovian) stochastic process into a standard (Markovian) diffusion process without memory. Furthermore, we provide a way to go back and forth.
More precisely, we embed, using kernels associated with different decay laws (including but not limiting ourselves to power laws, corresponding to rough models), what we call the Markovian \emph{volatility memory process} $\xi$, in the dynamics of the relevant non-Markovian process $X$. 
Our framework can be seen as a special case of an $\mathbb R^d$-valued Volterra stochastic process as introduced by \cite{BM1980} and recently revisited by \cite{abi2019affine}. For further details see Remark~\ref{rem:gen_vol}. 

The new framework requires a careful investigation in terms of existence and path-wise regularity of the solutions of the stochastic Volterra equations therein introduced. 
This is performed in the first part of the  paper, where we also illustrate the methodology that allows one to transform the Volterra process into a Markovian diffusion process and back. 
Then, we propose a numerical scheme for the simulation of this couple of processes, together with a thorough analysis of its numerical error and strong rate of convergence. 
In the fractional-kernel case, our numerical scheme approximation satisfies a strong error bound of order $\gamma\wedge\frac12$, where $\gamma$ is related to the time regularity
of the coefficients, independently of the Hurst parameter $H$; hence the order is $\frac12$ whenever the coefficients are at least $\frac12$-H\"older continuous in time. Classical Euler schemes for general stochastic Volterra equations may instead have an order governed by $H$, see, e.g., \cite{li2022numerical, richard2021discrete} and the introduction of~\cite{bonesini2023rough} for a complete overview on the state of the art on the topic.
As stressed above, this is a class-to-class comparison made possible by the Markovian-memory structure of our model.
An additional feature of our methodology is its flexibility, as it allows Volterra processes to be transformed into processes that do not necessarily possess Markovian characteristics. 
The determinative factor lies in the definition of the (pseudo)-inverse of the kernel employed in the transformation. While it is preferable, particularly for practical applications, to define the pseudo-inverse so that the transformed process becomes Markovian, as exemplified in the paper, this choice is specific to the situation at hand. The main point is to be able to exploit the structure of the transformed process.

Our approach can be seen as a generalization of the fractional transformation mentioned in \cite[page 1020]{guennoun2018asymptotic} in the case of fractional Heston model (and further discussed in \cite{abi2019affine}), in that we use a more general family of kernels together with the introduction of a non-deterministic initial condition depending on time and of the co-kernel.
The contribution of this quantity is not just in terms of adding some technicalities in the proofs,  but it can be seen as a burst of memory at $t=0$, needed to restore the memory of the process of what was somehow happening before the initial time.
Another attempt of generalising \cite{guennoun2018asymptotic} can be found in \cite{horvath2024functional}. Our approach is somehow reminiscent of theirs but the purpose, as well as the tools exploited therein, differ from ours. 
In particular, our theoretical study and the simulation scheme we propose are completely different. Furthermore, the formalisation yields a path-dependent Volterra equation that remains numerically tractable because its memory state is finite-dimensional and Markovian.
Last but not least, we get strong existence and uniqueness of solutions for the equation of the memory process, $\xi$, and, by fractional integration of it, of the process $X$ itself. Our work is also related to \cite{hamaguchi2024markovian}, who independently investigated a similar stochastic Volterra integral equation with 
power kernel and proved the existence, uniqueness and Markov property for the lifted stochastic evolution equation defined on an (infinite 
dimensional) Hilbert space. We emphasize that the performance of our numerical results is due to the fact we are able to work in finite dimension.

The rest of the paper is organized as follows.
In Section \ref{Sec:preliminaries}, we develop a preliminary yet detailed investigation of the properties of the Volterra-type processes of interest for us.
In particular, after establishing the notation used in the reminder of the paper, we define the \emph{convolution transform} as well as the concept of \emph{$\rho$-pseudo-inverse co-kernel}. Finally, we present some preliminary existence and regularity results to guarantee the well-posedness of the processes we introduce.
Then, in Section \ref{Sec:path_dependent_to_SDEs} we prove and comment the two-way connection between our path-dependent Volterra Equation and an associated standard SDE.
We first present an informal argument and then the detailed proof is discussed in Section \ref{Sec:proof_thm:existence}. Therein a thorough analysis of the mathematical features of the approach is developed, proving in particular strong existence and uniqueness of solutions and the H\"older regularity of sample paths for our processes.
In Section \ref{sec:Euler} we study the continuous time Euler scheme and their convergence for both the Markovian and the non-Markovian processes.
In Section \ref{sec:rel_example} we deal with the fractional kernel case, in the case of interest for the financial literature, i.e., when the Hurst index $H \in (0, \frac12)$. We adapt the theorem obtained in the previous sections to this setting: we first provide well-posedness results and path-wise regularity results and we then study the Euler scheme and its convergence. The main result concerns the strong rate of convergence.
The resulting bound is of order $\gamma\wedge\frac12$, independently of $H$, and therefore has order $\frac12$ when $\gamma\ge\frac12$. Finally, in Section \ref{sec:numerics} we provide some simulations and numerically display the results theoretically proved in the previous sections.

We gather in Appendix \ref{app:frac_Laplace} some material on fractional calculus and Laplace transforms for the reader's convenience and in Appendix \ref{app:proofs} the technical proofs.
Appendix \ref{app:simu_det} concludes the paper with a deep insight into simulating the Euler scheme for the non-Markovian process.

\bigskip



\section{Mathematical preliminaries and prerequisites}\label{Sec:preliminaries}

\subsection{Notation}

    Let us consider a fixed finite time horizon $T>0$ and a complete filtered probability space $(\Omega,\F, \mathbb{P};\bF)$, with $\bF= (\mathcal{F}_t)_{t \in [0,T]}$. 
    On this probability space we consider an $\mathbb R$-valued random variable $X$ and an $\mathbb R$-valued random process, $ Y=(Y_t)_{t \in [0,T]}$.
    We denote by $\textrm{AC}([0,T])$ the space of absolutely continuous functions on $[0,T]$.
    For $x \in \mathbb{R}$, we denote by $\delta_x$ the Dirac delta distribution centred at $x$.
    Given a function $f: (t,x) \in [0,T] \times \mathbb R \rightarrow \mathbb R$ which is Lipschitz in $x$, uniformly in time, $t$, we denote by $[ f]_{{\rm Lip},x}$ its Lipschitz constant. 
    The set of functions $f :\R_+\to {\mathbb R}$ such that
    $\int_0^T |f(t)|^p dt <+\infty$,   for every $T>0$, 
    is denoted by ${L}^p_{loc}(\R_+, {\rm Leb}_1)$, $1 \le p < \infty$.
    In case of no ambiguity, we use the notation ${L}^p_{loc}(\R_+)$ or simply ${L}^p_{loc}$.
    For $1 \le p < \infty$, we denote by $\| X \|_p$ the $p$-norm of $X$, i.e., $\| X \|_p:= \mathbb{E}[|X|^{p}]^{\frac{1}{p}}$,
    and by $\| Y \|_{\mathcal{L}^p}$ the $\mathcal{L}^p$-norm of the stochastic process $Y$, i.e.
    $\| Y \|_{\mathcal{L}^p}
    	:= \mathbb{E}\left[\int_0^T |Y_s|^{p} ds \right]^{\frac{1}{p}}$. For $p>0$ we denote by $L^p(\mathbb{P})$ the set of real-valued random variables $X$ such that $\mathbb{E}[|X|^{p}] < \infty$.
    Finally,  $\mathcal{L}^2:= \mathcal{L}^2([0,T])$ represents the set of real-valued stochastic processes $Y=(Y_t)_{t \in [0,T]}$ which are $\bF-$progressively measurable and such that $ \| Y \|_{\mathcal{L}^2}^2 < + \infty$.

    Let us stress that the stochastic basis $(\Omega,\F, \mathbb{P};\bF)$, with $\mathbb F = {(\mathcal F_t)}_{t \in [0,T]}$ is assumed to be rich enough to model all the randomness in our model and such that a standard Brownian motion $\{W_t, t \in [0,T]  \}$ is defined on it.


\subsection{Convolution transforms}\label{sec:gen_app}

In order to introduce our \emph{convolution transform}, first we have to recall the notion of \emph{convolution kernel}. 

\begin{definition}\label{def:conv_K}
    A \emph{convolution kernel} is a function $K: \R_+\to \R_+$  such that, for $0 \le s \le t \le T$, $K(t,s)=K(t-s,0)=K(t-s)$ and $\int_0^T K(t) dt >0$, for every $T>0$.
\end{definition}

Given a convolution kernel, $K$, and a function, $f$, on $\mathbb R_+$, their convolution $(K \star f)$ is defined as
\begin{equation}\label{eq:conv_std}
    (K \star f) (t) :=  \int_0^t K(t-s)f(s) ds, 
\end{equation}
for all $t < T$ such that the above integral exists finite. A complete list of properties of $(K \star f)$ can be found in \cite[Chapter 2, Section 2.2]{Grip90}.  In particular, in \cite[Theorem II.2.2 and Corollary II.2.3]{Grip90} the authors prove that, for $K \in L^1_{\textrm{loc}}(\mathbb R_+)$ and $f \in L^p_{\textrm{loc}}(\mathbb R_+)$, for some $p \in [1, \infty]$, the convolution $(K\star f) \in L^p_{\textrm{loc}}(\mathbb R_+)$.
Now, with a slightly different notation, we introduce a stochastic version of the Lebesgue integral in Equation~\eqref{eq:conv_std}. Given a square-integrable stochastic process $Y \in \mathcal L^2$, we define the family of random variables:
\begin{equation}\label{eq:conv_X}
    {(K \star Y)}_t :=  \int_0^t K(t-s) Y_s ds, \quad 0\le t \le T,
\end{equation}
such that the above integral exists finite $\mathbb P$-a.s. Moreover, we introduce the following stochastic It\^o integral of convolution type:        
\begin{equation}\label{eq:conv_X_Ito}
    {(K \stackrel{W}{\star} Y)}_t := \int_0^t K(t-s) Y_s dW_s, \quad 0 \le s \le t \le T,
\end{equation}	
such that the above integral is well defined (e.g., when the integrand is in $\mathcal{L}^2$).
In Table~\ref{Table_K} we list the most relevant kernels that appear in the literature. In particular, the Gamma kernel is denoted here by $K_{c, \alpha,\rho}$, for positive $c, \alpha, \rho$.
 
\begin{table}[h!]
\begin{tabular}{| l c c c |}
\hline
& $K(t)$ & Notation & Domain\\
\hline
Constant & $c$  & $K_{c, 1,0}$ & $c >0$\\
Fractional &  $c \frac{t^{\alpha -1}}{\Gamma(\alpha)}$  & $K_{c, \alpha,0}$ & $c >0, \alpha >0$\\
Exponential & $ ce^{- \rho t} $  &  $K_{c, 1,\rho}$ & $c >0, \rho >0$\\
Gamma & $ c e^{- \rho t} \frac{t^{\alpha -1}}{\Gamma(\alpha)}$ &  $K_{c, \alpha,\rho}$ & $c >0, \alpha >0, \rho >0$\\
\hline
\end{tabular} \vspace{0.3 cm}
\caption{\label{Table_K} Some examples of kernels. }
\end{table}


\subsection{First existence and path-regularity results}
Before introducing our new class of stochastic Volterra-type equations, we prove existence and path-regularity for the stochastic process $X$ defined as
\begin{equation}\label{eq:X_base}
    X_t = \xi^0 + \int_0^t K(t-s) Y_s dW_s, \quad t \in [0,T],
\end{equation}	
where $K$ is a convolution kernel, the process $Y \in \mathcal L^2$ is an integrand stochastic process, $W$ is the Brownian motion defined on our filtered probability space $(\Omega,\F, \mathbb{P};\bF)$ and $\xi^0$ is a random variable defined on the same probability space and assumed to be  independent of $W$. So, $\xi^0 \in \mathcal F_0$ and, when useful, we use the notation $\mathbb F = \mathbb F^{\xi^0, W}$.
For simplicity, as it is standard in the literature on SDEs with random initial condition (see \cite[Chapter 3]{KS}), we assume that $\xi^0 \in L^p(\mathbb{P})$ with $p\geq 2$. 
Let us notice that a more general existence and uniqueness result for stochastic Volterra equations is given in \cite[Thm 1.1]{jourdain2025convex} for $\xi^0 \in L^0(\mathbb{P})$. 
It is also worth mentioning \cite[Thm. 3.1]{Zha10}, where the author proves a global existence and uniqueness result for stochastic Volterra equations in Banach spaces with $\xi^0\in L^{p}(\mathbb{P}), p \ge 2$.
Nevertheless, for the reader's ease we detail its proof in Appendix~\ref{sec_proof_lem:regularityX}.

{
\begin{lemma}\label{lem:regularityX}
    Let $\xi^0\in L^2(\mathbb P)$ and let $Y$ be any progressively measurable process $Y$, satisfying: 
    \begin{itemize}
        \item[i)] $\sup_{t\in[0,T]}\|Y_t\|_2<+\infty$;
        \item[ii)] there exists $\beta>1$ such that $\int_0^T K(t)^{2\beta}\,dt<+\infty$ for every $T>0$.
    \end{itemize}
    Then the stochastic convolution
    \[
       X_t=\xi^0+\int_0^t K(t-s)Y_s\,dW_s
    \]
    is well defined as an $(\mathcal F_t)_{t\in[0,T]}$-adapted process and $X_t\in L^2(\mathbb P)$ for every $t\in[0,T]$.

    Moreover, assume that there exist $p\ge2$ and $\theta\in(0,1)$ such that $\xi^0\in L^p(\mathbb P)$,
    \[
       p>\frac{1}{\theta\wedge\frac{\beta-1}{2\beta}},
    \]
    and:
    \begin{itemize}
        \item[iii)] $\sup_{t\in[0,T]}\|Y_t\|_p<+\infty$;
        \item[iv)] there exists $C_K>0$ such that, for every $\delta\in(0,T)$,
        \[
        \left(\int_0^T\left[K((\delta+v)\wedge T)-K(v)\right]^{2\beta}dv\right)^{1/(2\beta)}\le C_K\delta^\theta.
        \]
    \end{itemize}
    Then $X_t\in L^p(\mathbb P)$ for every $t\in[0,T]$, and $X$ has a pathwise continuous modification whose trajectories are locally $a$-H\"older continuous for every
    \[
       a\in\left(0,\left(\theta\wedge\frac{\beta-1}{2\beta}\right)-\frac1p\right).
    \]
\end{lemma}
}

{
\begin{example}
    Consider $\alpha\in(\frac12,1)$. For the Gamma kernel $K_{c,\alpha,\rho}$, condition ii) in Lemma~\ref{lem:regularityX} holds for every
    \[
       \beta\in\left(1,\frac{1}{2(1-\alpha)}\right),
    \]
    and condition iv) holds with $\theta=\alpha-1+\frac{1}{2\beta}$. Consequently, if $\xi^0\in L^p(\mathbb P)$ and $\sup_{t\le T}\|Y_t\|_p<\infty$ for
    \[
       p>\frac{1}{\left(\alpha-1+\frac{1}{2\beta}\right)\wedge\frac{\beta-1}{2\beta}},
    \]
    Lemma~\ref{lem:regularityX} yields locally $a$-H\"older continuous trajectories for every
    \[
       a<\left(\alpha-1+\frac{1}{2\beta}\right)\wedge\frac{\beta-1}{2\beta}-\frac1p.
    \]
\end{example}
}


\section{From path-dependent to standard SDEs}\label{Sec:path_dependent_to_SDEs}

The aim of this section is to investigate a class of path-dependent Volterra-type SDEs that is associated to a convolution kernel $K$ when it  admits what we call a $\rho$-pseudo-inverse co-kernel $\widetilde K$.
As stated in the introduction they are particularly relevant because of the direct link with a standard SDE that can be associated to them in a very natural way.
Therefore, in this section not only we present such a class of SDEs but we also detail the two-fold connection between the non-Markovian process and Markovian one.

\subsection{The $\rho$-pseudo-inverse co-kernel}

Let us start by providing the definition of $\rho$-pseudo-inverse co-kernel, the key ingredient in the formal definition of class of SDEs we are interested in.

\begin{definition}\label{def:coker}
    A $\rho$-pseudo-inverse co-kernel with respect to $K:=K_{\cdot, \cdot,\rho}$, with $\rho \ge 0$, is a {{function $\widetilde K: \R_+\to \R_+$, continuous on $(0,+\infty)$, satisfying, for every $t> 0$,}}
    \begin{equation}\label{eq:KKtilde}
        (K\star \widetilde K) (t) = (\widetilde K \star K) (t) =  e^{-\rho t}.
    \end{equation}
\end{definition}

{{
\begin{example}
We here focus on the $\rho$-pseudo inverse co-kernel $\widetilde K$ in the cases listed in Table \ref{Table_K} and for $0<\alpha <1$. An elegant derivation of $\widetilde K$ is provided in Appendix \ref{app:frac_Laplace}.
    \begin{itemize}
        \item (Fractional) We have $\widetilde K = K_{\frac{1}{c},1-\alpha,0}$, namely $\widetilde K(t)=\frac1c \frac{t^{- \alpha }}{\Gamma( 1 - \alpha)}$. See also Example \ref{ex:Ktilde_Laplace}$b)$ in the case $\rho=0$.  
        \item (Gamma) We have
 $\widetilde K = K_{\frac{1}{c},1-\alpha,\rho}$, namely 
$\widetilde K(t)=  \frac1c e^{- \rho t} \frac{t^{ - \alpha}}{\Gamma( 1 - \alpha)} $. See also Example \ref{ex:gammaKtilde_Laplace}.
    \item (Constant) Here Defintion \ref{def:coker} extends to the distributional sense: $\widetilde K = \frac1c \delta_0 -\frac{\rho}{c} e^{- \rho \cdot}$, where $\delta_0$ is the Dirac delta at $0$.
    \item (Exponential) Similarly, in the distributional sense: $\widetilde K = \frac 1c \delta_0$.
\end{itemize}
\end{example}
}} 
    
\begin{remark}\label{rem:resolvent}
    We recall here the definition of one dimensional \emph{functional resolvent of the first kind}, which can be found e.g. in \cite[Def. 5.5.1]{Grip90} (see also  \cite{abi2019affine}). Given $K \in {L}^1_{loc}(\R_+) $, a function $r$ belonging to ${L}^1_{loc}(\R_+)$ is called \emph{functional resolvent of the first kind  of $K$} if 
    $$
    (K \star r)(t) = (r \star K)(t) = 1,
    $$
    for all $t \in \mathbb R_+$.
    So, when $\rho$ is equal to zero, $\widetilde K$ is the functional resolvent of the first kind of $K$.\\     
    For $\rho \neq 0$, we find that $\widetilde K$ in Definition \ref{def:coker} is $\widetilde K(u) = e^{- \rho u} \overline r(u)$, where $\overline r$ is the functional resolvent of the first kind of $\overline K(u) := e^{\rho u} K(u)$. Indeed, $\overline r$ by definition satisfies 
    $$
    \int_0^t \overline K(t-s) \overline r(s) ds = \int_0^t K(t-s) e^{\rho (t-s)} \overline r(s) ds = 1,
    $$
    which is equivalent to
    $
    \int_0^t K(t-s)\widetilde K(s) ds = e^{- \rho t}, 
    $
    as $\widetilde K (s) := e^{- \rho s} \overline r(s)$.
\end{remark}

As we are going to make an extensive use of kernels and co-kernels, { we now highlight a few properties of them.}

First of all, {{since $K\ge 0, \widetilde K \ge 0$}}, by applying Fubini-Tonelli's Theorem, we find
        \begin{align*}
        \int_0^{2T}  e^{-\rho u} du 
        & = \int_0^{2T} (\widetilde K \star K) (u) du   
        =  \int_0^{2T} \left( \int_0^u  K(u-s) \widetilde K(s)  ds \right) du = \int_0^{2T} \widetilde K(s)\int_0^{2T-s}  K(v)dv ds
    \end{align*}
    {{and  exploiting once more the positivity of $K$ and $\widetilde K$, we obtain}}
    \begin{align*}
        \int_0^{2T}  e^{-\rho u} du 
        & 
        \ge  \int_0^T \widetilde K(s)ds\int_0^{T}  K(v)dv,
    \end{align*}
    so that if (the restriction to $[0,T]$ of) $K\!\in L^1([0,T], {\rm Leb}_1)$ and $K$ is a convolution kernel (recall Definition~\ref{def:conv_K}), then, for every $T>0$,  $\int_0^T \widetilde  K(s)  ds < \infty$.
    
    Moreover, exploiting the positivity of $K$, we obtain that $\widetilde K$ is a convolution kernel, too, as it also satisfies $\int_0^T \widetilde  K(s)  ds >0$.
    
    Finally, for notational homogeneity,  we often denote $\int_0^t \widetilde K(s)ds$ as the convolution product: 
        \begin{equation}
        \int_0^t \widetilde K(s)ds = (\mbox{\bf 1} \star \widetilde K )(t) = ( \widetilde K \star \mbox{\bf 1} )(t)=:\widetilde{\varphi}(t).
    \end{equation}
    Being $\widetilde K$ continuous on $(0, +\infty)$ and non-negative, $\widetilde \varphi(t)=(\widetilde K\star\mbox{\bf 1})(t)$ is differentiable and non-decreasing. Moreover, if $\widetilde K$ is non-increasing, $\widetilde \varphi $ is concave.

\subsection{Transformation into a Markovian process (and back)}

We are finally ready to introduce the family of processes we are interested in.
We always work under the following assumption.
\begin{assumption}[\textbf{Stochastic Fubini}]\label{ass:Fubini}
	We assume interchangeability of Lebesgue and stochastic integration.
\end{assumption}
{{Sufficient conditions for this to hold in our setting will be provided below in Remark \ref{rem:fubini_check}. A version of stochastic Fubini theorem, to interchange the order of integration with respect to a finite measure and stochastic integration with respect to a semi-martingale, are given in \cite[Thm. IV.65]{Protter}}}.

The class of Volterra-type SDEs we focus on formally reads, for $t \in [0,T]$,
\begin{equation*}
    X_t = \xi^0     + \int_0^t K(t-s) b\left(s, (\widetilde K\star X)_s \right) ds +   \int_0^t K(t-s) \sigma \left( s, (\widetilde K\star X)_s \right)dW_s,
\end{equation*}
where $\xi^0$ is a random variable assumed to be independent of $W$ and $b,\, \sigma : [0,T]\times\R\rightarrow  \R$ are Borel measurable functions. 
The key feature of the non-Markov stochastic process $X$ above is the nested dependency in the coefficients appearing in the SDEs. 
Indeed, the transformation from non-Markov to Markov, and back, is achieved by embedding a Markovian "memory process", which will be denoted by $\xi$, within the dynamics of the non-Markovian process $X$.

The following theorem, whose proof is postponed to Appendix \ref{Sec:proof_thm:existence}, emphasizes the connection between the class of path-dependent SDEs, represented by the process $X$ above, and standard Brownian SDEs, represented by the process $\xi$ which we will introduce in a few lines. Moreover, it provides existence and uniqueness results.

\begin{theorem} \label{thm:existence} 
    Let $\rho\ge 0$ and consider Equation~\eqref{eq:pathdepVolt}, namely
    \begin{equation}\label{eq:pathdepVolt}
        X_t = \xi^0     + \int_0^t K(t-s) b\left(s, (\widetilde K\star X)_s \right) ds +   \int_0^t K(t-s) \sigma \left( s, (\widetilde K\star X)_s \right)dW_s,
    \end{equation}
    where the kernels $K$ and $\widetilde K$ satisfy the condition in Equation \eqref{eq:KKtilde}. If there exist $\beta >1$ and $\theta\! \in (0,1)$, such that: 
    \begin{itemize}
        \item[a)] the kernel $K$ satisfies, for some $C_K >0$ and {{for any}} $\delta \in (0,T)$:
    	$$
    	\int_0^T K(t)^{2 \beta} dt < +\infty\quad \mbox{ and }\quad \left( \int_0^T \left[ K((\delta +v ) \wedge T) - K(v )\right]^{2\beta} dv \right)^\frac{1}{2\beta} \le C_K \delta^\theta;
    	$$
    	 
    	 \smallskip
    	 \item[b)] the Borel functions $b$ and $\sigma$ defined on $[0,T]\times \R$ are  Lipschitz in $x$ uniformly in $t\!\in [0,T]$, and uniformly bounded at $x=0$:
    	 \begin{align}\label{eq:LipxUt}
           \nonumber  & (i)\;\text{for all }\, t\!\in [0,T], \; \text{for all }\, x,\, y \!\in \R, \\ 
    	   & \qquad \qquad |b(t,x)-b(t,y)| \le  [b]_{{\rm Lip}, x} |x-y| \quad \mbox{and} \quad |\sigma(t,x)-\sigma(t,y)| \le [\sigma]_{{\rm Lip}, x} |x-y| \\
    	 &(ii) \;\sup_{t\in [0,T]} |b(t,0)|+|\sigma(t,0)| < +\infty;
    	\end{align}
    \item[c)]   
    {$\xi^0\in L^0(\mathbb P)$, i.e., $\xi^0$ is finite $\mathbb P$-a.s.} 
    \end{itemize}
    {
then the Markovian diffusion SDE
\begin{equation}\label{eq:Wdiff}
\xi_t=\xi^0 e^{\rho t}(\mathbf 1\star\widetilde K)(t)+\int_0^t\widetilde b(s,\xi_s)\,ds+\int_0^t\widetilde\sigma(s,\xi_s)\,dW_s,
\end{equation}
with
\begin{align}\label{Eq:def_over_b_sigma}
\widetilde b(u,x)&=e^{\rho u}b(u,e^{-\rho u}x), &
\widetilde\sigma(u,x)&=e^{\rho u}\sigma(u,e^{-\rho u}x),
\end{align}
admits a unique pathwise continuous strong solution $(\xi_t)_{t\in[0,T]}$. On every finite horizon, $\widetilde b$ and $\widetilde\sigma$ inherit the spatial Lipschitz and linear-growth properties of $b$ and $\sigma$.

Furthermore, Equation~\eqref{eq:pathdepVolt} admits a unique pathwise continuous strong solution $(X_t)_{t\in[0,T]}$ satisfying
\[
   (\widetilde K\star X)_t=e^{-\rho t}\xi_t,
\]
and this solution admits the representation
\begin{equation}\label{eq:Xrepresentation}
X_t=e^{-\rho t}\frac{d}{dt}\left[((e^{\rho\cdot}K)\star\xi)_t\right],\qquad t\in[0,T].
\end{equation}
}
\end{theorem}

Theorem \ref{thm:existence} provides a new stochastic working framework, which makes a solid and clear connection between the Markovian and the non-Markovian worlds. Moreover, it represents a real theoretical bridge, since it is possible to move both forward and backward.
A peculiar feature of Equation \eqref{eq:Wdiff} is the term $\xi^0\,e^{\rho t} (\widetilde K\star\mbox{\bf 1})(t) $, which vanishes only at time $t=0$ (implying that the starting value for the process $\xi$ is $\xi_0=0$), while, immediately after time $t=0$, it has an instantaneous effect on the trajectories of $\xi$, which bump up. {It may be interpreted as incorporating, from the initial time onward, a memory of the history preceding $t=0$.}

{ 
\begin{remark}\label{rem:gen_vol}
    As previously mentioned, considering $X$ and $\xi$ together, we can rewrite their joint dynamics, for $t \in [0,T]$, as
    \begin{align*}
        \begin{pmatrix}
             X^1_t  \\
             X^2_t
        \end{pmatrix}
        = \begin{pmatrix}
            1  \\
             e^{\rho t} {( 1 \star \widetilde{K})}_t
        \end{pmatrix}\xi_0
        + \int_0^t 
        \bold{K}(t-s) 
        \begin{pmatrix}
            e^{-\rho s}\widetilde{b}\left(s, X^2_s \right)\\
            \widetilde{b}\left(s, X^2_s \right)
        \end{pmatrix} ds
        + \int_0^t 
        \bold{K}(t-s)
        \begin{pmatrix}
            e^{-\rho s}\widetilde{\sigma}\left(s, X^2_s \right)\\
            \widetilde{\sigma}\left(s, X^2_s \right)
        \end{pmatrix} dW_s.
    \end{align*}
    where $\bold{K}(t-s):=\begin{pmatrix}
            K(t-s) & 0\\
            0 & 1
        \end{pmatrix}$.
    The resulting dynamics belong to a specific subclass of stochastic Volterra equations: while retaining non-Markovian and potentially rough trajectories (in their first component), the coefficients are driven by the second component which is a finite-dimensional Markovian process that encodes the relevant memory. 
    This structural feature is precisely what makes the proposed simulation approach effective. Accordingly, the convergence results discussed below are compared with those available for other classes of stochastic Volterra equations exhibiting similar path roughness.
\end{remark}
}

{
\begin{remark}[Sufficient conditions for stochastic Fubini]\label{rem:fubini_check}
Condition a) of Theorem~\ref{thm:existence} implies $K\in L^2(0,T)$, while condition b) gives linear growth of $\sigma$. For a solution $\xi$ of Equation~\eqref{eq:Wdiff}, define
\[
   \tau_m:=\inf\{t\in[0,T]:|\xi_t|\ge m\}\wedge T.
\]
For the stopped integrand, linear growth yields a deterministic constant $C_m$ such that
\[
\int_0^T\left(\int_0^t |K(t-s)\sigma(s,e^{-\rho s}\xi_s)|^2\mathbf 1_{\{s\le\tau_m\}}\,ds\right)^{1/2}dt
\le C_m T\|K\|_{L^2(0,T)}<\infty.
\]
The same stopping argument gives absolute integrability for the deterministic convolution term. Protter's stochastic Fubini theorem \cite[Thm.~IV.65]{Protter} applies to every stopped stochastic convolution. Since $\xi$ has continuous paths, $\tau_m\uparrow T$ almost surely, and the desired identities follow by localization. In particular, no global $L^2$ moment assumption on $\xi^0$ is required. If $\xi^0\in L^2(\mathbb P)$, the same conclusion may alternatively be obtained directly from the corresponding expectation-based integrability condition.
\end{remark}
}

\subsection{An informal intuition}

Since the proof of Theorem \ref{thm:existence} is postponed to Appendix \ref{Sec:proof_thm:existence}, we provide here an informal argument to provide some intuition without going into the details and technicalities of the proper proof. 
Let us start by noticing that the path-dependent SDE in Equation~\eqref{eq:pathdepVolt} may be rewritten in terms of convolution as
\begin{align*}
    X= \xi^0   + K\star b(\cdot,(\widetilde K\star X)_{\cdot}) +  K \stackrel{W}{\star} \sigma(\cdot, (\widetilde K\star X)_{\cdot}).
\end{align*}
As a consequence, when working under Assumption~\ref{ass:Fubini}, convoluting $X$ with $\widetilde K$, using the associativity of the convolution operators together with the identity in  Equation~\eqref{eq:KKtilde}, we find
\begin{align*}
    \widetilde K\star  X & =   \xi^0 \,     (\widetilde K\star\mbox{\bf 1})+ (\widetilde K\star K)\star b(\cdot,(\widetilde K\star X)_{\cdot}) +  (\widetilde  K  \star K)\stackrel{W}{\star} \sigma(\cdot,  (\widetilde K\star X)_{\cdot})\\
    &  =    \xi^0     (\widetilde K\star\mbox{\bf 1})+ e^{-\rho \cdot}\star b(\cdot,(\widetilde K\star X)_{\cdot}) +  e^{-\rho \cdot} \stackrel{W}{\star} \sigma(\cdot, (\widetilde K\star X)_{\cdot}) .
\end{align*}
Let us introduce the memory process $\xi$ as 
$$
    \xi_t= e^{\rho t}( \widetilde K\star  X)_t, \quad t \in [0,T].
$$
Furthermore, let use note that, 
$$
    e^{\rho t} \left( e^{-\rho \cdot}\star b(\cdot,(\widetilde K\star X)_{\cdot}) \right)_t 
    = e^{\rho t} \int_0^t e^{-\rho(t-s)} b(s,(\widetilde K\star X)_s ) ds 
    =  \int_0^t e^{\rho s} b(s,(\widetilde K\star X)_s ) ds,
$$ 
{{and
$$
    e^{\rho t} \left( e^{-\rho \cdot}\stackrel{W}{\star}  \sigma(\cdot,(\widetilde K\star X)_{\cdot}) \right)_t 
    = e^{\rho t} \int_0^t e^{-\rho(t-s)} \sigma(s,(\widetilde K\star X)_s ) dW_s 
    =  \int_0^t e^{\rho s} \sigma(s,(\widetilde K\star X)_s ) dW_s.
$$ 
}}
Thus, exploiting the identities in Equation~\eqref{Eq:def_over_b_sigma}, we immediately obtain the SDE satisfied by $\xi$:  
\begin{equation*}
    \xi_t = \xi^0e^{\rho t}   (\widetilde K\star\mbox{\bf 1})(t) +  \int_0^{t}  \widetilde  b(s, \xi_{s})ds  + \int_0^{t} \widetilde \sigma(s, \xi_{s})dW_s,
\end{equation*}
that corresponds to Equation~\eqref{eq:Wdiff}. 
{
\begin{remark}
When $\rho=0$, Equation~\eqref{eq:Wdiff} gives
\[
d\xi_t=\xi^0\widetilde K(t)dt+b(t,\xi_t)dt+\sigma(t,\xi_t)dW_t.
\]
Since $K\star\widetilde K=\mathbf 1$, the process $X$ admits the compact and intuitive representation $X=K\star d\xi,
X_t=\int_0^tK(t-s)d\xi_s.$
The contribution $\xi^0$ is already contained in $K\star d\xi$ through the term $\xi^0(K\star\widetilde K)(t)$ and must not be added a second time.
\end{remark}
}


\section{Euler schemes: $L^p(\mathbb P)$-convergence rate}\label{sec:Euler}

In this section we discuss one of the main results of this paper. In particular, we study the convergence for the so-called genuine Euler scheme with step $h:=\frac{T}{n}$, $n\!\in {\mathbb N_{>0}}$, which is the continuous time Euler scheme (as opposed to the discrete time Euler scheme and the step-wise constant Euler scheme, for a reference see \cite[Section 7.1]{GilPag}), for $\xi$ and for a continuous time Euler scheme for $X$.
{The connection with the diffusion $\xi$ makes it possible to obtain a convergence rate which inherits a Euler contribution $h^{\gamma\wedge1/2}$. For a general kernel, the convergence rate also contains the residual term in Equation~\eqref{thm:err_Euler}; in the fractional case this term contributes $h^{1/2}$, so the overall rate is $h^{\gamma\wedge1/2}$.}

More precisely, in Section \ref{sec_sch_xi},  Theorem \ref{thm:RateEulercont}, we first describe the continuous time Euler scheme for $\xi$, proving a strong rate  convergence in $L^p(\mathbb P)$ norm of order $\left( \frac12 \land \gamma \right)$, with $\gamma$ being the H\"older-Lipschitz regularity of the coefficients $b$ and $\sigma$. 
As a consequence of this result, we also get some error bounds for the step-wise constant Euler scheme (Corollary \ref{cor:L1RateEulercstpm}).
Then, in Section \ref{sec_sch_X} we focus on the process $X$: after introducing three possible continuous time Euler schemes for it, that we call the Smart-Euler, the projected Euler and the naive Euler, we show in Theorem \ref{thm:est_eu_sch_X} that, in case of the Smart scheme, the order $\left( \frac12 \land \gamma \right)$ for the strong rate  convergence in $L^p(\mathbb P)$ norm is preserved.
Finally,  some error bounds for the step-wise constant Euler scheme for $X$ are presented (Corollary \ref{cor:Euler_piece_c}).
To alleviate notation, we focus on the case $\rho =0$. The extension to the case $\rho \neq0$ straightforward.

\subsection{Genuine (continuous) Euler scheme for $\xi$.}\label{sec_sch_xi}

When $\rho=0$ we have $\widetilde b(\cdot,\cdot) = b(\cdot,\cdot)$ and $\widetilde \sigma(\cdot,\cdot) = \sigma(\cdot,\cdot)$,  and so the dynamics of $\xi$ in Equation \eqref{eq:Wdiff} reads as 
\begin{equation}\label{Xtilde_rho0}
    \xi_t = \xi^0\,\widetilde \varphi(t)+\int_0^t \  b(s, \xi_s)ds +\int_0^t   \sigma(s, \xi_s) dW_s, \quad t\!\in [0,T],  
\end{equation}
with $\widetilde \varphi(t) = (\widetilde K\star\mbox{\bf 1})(t)$. 
Its continuous time Euler scheme $\overline{\xi}= \overline{\xi}^h$, with time step $h = \frac Tn$, $n\!\in {\mathbb N_{>0}}$, is defined via the pseudo-SDE with frozen coefficients 
\begin{equation}\label{eq:Euler_cont_xi}
    \overline{\xi}_t = \xi^0 \,\widetilde \varphi(t) +\int_0^t  b(\underline s, \overline{\xi}_{\underline s})ds +\int_0^t   \sigma(\underline s, \overline{\xi}_{\underline s}) dW_s, 
\end{equation}
where $\underline s= t_k = t^n_k:= \frac{kT}{n}$, for $s\!\in \big[ \frac{kT}{n},  \frac{(k+1)T}{n}\big)$, $k \in \{0,\dots, n-1\}$.  
{{So, we have
\[
    \xi_t - \overline{\xi}_{t} = \int_0^t (b(s, \xi_s) - b(\underline s, \overline{\xi}_{\underline s})) ds + \int_0^t (\sigma(s, \xi_s) - \sigma(\underline s, \overline{\xi}_{\underline s})) d W_s
\]
and 
\[
    \overline{\xi}_{t} - \overline{\xi}_{\underline t} = \xi^0\big(\widetilde \varphi(t) -\widetilde \varphi(\underline t) \big)   + (t-\underline t) b(\underline t, \overline{\xi}_{\underline  t}) +\sigma(\underline t, \overline{\xi}_{\underline  t})(W_t-W_{\underline t}).
\]
Hence, by pathwise continuity of the Brownian motion, one has the simulation scheme
\[
    \overline{\xi}_{t_{k+1}} -\overline{\xi}_{t_k} = \xi^0  \big(\widetilde \varphi(t_{k+1}) -\widetilde \varphi(t_{k}) \big) + h \cdot b(t_k, \overline{\xi}_{t_k}) + \sigma(t_k, \overline{\xi}_{t_k}) \Delta W_{t_{k+1}},
\]
where $ \Delta W_{t_{k+1}} = W_{t_{k+1}}-W_{t_k}$, $k= 0, \ldots, n-1$.
}}

The following bound for the strong rate of convergence of the continuous Euler scheme holds.

\begin{theorem}\label{thm:RateEulercont} 
    Let $p\!\in (0, +\infty)$ and let us assume that the following conditions are satisfied:
    \begin{itemize}
        \item[i)]  there exists a positive real constant $C$  such that $b$ and $\sigma$ satisfy the H\"older-Lipschitz time-space condition  with $\gamma \in (0,1]$, namely  for every $ s,\, t\!\in [0,T]$ and every $ x,\, y \!\in \R$, 
        \begin{equation}\label{eq:bsigHolLip}
            |b(s,x)-b(t,y)|+|\sigma(s,x)-\sigma(t,y)| \le C \big((1+|x|+|y|)|t-s|^{\gamma} + |x-y| \big);
        \end{equation}
        \item[ii)]  the function $\widetilde \varphi$ is non-decreasing, concave and  $\widetilde \varphi(0)=0$;
        \item[iii)]  the initial condition $\xi^0 \in L^p(\mathbb{P})$. 
    \end{itemize}
    Then,
     \begin{equation}\label{eq:Eulercont}
     \Big\|\sup_{t\in [0,T]} |\xi_t -\overline{\xi}_t| \Big\|_p  \le \,C_{b,\sigma, T, p}\, h^{\frac 12\wedge\gamma}(1+ \|\xi^0\|_p). 
     \end{equation}
\end{theorem}
The proof of this theorem is postponed to Appendix \ref{sec_proof_thm:RateEulercont}.
The key point in the proof is dealing with the non-null starting condition of the process $\xi$. In particular, in the second step of the proof, it is crucial to get rid of the impact of the ``H\"olderianity''  (if any) of $\widetilde \varphi$ at $0$ in the rate of convergence, exploiting the fact that $\widetilde \varphi $ is Lipschitz on any interval $[\varepsilon, T]$, if $\varepsilon\!\in (0,T)$. Indeed, thanks to the monotonicity and concavity of $\widetilde \varphi $ we have, for every $s, t \in [\varepsilon, T]$, $|\widetilde \varphi(t)-\widetilde \varphi(s)|\le \widetilde\varphi'_{\textrm{left}}(\varepsilon) |t-s|$.

\begin{remark}
	The strong convergence result proved in Theorem \ref{thm:RateEulercont} holds in the case where $\widetilde \varphi$ can be replaced by a more general function $g$ only satisfying assumption $ii)$ in the statement.
\end{remark}

  We conclude this subsection by providing the error bounds relative to the step-wise constant c\`adl\`ag scheme.
  For the sake of readability, the proof of this corollary is postponed to the Appendix \ref{sec_proof_cor:L1RateEulercstpm}.
  
\begin{corollary}[Step-wise constant Euler scheme $\overline{\xi}_{\underline t}$: $L^r(dt)$-$L^p(\mathbb{P})$-error] \label{cor:L1RateEulercstpm} 
    Under the assumptions of Theorem~\ref{thm:RateEulercont}, for every $r>0$ and $p>0$ there exists a real constant $C_{b,\sigma,T,p,r}>0$, such that 
    \begin{equation}\label{eq:Eulercstpm}
         \left( \int_0^T \|\xi_t -\overline{\xi}_{\underline t}\|^r_pdt\right)^{1/r}  \le C_{b,\sigma, T,p,r}(1+\|\xi^0\|_p)\left( \Big(\frac Tn\Big)^{\gamma\wedge \frac 12}+\Big( \int_0^T(\widetilde \varphi(t)-\widetilde \varphi(\underline t))^rdt \Big)^{1/r} \right).
    \end{equation}
\end{corollary}  

We now move to the non-Markovian process $X$.

\subsection{Euler schemes for $ X$}\label{sec_sch_X}
In this section we investigate the convergence of the (continuous time, semi-integrated) Smart-Euler scheme for the path-dependent process $X$, with time step $h= \frac Tn$, which we will denote by $\overline X^h$ or equivalently by $\overline X^n$, or $\overline X$.
Let us notice that, since $\rho=0$, we have $(\widetilde K \star X)_s = \xi_s$ and so the dynamics of $X$ reads as
\[
  X_t = \xi^0 + \int_0^t K(t-s)  \big(b(s,\xi_s) ds +\sigma(s,\xi_s)dW_s \big), \qquad t \in [0,T].
\]
We introduce the time grid $t_k =t^n_k =\frac{kT}{n}, k \in \{0, \dots,n \}$ and three continuous time Euler schemes we are going to study. Recall that $\underline s= t_k$, for $s\!\in \big[ \frac{kT}{n},  \frac{(k+1)T}{n}\big)$, $k \in \{0,\dots, n-1\}$.
\begin{itemize}
	\item [I)][\textbf{Smart Euler}] The first scheme is: 
\begin{equation}\label{eq:EulerXcont}
	\overline X_t 
	:= \xi^0 + \int_0^t K(t-s) \big(b(\underline s,\overline \xi_{\underline s}) ds +\sigma(\underline s,\overline \xi_{\underline s})dW_s \big),
	\end{equation}
	so that, in particular, at the discretization points $t_k, k \in \{0, \dots, n-1\}$, we have $\overline{X}_{0}= \xi^0$ and
	\begin{equation} \label{eq:EulerXdisc1}
	\overline X_{t_{k+1}} 
	= \xi^0 +\sum_{\ell=0}^{k} \left( b(t_{\ell}, \overline \xi_{t_{\ell}})  \int_{t_{\ell}}^{t_{\ell+1}}K(t_{k+1} -s) ds + \sigma(t_{\ell}, \overline{\xi}_{t_{\ell}}) \int_{t_{\ell}}^{t_{\ell+1}} K(t_{k+1} -s)\, dW_s \right). 
	\end{equation}
    {{The delicate part in the simulation will be handling the random terms $\int_{t_\ell}^{t_{\ell +1}} K(t_{k+1} - s) dW_s$, for $k \in \{0, \dots, n-1 \}, \ell \in \{0,\dots,k\}$. This will be the focus of Appendix \ref{app:simu_det}.}}
	\item [II)][\textbf{Projected Euler}] Here we replace the Wiener integral $\int_{t_\ell}^{t_{\ell+1}} K(t_{k+1}-s)dW_s, \ell \in \{0, \dots, k\}$, with its $L^2$-projection on the $\sigma$-algebra generated by $\Delta W_{\ell+1} = (W_{\ell+1}-W_{\ell})$:
	$$ 
	\mathbb{E} \left[ \int_{t_\ell}^{t_{\ell+1}} K(t_{k+1}-s)dW_s \big\vert \sigma\left( \Delta W_{\ell+1}\right) \right] =: C_{\ell,k} \Delta W_{\ell+1}. 
	$$
	So, {{we obtain}}
	$ C_{\ell,k}= \left(\mathbb{E}\left[\int_{t_\ell}^{t_{\ell+1}} K(t_{k+1}-s)dW_s \Delta W_{\ell+1} \right]\right) / \mathbb{E}\left[\Delta W_{\ell+1}^2\right]$,
	which yields 
	$ C_{\ell,k}= \frac{n}{T} \int_{t_\ell}^{t_{\ell+1}} K(t_{k+1}-s)ds $
	and the scheme reads 
	\begin{equation}\label{eq:proj_Euler}
	{  \overline{\overline{X}}_{t_{k+1}}= \xi^0+\sum_{\ell=0}^{k} \left( b(t_{\ell}, \overline{\xi}_{t_{\ell}})  + \frac{n}{T}\sigma(t_{\ell}, \overline{\xi}_{t_{\ell}})  \Delta W_{\ell+1}\right) \int_{t_\ell}^{t_{\ell+1}} K(t_{k+1}-s)ds.}
	\end{equation}
	\item [III)][\textbf{Naive Euler}] Here we freeze the kernel and we consider, for $t \in [0,T]$,
\begin{equation}\label{eq:EulerXcont_freeze}
	\widetilde {\overline X}_t 
	:= \xi^0 + \int_0^t K(t-{\underline s}) \big(b(\underline s,\overline \xi_{\underline s}) ds +\sigma(\underline s,\overline \xi_{\underline s})dW_s \big),
	\end{equation}
	so that, at the discretization points $t_k =t^n_k =\frac{kT}{n}$, $k \in \{0, \dots, n-1\}$, the scheme reads \begin{equation}\label{eq:EulerXdisc2}
	\widetilde {\overline{X}}_{t_{k+1}} =  \xi^0 + \sum_{\ell=0}^{k}K(t_{k+1}-t_\ell) \left( \tfrac Tn  b(t_{\ell}, \overline{ \xi}_{t_{\ell}}  )+\sigma(t_{\ell}, \overline{\xi}_{t_{\ell}}  ) \Delta W_{\ell + 1}\right), \quad \widetilde {\overline{X}}_{0}= \xi^0. 
	\end{equation}
\end{itemize}

We do not discuss the theoretical convergence of the Projected and Naive schemes here but we test them numerically in Section \ref{sec:numerics} by comparing their performance with the one of the Smart-Euler scheme.

{
\begin{remark}
For $\rho\neq0$, the two stages of the algorithm must be distinguished. The Euler approximation of the memory process $\xi$ uses the transformed coefficients $\widetilde b$ and $\widetilde\sigma$ defined in Equation~\eqref{Eq:def_over_b_sigma}. Once $\xi$ has been approximated, the reconstruction schemes for $X$ use the original coefficients evaluated at $\xi$, namely $b(u,e^{-\rho u}x)$ and $\sigma(u,e^{-\rho u}x)$. Thus, the extension to $\rho\neq 0$ is straightforward, but not readily available in closed form for the numerical implementation, which would require additional care.
\end{remark}
}

Notice that our continuous time semi-integrated scheme for $X$ is, of course, combined with the Euler scheme $\overline {\xi}$ for the underlying diffusion process $\xi$, defined in Equation~\eqref{eq:Euler_cont_xi}, and its discrete time ``sub-scheme''. 
Any discrete time scheme can be extended as step-wise constant c\`adl\`ag scheme by setting, here in the case of $X$, $\overline X^{\dag}_t := \overline X_{\underline t}$, $t\!\in [0,T]$. 
We provide error bounds for this step-wise constant scheme in Corollary \ref{cor:Euler_piece_c}. 

\begin{remark}[Complexity of the schemes]\label{rem:complexity}
    \begin{itemize}
        \item[i)] If the objective is to simulate $\overline X^n_{T}= \overline X^n_{t_n}$, the complexity of the scheme is proportional to $n$ whereas the  simulation of its counterpart in a standard Volterra SDE is  $O(n^2)$.
        This is clearly crucial from a numerical viewpoint.

        \item[ii)] Nevertheless, the complexity to simulate the whole path $(\overline X^n_{t_k})_{k=0,\ldots,n}$ is formally $O(n^2)$, seemingly as for standard Volterra SDEs. Let us stress that, for the discrete time scheme in Equation \eqref{eq:EulerXdisc2}, this is only due to the deterministic weights induced by the kernel, the other  terms being computed from $(\overline{\xi}_{t_k})_{k=0,\ldots,n}$, which has a complexity $O(n)$, and an already existing sequence of Brownian increments.     \end{itemize}
\end{remark}

In what follows we focus on the Euler scheme in Equation \eqref{eq:EulerXcont} (and its discrete time counterpart in Equation \eqref{eq:EulerXdisc1}) to keep the length of the paper reasonable. 
We do not comment here practical aspects of the simulation of its  discrete counterpart, which are dealt exhaustively with elsewhere, see, e.g., ~\cite[Practitioner's corner]{jourdain2025convex}. {Nonetheless we provide more details on this in Section~\ref{sec:rel_example}, in the fractional kernel case.}

\begin{theorem}\label{thm:est_eu_sch_X}
    Let $p \ge 1$ and assume that the following conditions are satisfied: 
    \begin{itemize}
        \item[i)] there exists $\beta >1$ such that  $\int_0^T K(t)^{2 \beta} dt < +\infty, \quad \text{for all }  \ T>0$;
        
        \smallskip
        \item[ii)]  the co-kernel $\widetilde  K$ exists and it is (continuous) non-negative and non-increasing;
        
        \smallskip
        \item[iii)]  the functions $b$ and $\sigma$ satisfy the time-space H\"older-Lipschitz assumption in Equation~\eqref{eq:bsigHolLip}.
    \end{itemize}
    Then, for every $\beta>1$, such that {{condition $i)$}} is satisfied, there exists a real constant $C_{K,b,\sigma,T,p,\beta}>0$, such that
    \begin{equation}\label{thm:err_Euler}
        \sup_{t\in [0,T]} \| X_t -\overline X_t\|_p\le C_{K,b,\sigma,T,p,\beta}\big( 1+ \|\xi^0\|_p\big)\left( \Big(\frac Tn\Big)^{\gamma\wedge \frac 12}+\Big( \int_0^T(\widetilde \varphi(s)-\widetilde \varphi(\underline s))^{\frac{2\beta}{\beta-1}}ds \Big)^{\frac{\beta-1}{2\beta}} \right).
    \end{equation}
\end{theorem}
The proof of this result is postponed to Appendix \ref{sec_proof_thm:est_eu_sch_X}.

{
The estimate in Equation~\eqref{thm:err_Euler} yields a conditional rate. More precisely, the overall order is the minimum between the Euler contribution $h^{\gamma\wedge 1/2}$ and the decay rate of
\[
\left(\int_0^T\big(\widetilde\varphi(s)-\widetilde\varphi(\underline s)\big)^{\frac{2\beta}{\beta-1}}ds\right)^{\frac{\beta-1}{2\beta}}.
\]
Consequently, Theorem~\ref{thm:est_eu_sch_X} does not by itself provide an unconditional order $1/2$ for a general kernel.

For example, consider the Gamma co-kernel
\[
\widetilde K(u)=\frac{e^{-\rho u}u^{-\alpha}}{\Gamma(1-\alpha)},\qquad \alpha\in(0,1),
\]
and set $q=2\beta/(\beta-1)$. Since $\widetilde K(u)\le C u^{-\alpha}$ and $\alpha q>1$ when $\alpha>1/2$, we find
\begin{equation}\label{eq:GammaResidual}
\left(\int_0^T\big(\widetilde\varphi(s)-\widetilde\varphi(\underline s)\big)^qds\right)^{1/q}
\le C h^{1-\alpha+1/q}
= C h^{1-\alpha+\frac{\beta-1}{2\beta}}.
\end{equation}
so, as $\beta$ approaches the upper integrability threshold $1/[2(1-\alpha)]$, the exponent in Equation~\eqref{eq:GammaResidual} approaches $1/2$. In the fractional case, Theorem~\ref{thm:partK} uses a sharper weighted estimate from the proof of Theorem~\ref{thm:est_eu_sch_X} and obtains the exact contribution $h^{1/2}$.
}

\begin{remark} \label{cor:Euler_piece_c}
{Under the assumptions of Theorem~\ref{thm:est_eu_sch_X}, one also has, for every $p\ge1$, the following bound for the error of the stepwise constant Euler scheme:}
\[
\sup_{t\in [0,T]} \| X_t -\overline X_{\underline t}\|_p\le C_{K,b,\sigma,T,p,\beta}\big( 1+ \|\xi^0\|_p\big)\left( \Big(\frac Tn\Big)^{\gamma\wedge \frac 12}+\Big( \int_0^T(\widetilde \varphi(s)-\widetilde \varphi(\underline s))^{\frac{2\beta}{\beta-1}}ds \Big)^{\frac{\beta-1}{2\beta}} \right).
\]
The proof is omitted as it is the same as that of Corollary~\ref{cor:L1RateEulercstpm} and relies on the $L^p(\mathbb P)$-path regularity of $\int _0^T\|X_s-X_{\underline s}\|_pds$.
\end{remark}


\section{The case of fractional kernels: a relevant example}\label{sec:rel_example}

{
In this section we specialize the preceding framework to the fractional kernel
\[
K(t)=K_{1,\alpha,0}(t)=:K_{1,\alpha}(t)=\frac{t^{\alpha-1}}{\Gamma(\alpha)},\qquad t\in[0,T].
\]
Motivated by the low path regularity appearing in rough-volatility modelling, we take $\alpha\in(1/2,1)$ and write $\alpha=H+1/2$, with $H\in(0,1/2)$. The kernel then belongs to $L^2(0,T)$. Our equations share this fractional kernel and the resulting $H$-type path regularity with stochastic Volterra equations used in rough-volatility models, see, e.g., \cite{Jaissonetal} for a setting where $\sigma$ is not Lipschitz, or~\cite{GaJuRo202}, where the  volatility follows a dynamics with a Lipschitz $\sigma$. They are, however, structurally different: here the coefficients in the dynamics of $X$ depend on $\xi= I^{1-\alpha}X$ (as it will be clear in a few lines), rather than directly on the Volterra process $X$. The comparison of convergence rates below is therefore a comparison across different classes of equations, not between two schemes applied to the same equation.

Still taking $\rho=0$, we work with
\[
\widetilde K(t)=K_{1,1-\alpha,0}(t)=\frac{t^{-\alpha}}{\Gamma(1-\alpha)},
\qquad
\widetilde\varphi(t)=(\widetilde K\star\mbox{\bf 1})(t)
=\frac{t^{1-\alpha}}{\Gamma(2-\alpha)}.
\]
In particular, $\widetilde\varphi$ is non-decreasing, concave, null at the origin and $(1-\alpha)$-H\"older continuous. Condition a) of Theorem~\ref{thm:existence} and conditions i)--ii) of Theorem~\ref{thm:est_eu_sch_X} hold for every
\[
\beta\in\left(1,\frac{1}{2(1-\alpha)}\right).
\]
Moreover,
\[
\xi_t=(\widetilde K\star X)_t
=\frac{1}{\Gamma(1-\alpha)}\int_0^t(t-s)^{-\alpha}X_sds
=I^{1-\alpha}(X)_t,
\]
where $I^{1-\alpha}$ denotes the Riemann--Liouville fractional integral from Definition~\ref{Def_st_int}. Hence Equations~\eqref{eq:pathdepVolt} and~\eqref{eq:Wdiff} become, respectively,
\begin{equation}\label{eq:pathdepVoltrough}
X_t=\xi^0+\int_0^t\frac{(t-s)^{\alpha-1}}{\Gamma(\alpha)}b\big(s,I^{1-\alpha}(X)_s\big)ds
+\int_0^t\frac{(t-s)^{\alpha-1}}{\Gamma(\alpha)}\sigma\big(s,I^{1-\alpha}(X)_s\big)dW_s,
\end{equation}
and
\[
\xi_t=\xi^0\frac{t^{1-\alpha}}{\Gamma(2-\alpha)}+\int_0^t b(s,\xi_s)ds+\int_0^t\sigma(s,\xi_s)dW_s.
\]
The second identity is the fractional specialization of Theorem~\ref{thm:existence}. At the level of fractional calculus, Lemma~\ref{lem:I_Ito} below justifies the composition rule used for the stochastic-convolution term.
}

We start by introducing the following It\^o process, as a special example of the integral in Equation~\eqref{eq:conv_X_Ito}:
$$
{(K_{1,\alpha} \stackrel{W}{\star} Y)}_t = \frac{1}{\Gamma(\alpha)} \int_0^t (t - s )^{\alpha - 1} Y_s dW_s, \qquad t \in [0,T],
$$
and we study its existence and path-regularity.

\subsection{Well-posedness results}

{The following corollary is a specialization of Lemma~\ref{lem:regularityX}.}
The corresponding proof is presented in Appendix~\ref{sec_proof_lem:continuous_version}.

{
\begin{corollary}\label{lem:continuous_version}
Let $\xi^0\in L^2(\mathbb P)$ and let $Y$ be progressively measurable with $\sup_{t\in[0,T]}\|Y_t\|_2<\infty$. Then the stochastic convolution
\[
X_t=\xi^0+\frac{1}{\Gamma(\alpha)}\int_0^t(t-s)^{\alpha-1}Y_s\,dW_s,
\qquad t\in[0,T],
\]
is well defined as an adapted $L^2(\mathbb P)$-process. If, in addition, for some
\[
p>\max\left\{\frac{1}{\alpha-1/2},2\right\}
\]
one has $\xi^0\in L^p(\mathbb P)$ and $\sup_{t\in[0,T]}\|Y_t\|_p<\infty$, then $X$ admits a pathwise continuous modification whose trajectories are locally $a$-H\"older continuous for every
\[
a\in\left(0,\alpha-\frac12-\frac1p\right).
\]
\end{corollary}
}

The degree of integrability required  for $Y$ in order to apply Kolmogorov criterion should compensate the irregularity of the stochastic integral $X$, which depends on $\alpha$. Recalling that $\alpha =H+\frac12$, we get $	p > \max \left\{\frac{1}{H} , 2 \right	\}$. In particular, for the well known case of interest in the rough volatility literature, corresponding to $H=0.1$, we find $p >10$.

We now consider a key lemma, whose proof is provided in Appendix \ref{app:Lem_Ito_frac}, that is crucial when dealing with stochastic Volterra integrals of convolution type with fractional kernel. 

{
\begin{lemma}\label{lem:I_Ito}
Let $\alpha>1/2$, let $\beta>0$, and let $Y$ be progressively measurable with $\sup_{u\in[0,T]}\|Y_u\|_2<\infty$. Then, for every $t\in[0,T]$,
\begin{equation}\label{eq:Fubini}
I^\beta\left(\int_0^\cdot\frac{(\cdot-u)^{\alpha-1}}{\Gamma(\alpha)}Y_u\,dW_u\right)(t)
=\int_0^t\frac{(t-u)^{\alpha+\beta-1}}{\Gamma(\alpha+\beta)}Y_u\,dW_u
={(K_{1,\alpha+\beta}\stackrel{W}{\star}Y)}_t,
\qquad \mathbb P\text{-a.s.}
\end{equation}
\end{lemma}
}

By defining 
	\begin{equation}
		I^{\alpha, W} Y :=  \int_0^\cdot \frac{ (\cdot - u )^{\alpha - 1}}{\Gamma(\alpha)} Y_u dW_u,
	\end{equation}
	Equation~\eqref{eq:Fubini} reads
	\begin{equation}
		I^\beta \circ  I^{\alpha, W} =I^{\alpha + \beta,W},
	\end{equation}
	{which is the stochastic analogue of the deterministic composition rule} (see Equation~\eqref{eq:I_prop} in Appendix \ref{sec:fractional_op}).

{
We now turn to the Smart-Euler scheme. The coefficients in the dynamics of $X$ depend on the Markov process $\xi=I^{1-\alpha}X$. Consequently, $\xi$ can be simulated at the cost of a standard diffusion, and $X$ can be reconstructed at a fixed terminal time with $O(n)$ operations. Reconstructing the complete path of $X$ still requires $O(n^2)$ operations because of the deterministic convolution weights, as explained in Remark~\ref{rem:complexity}.
}

\subsection{Numerical analysis: Smart Euler scheme}\label{sect_Euler_frac}

{
In the fractional case, the weighted residual term appearing in Equation~\eqref{thm:err_Euler} in Theorem~\ref{thm:est_eu_sch_X} is of order $h$ before taking the square root, and therefore contributes $h^{1/2}$. The resulting overall rate is $h^{\gamma\wedge1/2}$ and equals $h^{1/2}$ whenever the coefficients are at least $1/2$-H\"older continuous in time. This result is specific to the present bi-dimensional Volterra SDE, whose second component is Markovian. 
The proof is given in Appendix~\ref{sec_proof_thm:partK}.

\begin{theorem}\label{thm:partK}
Let $K=K_{1,\alpha}$ with $\alpha\in(1/2,1)$, let $p\ge1$, and assume $\xi^0\in L^p(\mathbb P)$. Suppose that $b$ and $\sigma$ satisfy the time-space H\"older--Lipschitz condition in Equation~\eqref{eq:bsigHolLip}, with H\"older parameter $\gamma\in(0,1]$. Then
\begin{equation}\label{Eq_strong_rate_X_frac}
\sup_{t\in[0,T]}\|X_t-\overline X_t\|_p
\le C_{\alpha,b,\sigma,T,p}(1+\|\xi^0\|_p)
\left(\frac{T}{n}\right)^{\gamma\wedge1/2},
\end{equation}
for some constant $C_{\alpha,b,\sigma,T,p}>0$.
\end{theorem}

\begin{remark}
Theorem~\ref{thm:partK} provides an upper strong-error bound. We do not establish a matching lower bound and therefore do not prove theoretically that the exponent $1/2$ is sharp. Numerical evidence concerning the attained rate is discussed separately in Section~\ref{sec:numerics}.
\end{remark}
}


\section{Numerical illustration} \label{sec:numerics}

{{In this section we provide numerical evidence for the $1/2$ strong-error behaviour of the Smart-Euler scheme predicted by Theorem~\ref{thm:partK} when the coefficients are sufficiently regular in time. We then compare it empirically with the Projected- and Naive-Euler alternatives introduced in Section~\ref{sec_sch_X}. Since no theoretical convergence rate is established here for these two alternative schemes, the corresponding numerical comparison is intended as a diagnostic rather than as a sharp-rate statement.}}

In what follows we consider the following Volterra-like SDE (which is reminiscent of the one in~\cite{GaJuRo202}) for the non-Markovian process:
\begin{align}\label{eq:def_Z}
	X_t= \xi^0 + \int_0^t \frac{(t-s)^{\alpha-1}}{\Gamma(\alpha)}  (\mu - \lambda \xi_s)\, ds + \eta \int_0^t  \frac{(t-s)^{\alpha-1}}{\Gamma(\alpha)}\,\sigma(\xi_s) \ dW_s, \quad t \in [0,T]
\end{align}
with $\mu, \lambda, \eta >0$ and $\xi_0\!\in L^1(\mathbb{P})$, while the Markovian process $(\xi_t)_{t\in{[0,T]}}$ satisfies
\begin{align}\label{eq:def_Y}
	\xi_t = \xi^0 \frac{t^{1-\alpha}}{\Gamma(2-\alpha)}+\int_0^t  (\mu - \lambda \xi_s) ds + \eta \int_0^t \sigma(\xi_s) dW_s, \quad t \in [0,T].
\end{align}

{{Before dealing with simulations, we state a lemma on the asymptotic behaviour of $X$ and $\xi$.}} For its proof we refer the interested reader to Appendix \ref{sec_proof_lem:asympt_average}.

\begin{lemma}\label{lem:asympt_average}
	Given the stochastic processes $X$ and $\xi$ defined, respectively, in Equations \eqref{eq:def_Z} and \eqref{eq:def_Y}, we have
	\begin{align}
	 \mathbb{E}\, [X_t] \to 0, \quad \mbox{ and } \quad \mathbb{E}\, [\xi_t] \to \frac{\mu}{\lambda}\quad \quad \mbox{ as }\quad t\to +\infty.
	\end{align}
\end{lemma}

\begin{figure}[htbp]
	\begin{minipage}[b]{0.5\linewidth}
		\centering
		\includegraphics[width=\linewidth]{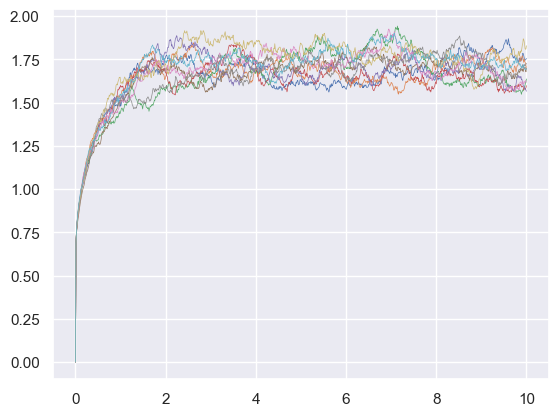}
	\end{minipage}
	\begin{minipage}[b]{0.5\linewidth}
		\centering
		\includegraphics[width=\linewidth]{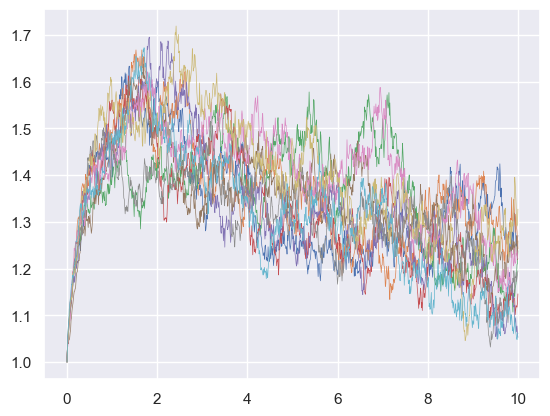}
	\end{minipage}
	\caption{{{Simulation of $10$ trajectories of $\xi$ (left) and $X$ (right), for $\alpha=0.9$, $\sigma= \sigma_1$, $T=10$, $\xi^0=1$ and $\eta=0.01$.}}}
	\label{fig:H=0.4}
\end{figure}

\begin{figure}[h]
	\begin{minipage}[b]{0.5\linewidth}
		\centering
		\includegraphics[width=\linewidth]{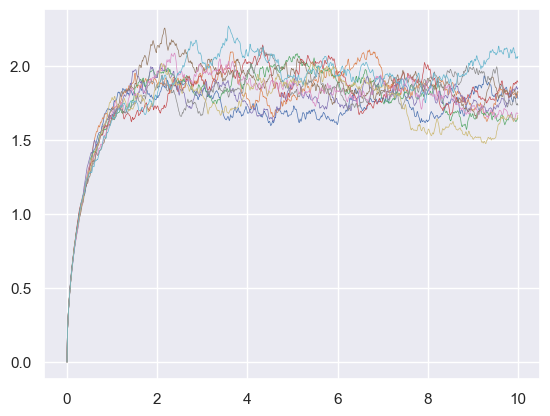}
	\end{minipage}
	\begin{minipage}[b]{0.5\linewidth}
		\centering
		\includegraphics[width=\linewidth]{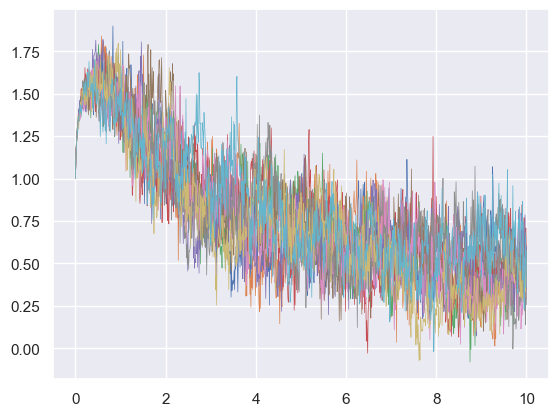}
	\end{minipage}
	\caption{{{Simulation of $10$ trajectories of $\xi$ (left) and $X$ (right), for $\alpha=0.6$, $\sigma=\sigma_1$, $T=10$, $\xi^0=1$ and $\eta=0.01$.}}}
	\label{fig:H=0.1}
\end{figure}

We consider the two settings in Table \ref{twomodels}, {{with $a,c,\mu,\lambda,\eta>0$, $C>1$ and $b\in\mathbb R$.
\begin{table}[h]
\begin{center}
	\begin{tabular}{ |c|c|c|} 
		\hline
		Setting &  Diffusion coefficient \\
		\hline 
		 $\sigma$ unbounded  & $\sigma_1 (x) =  \sqrt{a(x-b)^2+c}$\\ 
		$\sigma$ bounded  & $\sigma_2(x) = \sqrt{C+\tanh(x)}$\\ 
		\hline
	\end{tabular}
\caption{Two settings for the volatility coefficient $\sigma$.}
\label{twomodels}
\end{center}
\end{table}
We use $\mu=2$, $\lambda=1.2$, $a=0.384$, $b=0.095$, $c=0.0025$ and $C=1.2$, values inspired by the numerical specifications used for the quadratic rough-Heston model in \cite{gatheral2020quadratic}. We perform two numerical experiments. For the path illustrations in Figures~\ref{fig:H=0.4}--\ref{fig:H=0.1}, we take $T=10$, $n=1000$, $\xi^0=1$ and $\eta=0.01$, and use the unbounded coefficient $\sigma_1$. For the convergence experiments later below, we take $T=1$, $\xi^0=0$ and $\eta=1$. The convergence plots focus on the unbounded coefficient $\sigma_1(x)=\sqrt{a(x-b)^2+c}$. The choice $\xi^0=0$ removes the deterministic initial-memory forcing from the convergence experiment but does not change the theoretical convergence exponent in Theorem~\ref{thm:partK}.}}

{{The Markovian memory process $\xi$ and the non-Markovian process $X$ are simulated as described in Sections~\ref{sec_sch_xi} and~\ref{sec_sch_X}. Figures~\ref{fig:H=0.4}--\ref{fig:H=0.1} illustrate the contrast between their path regularities for $\alpha=0.9$ ($H=0.4$) and $\alpha=0.6$ ($H=0.1$). In these path illustrations $X_0=\xi^0=1$, whereas the associated memory process starts from $\xi_0=0$ and it is immediately affected by the deterministic memory term in Equation~\eqref{eq:def_Y}.}}

\subsection{Numerical convergence of the Smart-Euler scheme}

We now numerically verify the theoretical results on the convergence of the continuous time Smart-Euler scheme introduced in Section \ref{sect_Euler_frac}. We show that the $L^2$ strong rate of convergence, $\sup_{t \in [0,T]}\mathbb{E}[|X_t-\overline{X}_t|^2]^{\frac12}$,  is $\frac{1}{2}$ as predicted.
The theoretical error in Equation \eqref{Eq_strong_rate_X_frac} is computed with respect to the exact solution, $X$, which is not available in closed-form here. Therefore, it is necessary to find an alternative method to show the convergence of the scheme.
To this purpose, we set $t^n_k=kT/n$, $k=0,\dots, n$ (notice that $t_{2k}^{2n}=t^n_k$) and we compare the numerical approximation  schemes $(\overline{X}^{n}_{t_k})_{k \in \{0, ..., n\}}$ and $(\overline{X}^{L\cdot n}_{L\cdot t_k})_{k \in \{0, ..., n\}}$, where $L$ is an integer, small with respect to $n$ (we take $L= 2$ here) and the superscripts $n$ and $L \cdot n$ are used to emphasise the grid dimension.
More precisely, we empirically show that there exists $M$ such that for every $k \in \{ 0, \dots, n\}$ and for sufficiently many values of $n$:
\begin{align}
\left\|\overline{X}^n_{t^n_k}-\overline{X}^{2n}_{t^{2n}_{{2k}}}\right\|_2 \leq M \sqrt{\frac{T}{n}} =: M_T\sqrt{\frac{1}{n}}. \label{error2N} 
\end{align}
{{This comparison avoids declaring a numerical simulation on a finite and ``very fine'' grid to be exact. If the estimate in \eqref{error2N} holds uniformly along all successive dyadic refinements, then the usual telescoping argument yields the same $n^{-1/2}$ bound with respect to the limiting process. In the numerical study below we therefore study this dyadic differences as an empirical rate diagnostic; a finite collection of Monte Carlo simulations is not, by itself, a proof of the asymptotic rate.}}

\begin{remark}
{{Inequality in Equation \eqref{error2N} is also necessary to numerically show the prescribed convergence: indeed, if $\gamma \ge 1/2$, Equation \eqref{Eq_strong_rate_X_frac} implies that for some $M>0$ we have $\max_{k \in \{0, ..., n\}} \mathbb{E}\left[|X_{t^n_k} - \overline{X}^{n}_{t^n_k}|^2\right] \leq \frac{M}{n}$ and $ \max_{k \in \{0, ..., n\}} \mathbb{E}\left[|X_{t^{2n}_{2k}} - \overline{X}^{2n}_{t^{2n}_{2k}}|^2\right] \leq \frac{M}{2n}$.
Therefore:
$
\max_{k \in \{0, ..., n\}} \left(\mathbb{E}\left[|\overline{X}^{n}_{t^n_k} - \overline{X}^{2n}_{t^{2n}_{2k}}|^2\right]\right)^{1/2} \leq \sqrt{\frac{M}{n}}+\sqrt{\frac{M}{2n}}=\left(1+\frac{1}{\sqrt{2}}\right)\sqrt{\frac{M}{n}}.
$
}}
\end{remark}

\begin{figure}[htbp]
  \centering
  \begin{subfigure}[b]{0.45\textwidth}
    \centering
    \includegraphics[width=\textwidth]{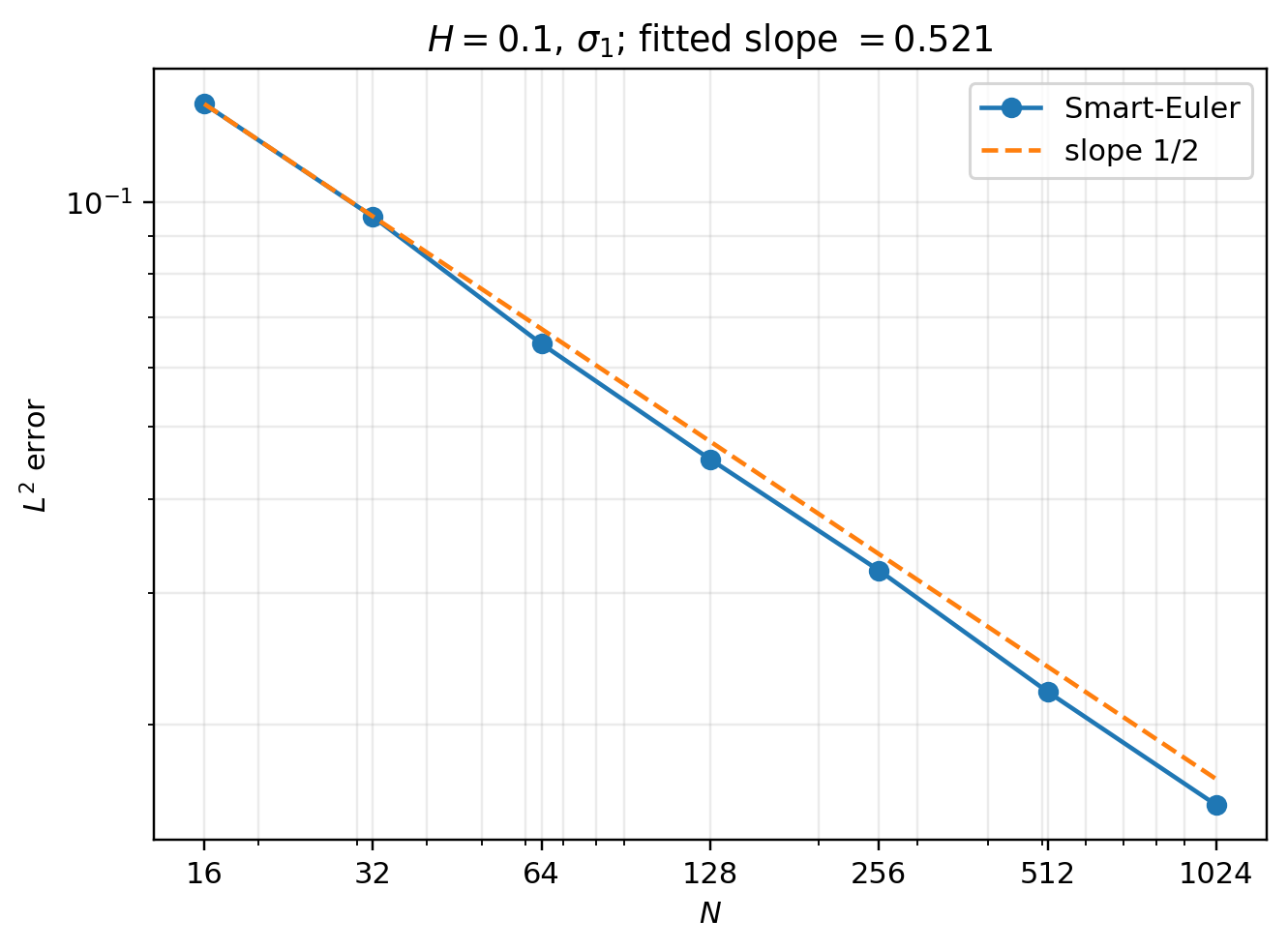}
    \caption{{{$H=0.1$, $\sigma=\sigma_1$.}}}
  \end{subfigure}
  \hfill
  \begin{subfigure}[b]{0.45\textwidth}
    \centering
    \includegraphics[width=\textwidth]{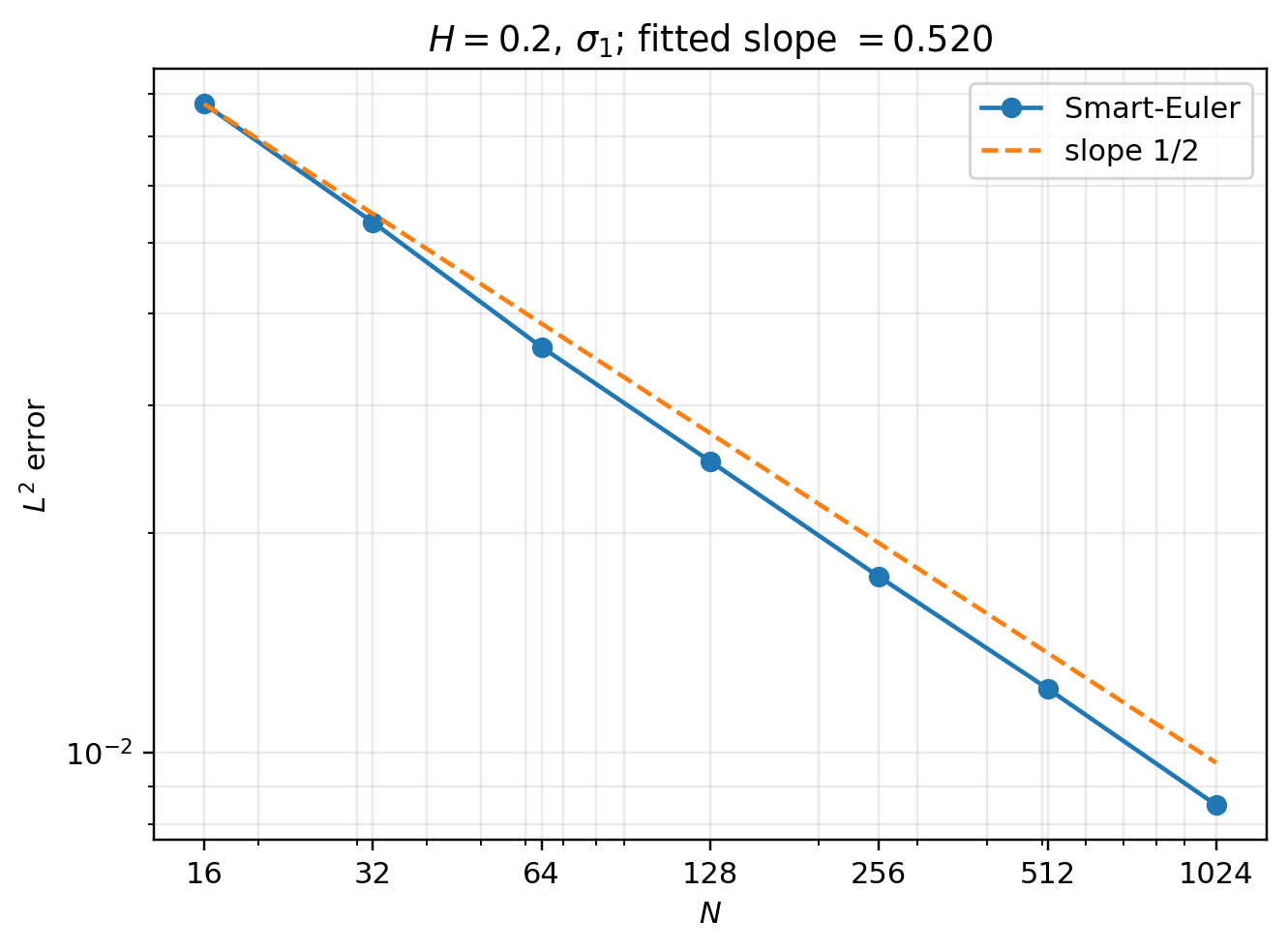}
    \caption{{{$H=0.2$, $\sigma=\sigma_1$.}}}
  \end{subfigure}
  \vspace{1em}
  \begin{subfigure}[b]{0.45\textwidth}
    \centering
    \includegraphics[width=\textwidth]{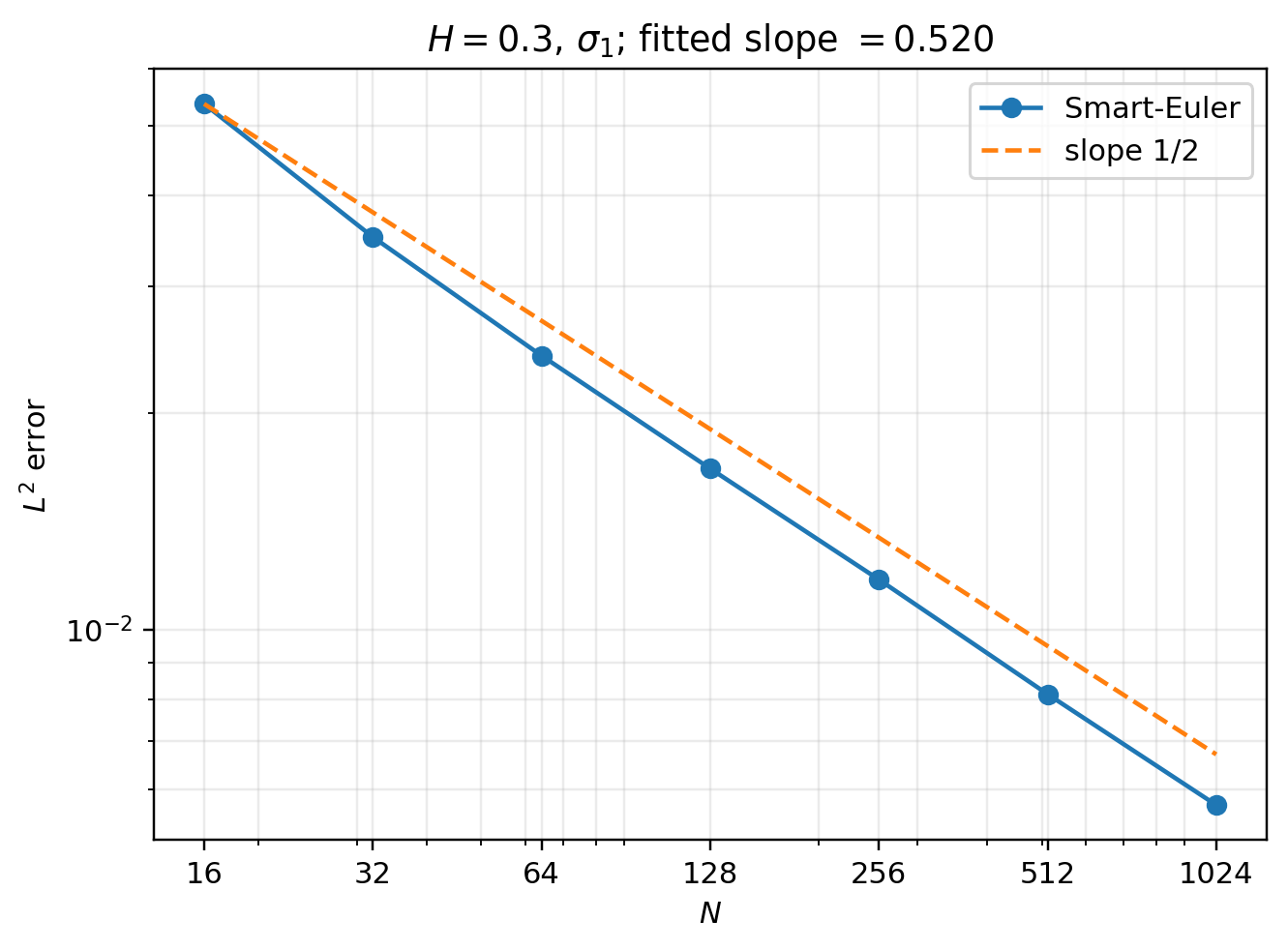}
    \caption{{{$H=0.3$, $\sigma=\sigma_1$.}}}
  \end{subfigure}
  \hfill
  \begin{subfigure}[b]{0.45\textwidth}
    \centering
    \includegraphics[width=\textwidth]{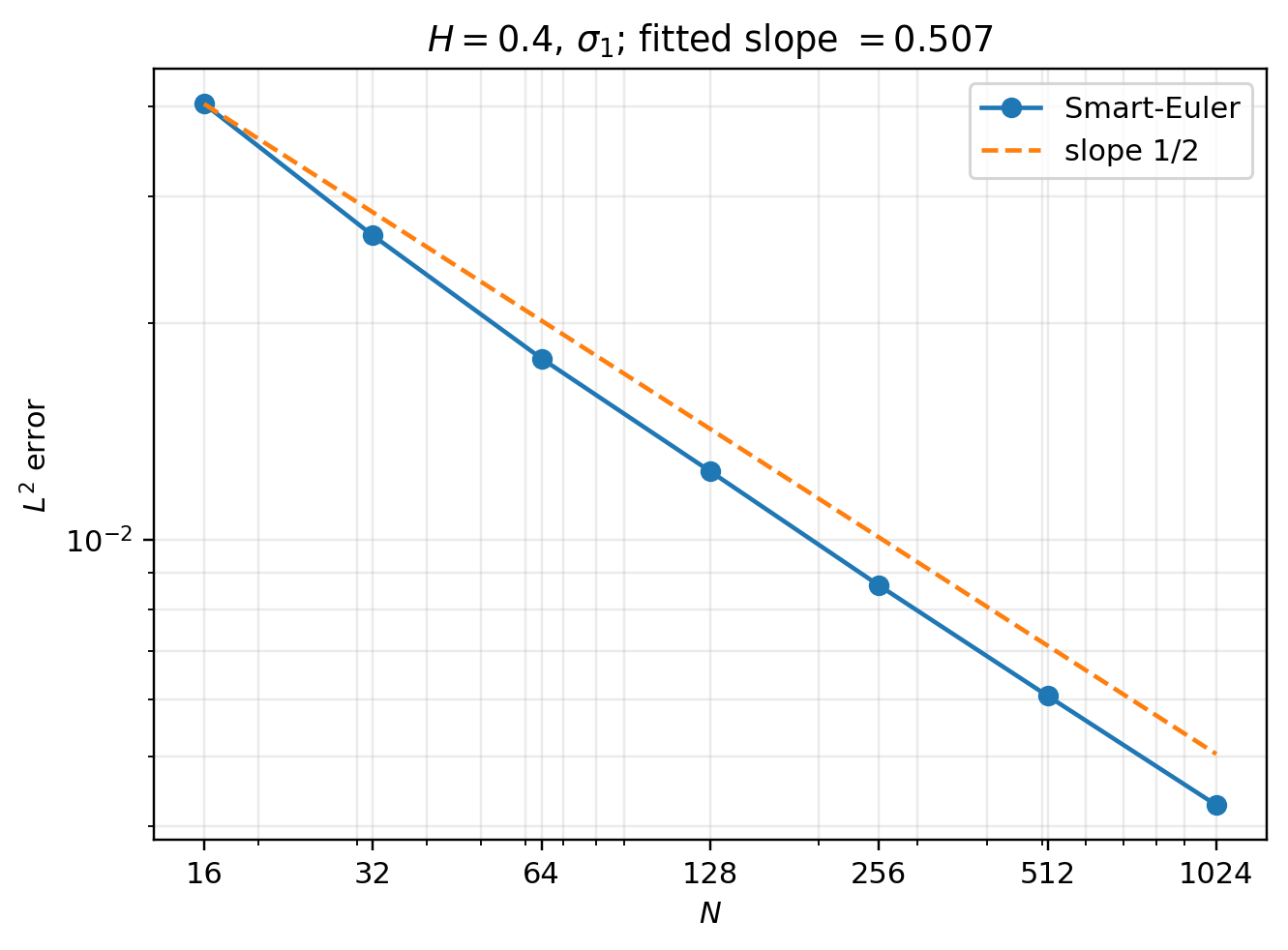}
    \caption{{{$H=0.4$, $\sigma=\sigma_1$.}}}
  \end{subfigure}
  \caption{{{Smart-Euler: dyadic maximum $L^2$ error for $N=16,\ldots,1024$, based on $30{,}000$ Monte Carlo paths. The dashed line has slope $1/2$; the fitted slopes displayed in the panels use the three finest grids.}}}
  \label{fig:H_01_max}
\end{figure}

{{In Figure~\ref{fig:H_01_max} we plot
\[
\max_{k\in\{0,\ldots,N\}}\left\|\overline X^N_{t_k^N}-\overline X^{2N}_{t_{2k}^{2N}}\right\|_2
\]
for $H\in\{0.1,0.2,0.3,0.4\}$, with $\eta=1$, $\xi^0=0$ and the unbounded coefficient $\sigma=\sigma_1$. The experiment uses $30{,}000$ Monte Carlo paths and extends the grid up to $N=1024$. On the three finest grids the fitted exponents are all close to $1/2$, in agreement with Theorem~\ref{thm:partK}.}}

\begin{figure}[htbp]
    \centering
    \begin{subfigure}[b]{0.45\textwidth}
        \centering
        \includegraphics[width=\textwidth]{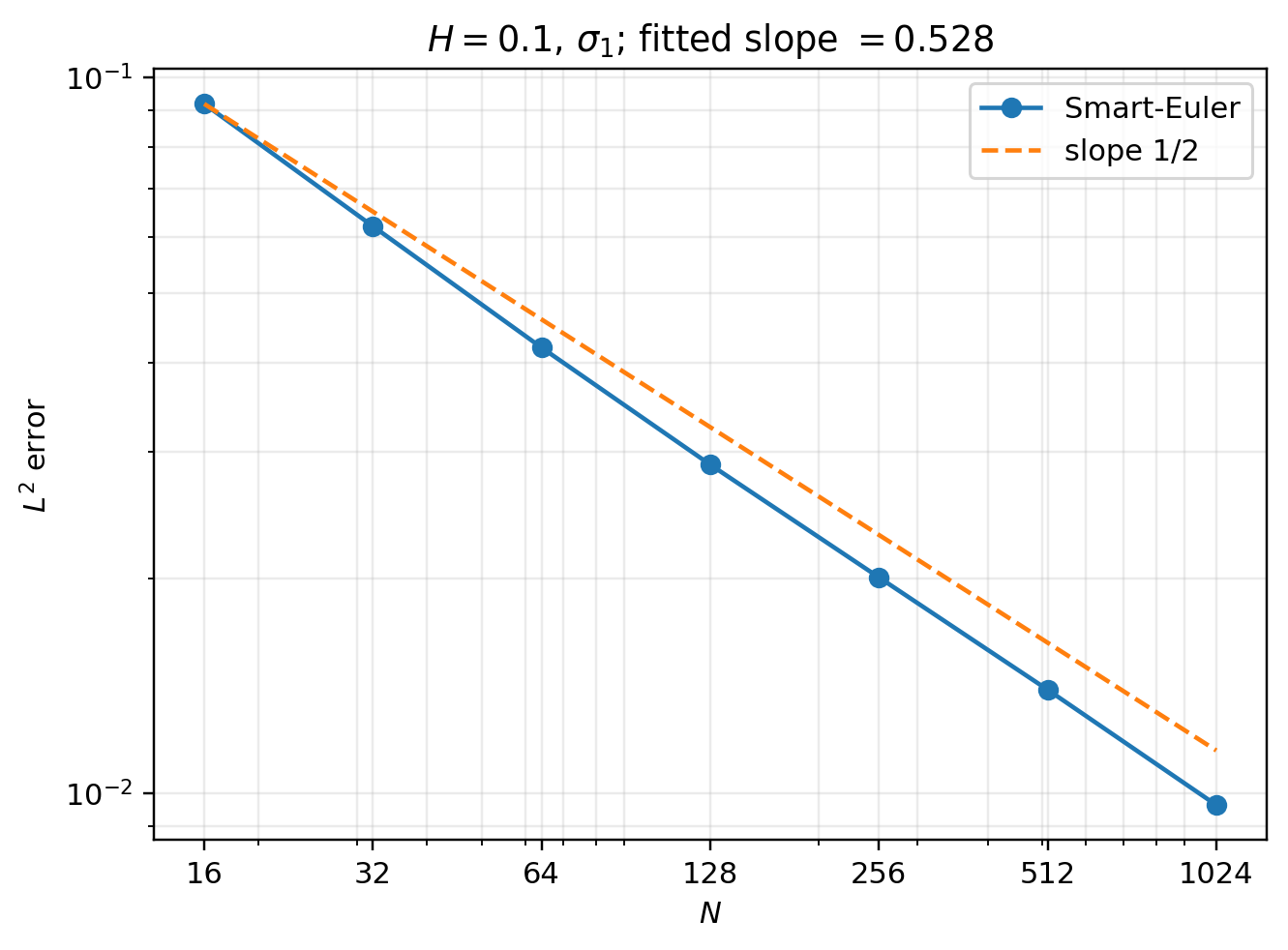}
        \caption{{{$H=0.1$, $\sigma=\sigma_1$.}}}
    \end{subfigure}
    \hfill
    \begin{subfigure}[b]{0.45\textwidth}
        \centering
        \includegraphics[width=\textwidth]{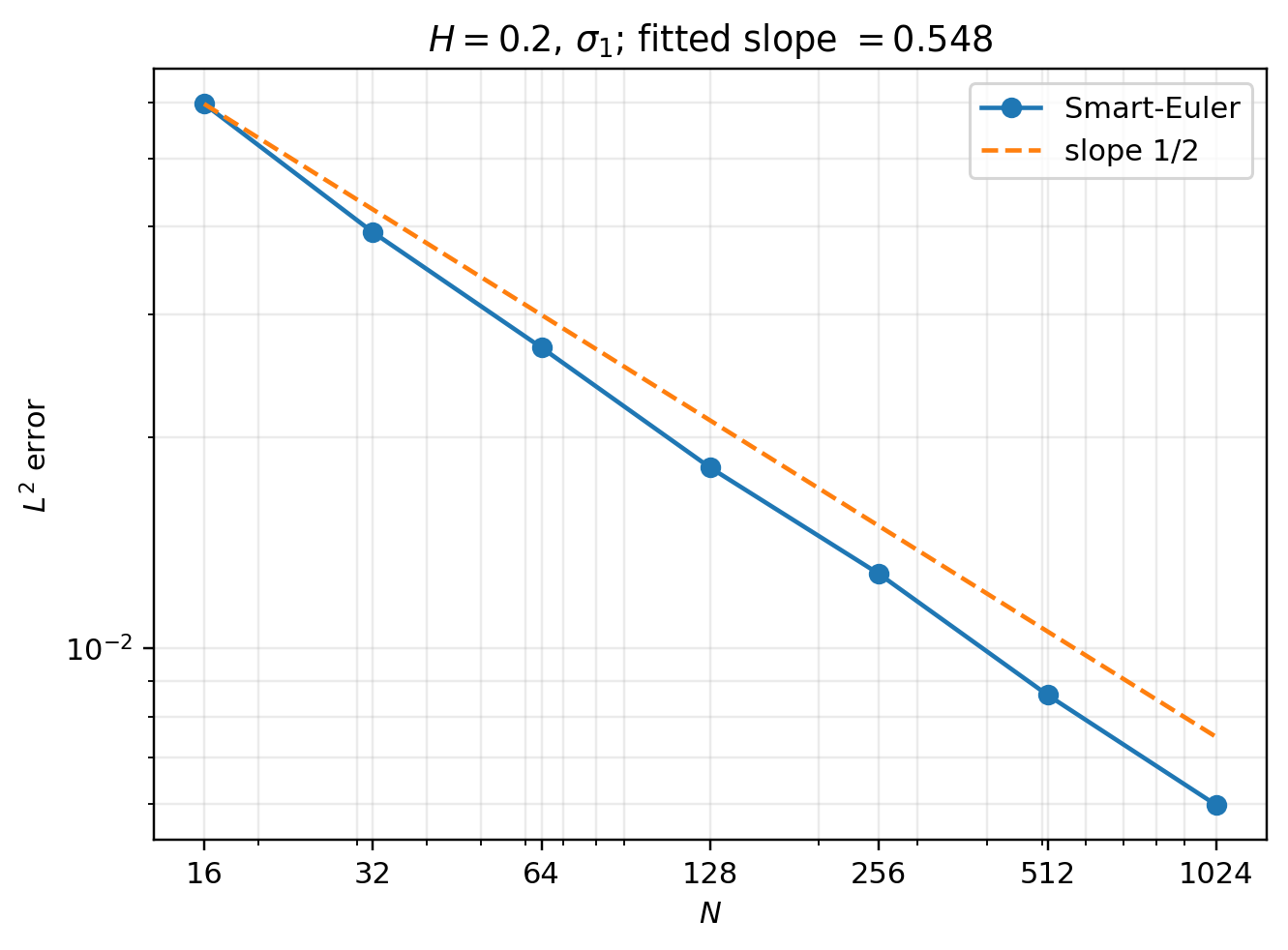}
        \caption{{{$H=0.2$, $\sigma=\sigma_1$.}}}
    \end{subfigure}
    \vfill
    \begin{subfigure}[b]{0.45\textwidth}
        \centering
        \includegraphics[width=\textwidth]{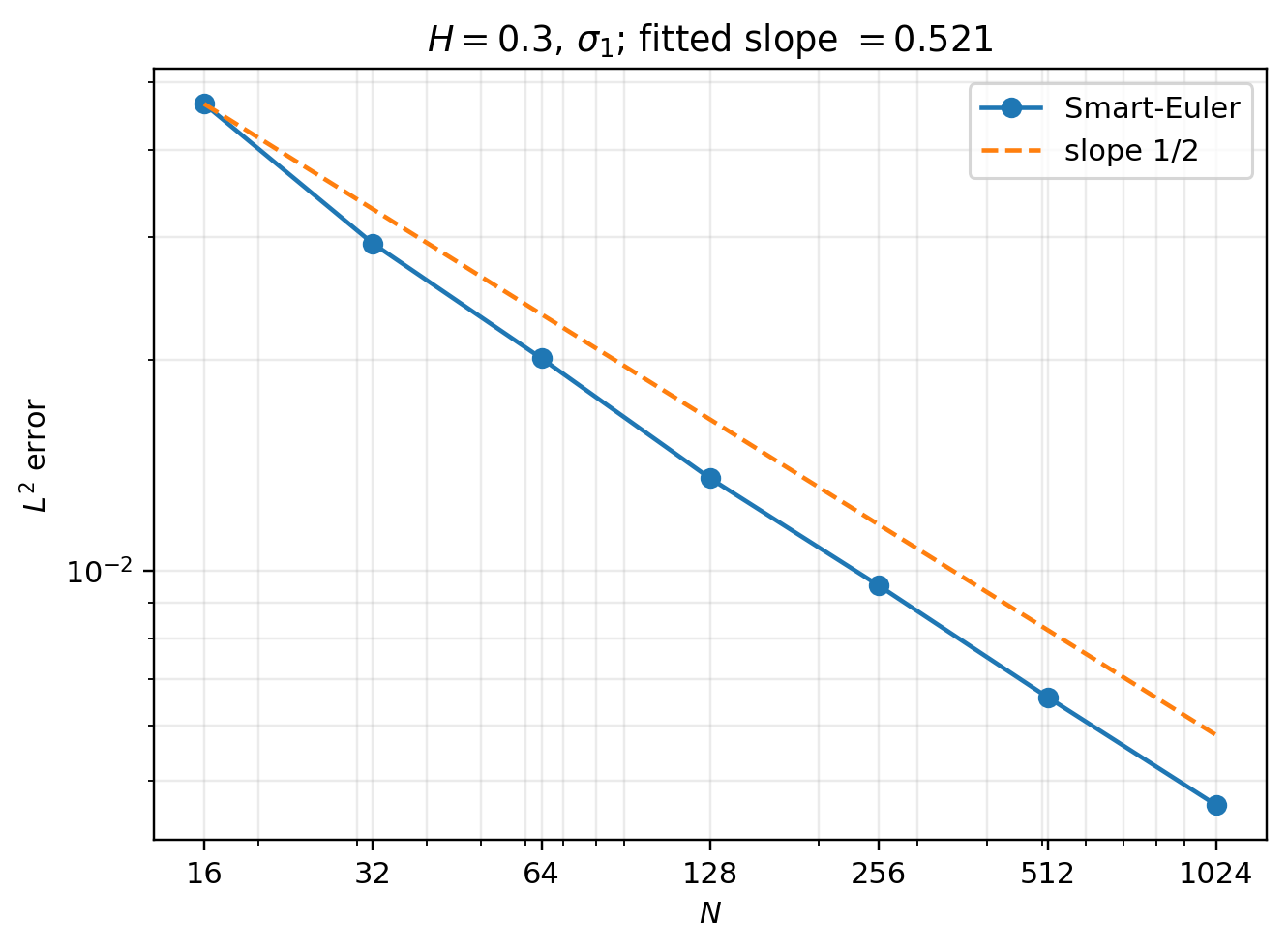}
        \caption{{{$H=0.3$, $\sigma=\sigma_1$.}}}
    \end{subfigure}
    \hfill
    \begin{subfigure}[b]{0.45\textwidth}
        \centering
        \includegraphics[width=\textwidth]{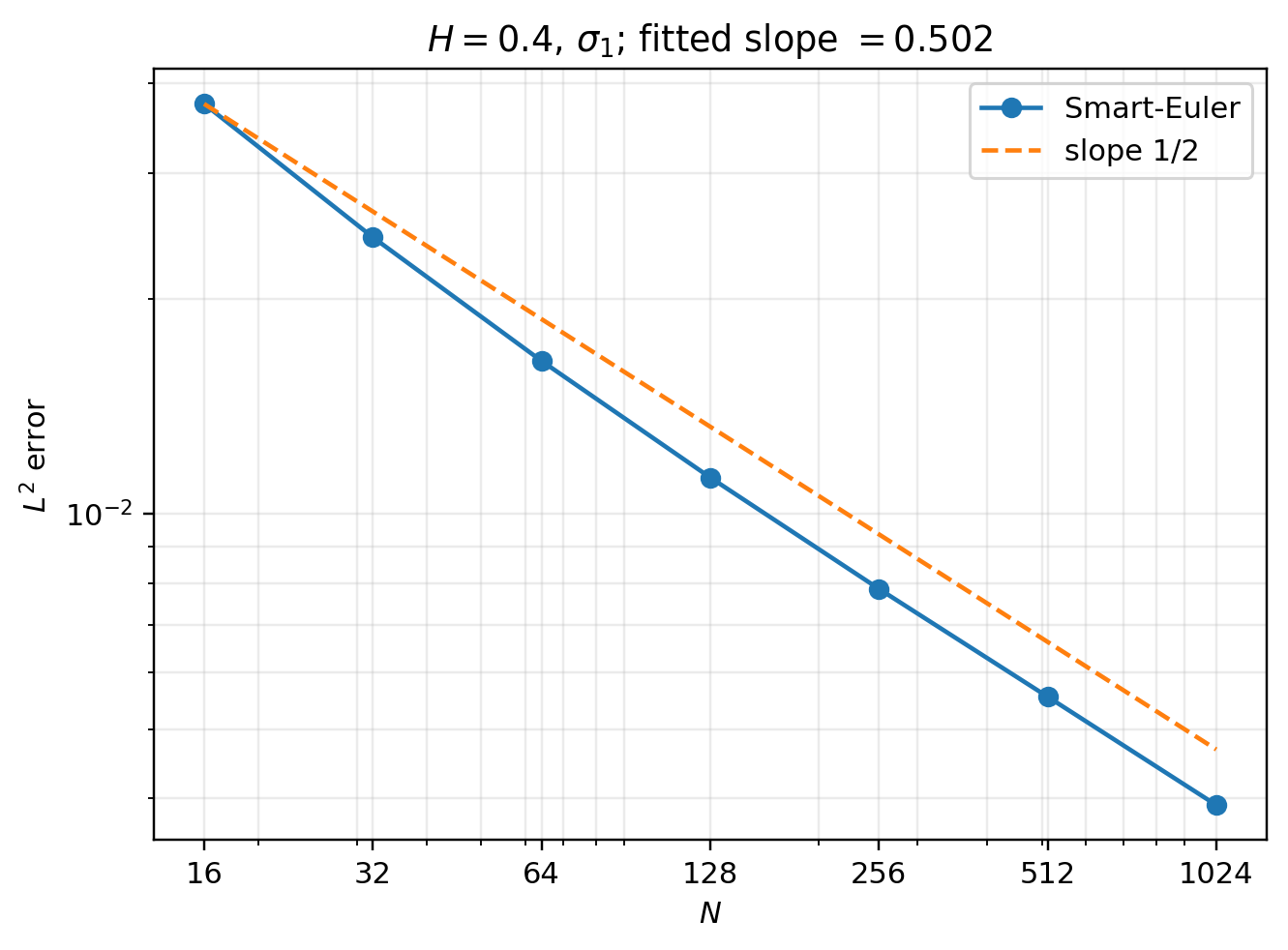}
        \caption{{{$H=0.4$, $\sigma=\sigma_1$.}}}
    \end{subfigure}
    \caption{{{Smart-Euler: dyadic endpoint error $\|\overline X_T^N-\overline X_T^{2N}\|_2$ for $N=16,\ldots,1024$, based on $30{,}000$ Monte Carlo paths. The dashed line has slope $1/2$.}}}
    \label{fig:H_04_endPoint}
\end{figure}

{{Figure~\ref{fig:H_04_endPoint} reports the corresponding endpoint errors. Their fitted slopes are again close to $1/2$, confirming that the behaviour seen for the pathwise maximum is not driven by the maximisation over the grid.}}

\subsection{Numerical convergence  for alternative schemes}\label{sec:alternative_Euler}

In this subsection we numerically study the convergence rate when applying the alternative schemes presented in Section \ref{sec_sch_X}. We recall below the three Euler schemes for the reader's ease, when the drift and the diffusion coefficients do not depend on time: for $k \in \{0, \dots, n-1\}$ and $\Delta W_{\ell} = W_{\ell+1} - W_\ell$
\begin{itemize}
	\item[I)][\textbf{Smart-Euler}] 
	\begin{equation} \label{eq:EulerXdisc_bis}
	\overline X_{t_{k+1}} ^n
	= \xi^0 +\sum_{\ell=0}^{k} \left( b(\overline{\xi}_{t_{\ell}}^n)  \int_{t_{\ell}}^{t_{\ell+1}}K(t_{k+1} -s) ds + {{\eta}} \sigma( \overline{\xi}_{t_{\ell}}^n) \int_{t_{\ell}}^{t_{\ell+1}} K(t_{k+1} -s)\, dW_s \right), {\overline{X}}_{0}^n= \xi^0.
	\end{equation}
	\item[II)][\textbf{Projected-Euler}] \begin{equation}\label{eq:proj_Euler_num}
	{{\overline{\overline{X}}_{t_{k+1}}^n= \xi^0+\sum_{\ell=0}^{k} \int_{t_\ell}^{t_{\ell+1}} K(t_{k+1}-s)ds \left( b(\overline{\xi}_{t_{\ell}}^n)  +{{\eta}}\frac{n}{T} \sigma( \overline{\xi}_{t_{\ell}}^n)  \Delta W_{\ell+1}\right)  , {\overline{\overline X}}_{0}^n= \xi^0.}}
	\end{equation}

	\item [III)][\textbf{Naive-Euler}] 
	\begin{equation*}
	\widetilde {\overline{X}}_{t_{k+1}}^n =  \xi^0 + \sum_{\ell=0}^{k}K(t_{k+1}-t_\ell) \left( \tfrac Tn  b( \overline{ \xi}_{t_{\ell}}^n  )+{{\eta}}\sigma(\overline{\xi}_{t_{\ell}}^n  ) \Delta W_{\ell + 1}\right), \quad \widetilde {\overline{X}}_{0}^n= \xi^0. 
	\end{equation*}
\end{itemize}

\begin{figure}[htbp]
    \centering
    \begin{subfigure}[b]{0.45\textwidth}
        \centering
        \includegraphics[width=\textwidth]{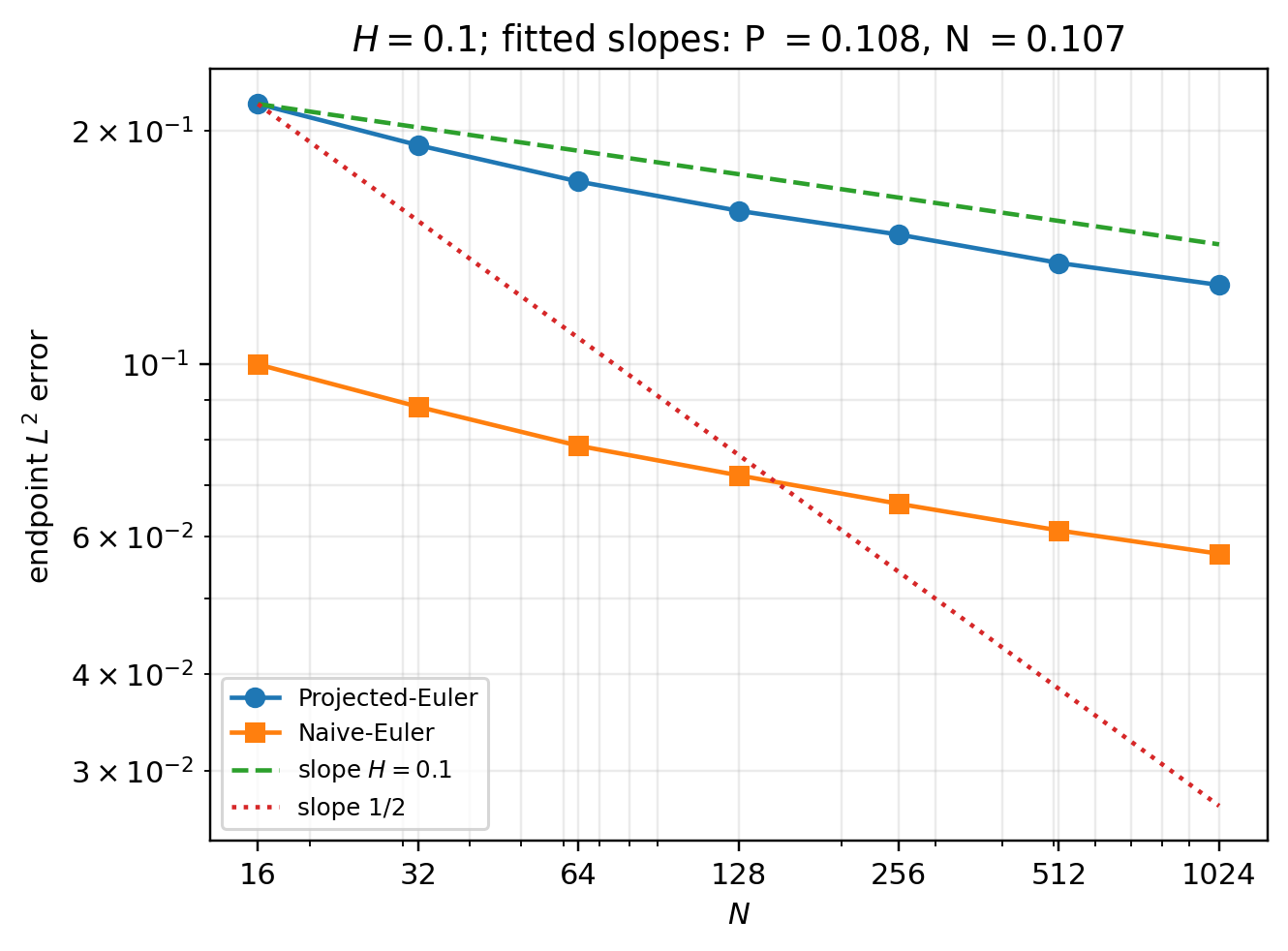}
        \caption{{{$H=0.1$, $\sigma=\sigma_1$.}}}
    \end{subfigure}
    \hfill
    \begin{subfigure}[b]{0.45\textwidth}
        \centering
        \includegraphics[width=\textwidth]{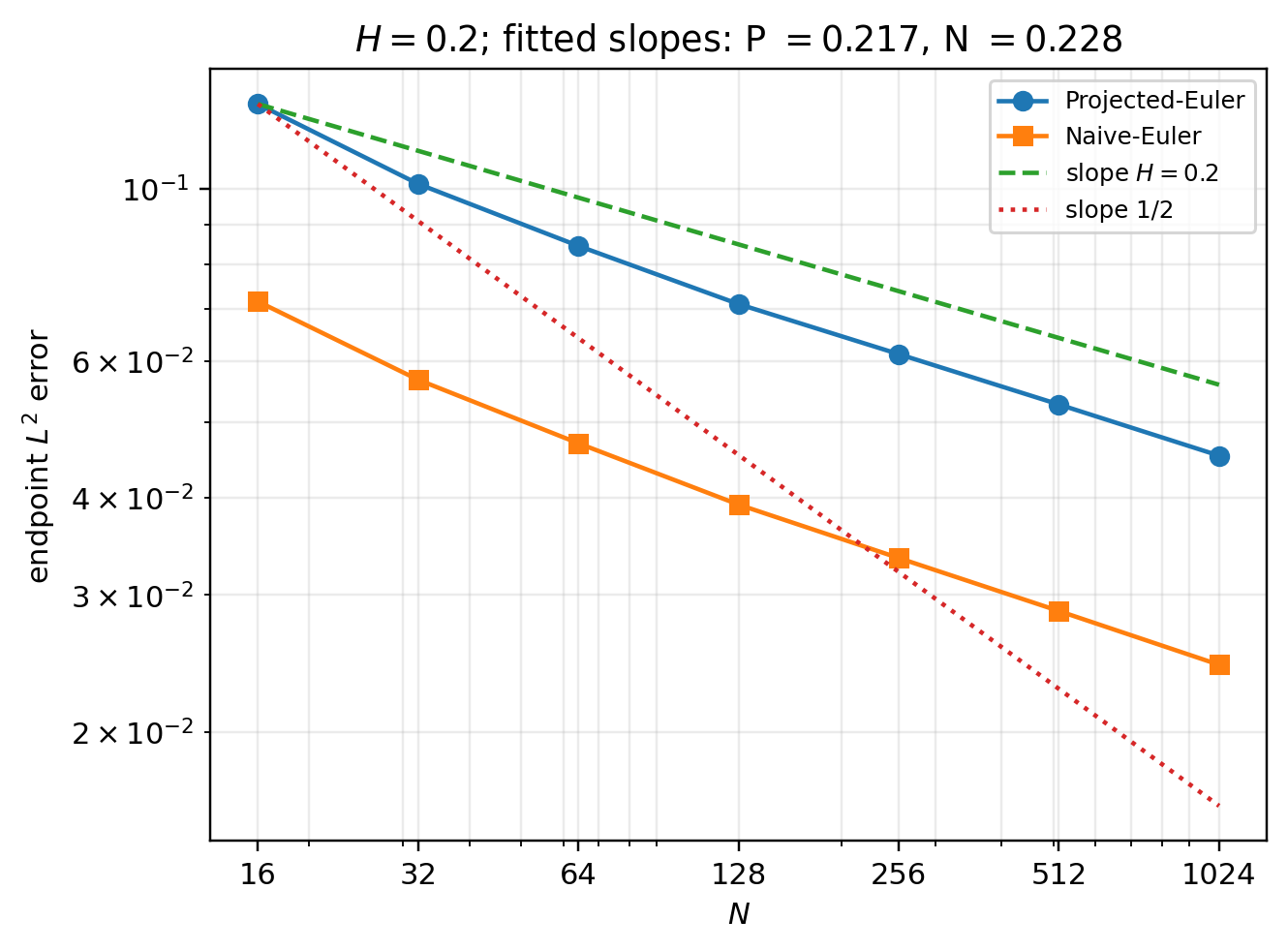}
        \caption{{{$H=0.2$, $\sigma=\sigma_1$.}}}
    \end{subfigure}
    \vfill
    \begin{subfigure}[b]{0.45\textwidth}
        \centering
        \includegraphics[width=\textwidth]{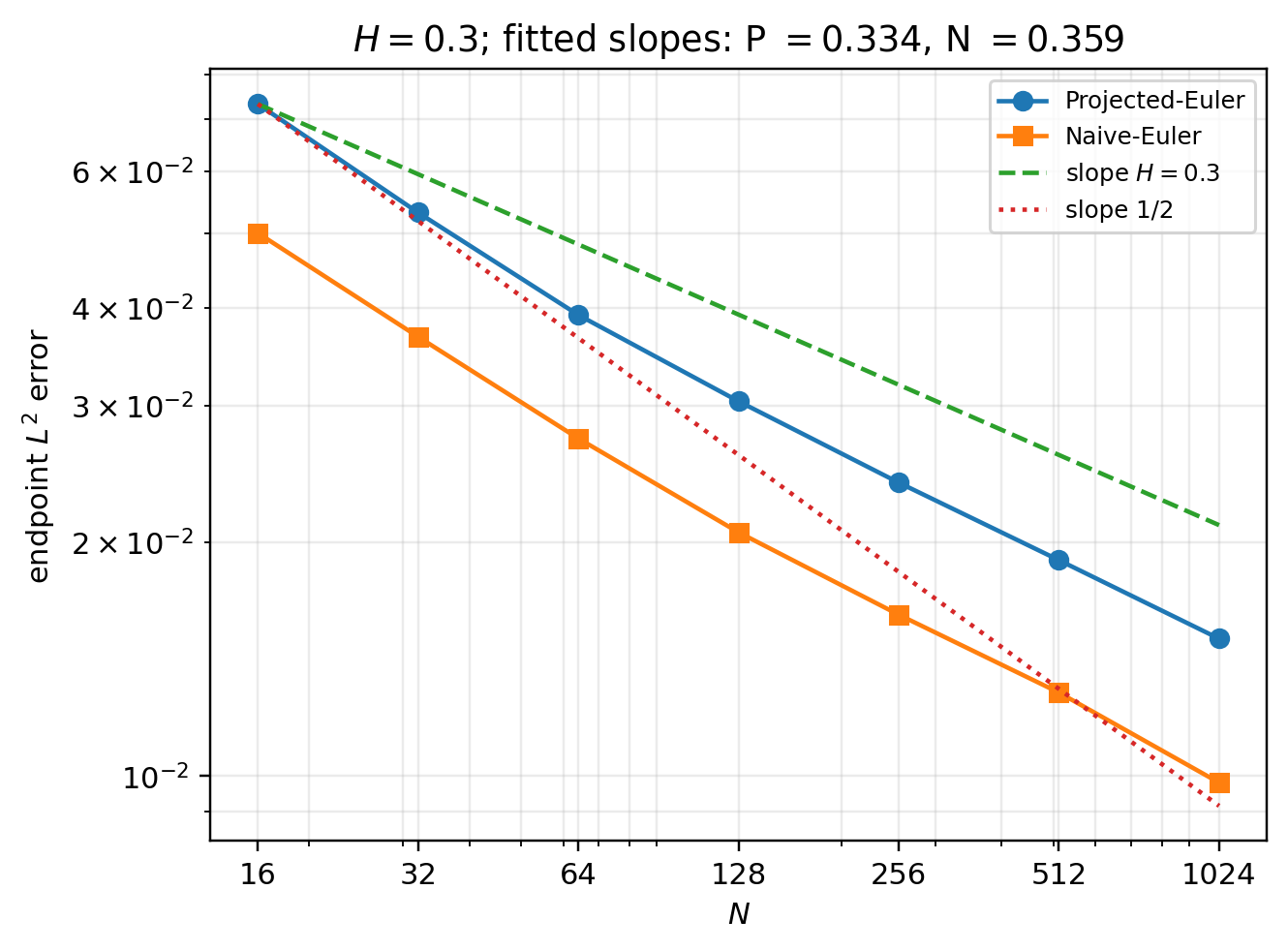}
        \caption{{{$H=0.3$, $\sigma=\sigma_1$.}}}
    \end{subfigure}
    \hfill
    \begin{subfigure}[b]{0.45\textwidth}
        \centering
        \includegraphics[width=\textwidth]{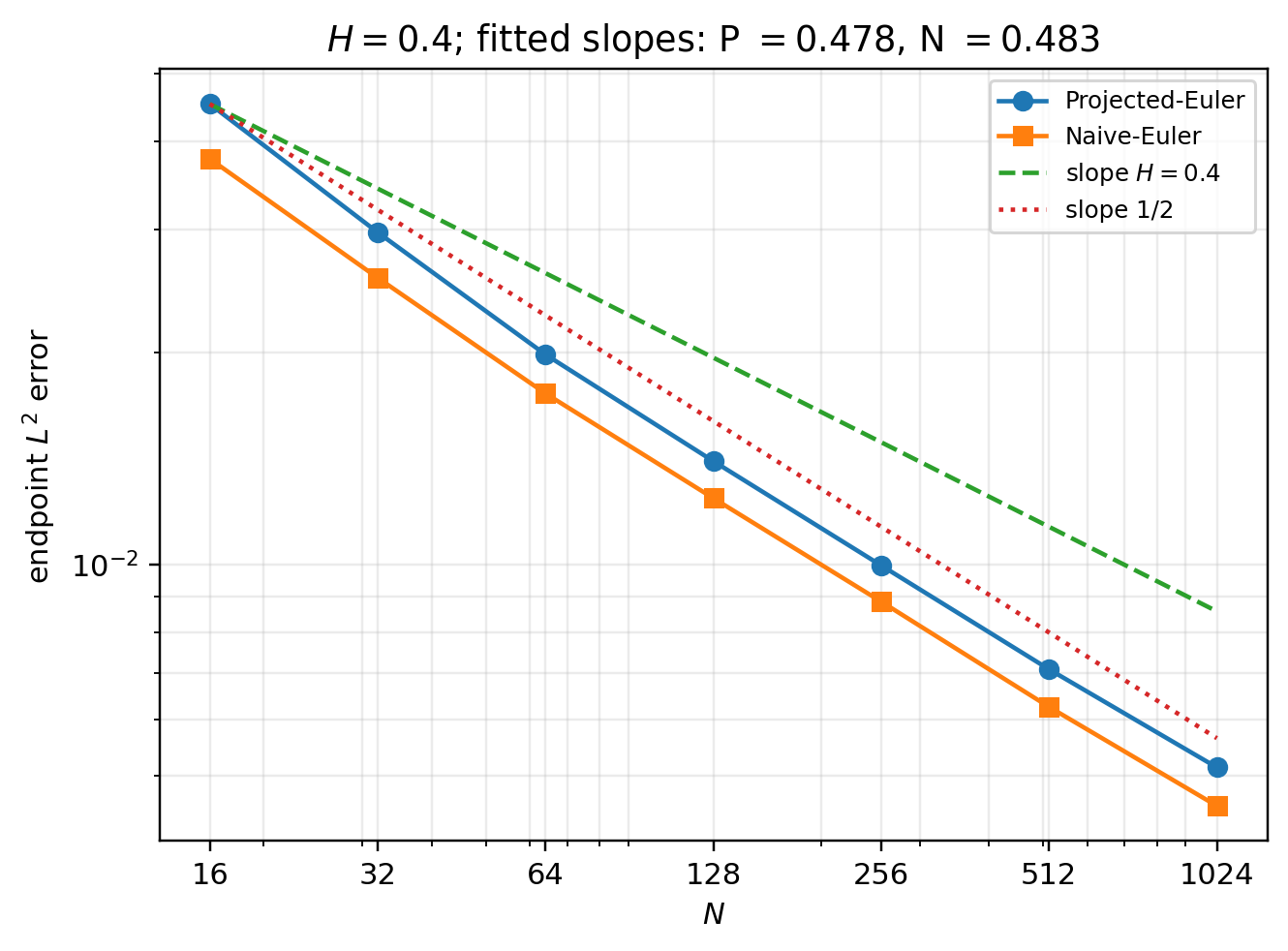}
        \caption{{{$H=0.4$, $\sigma=\sigma_1$.}}}
    \end{subfigure}
    \caption{{{Projected- and Naive-Euler endpoint dyadic errors for $N=16,\ldots,1024$, based on $30{,}000$ Monte Carlo paths. In each panel the dashed reference line has slope $H$ and the dotted line has slope $1/2$. The fitted slopes shown in the panel titles use $N=256,512,1024$.}}}
    \label{fig:alternativeEulerConv}
    \label{fig:ProjEulerConv}
    \label{fig:EulerConv}
\end{figure}

{{Figure~\ref{fig:alternativeEulerConv}  displays the Projected- and Naive-Euler schemes in the same panels and includes the reference rate $(T/N)^H$. We report the endpoint dyadic error $\|X_T^N-X_T^{2N}\|_2$, using the same coupling of the Brownian increments on the $N$ and $2N$ grids. The simulations use $30{,}000$ Monte Carlo paths and $N$ up to $1024$. When $H=0.1$, the fitted exponents on the three finest grids are approximately $0.108$ for the Projected-Euler and $0.107$ for the Naive-Euler, close to the reference exponent $H=0.1$. On the other hand, when $H=0.2,0.3,0.4$, the corresponding fitted exponents are $(0.217,0.228)$, $(0.334,0.359)$ and $(0.478,0.483)$ for (Projected, Naive)$=(P,N)$, respectively. These data support convergence of both alternative schemes, but they do not justify a universal claim that their exact asymptotic rate is $H$ for every $H$. In contrast, the Smart-Euler rate is theoretically controlled by Theorem~\ref{thm:partK} and is numerically close to $1/2$ throughout the experiments.}}

\begin{remark}[On the Python code and reproducibility]
{{The simulations were implemented in Python, with Numba used for the computationally intensive parts of the Smart-Euler implementation. In the  convergence study we use $30{,}000$ Monte Carlo paths and grids up to $N=1024$. The Brownian increments on the coarse grid are obtained by summing adjacent increments of the fine grid, so that the $N$ and $2N$ approximations are coupled pathwise. The code used to reproduce the numerical experiments are publicly available at \texttt{https://github.com/Edgar-Neboit/New-Volterra-Type-SDEs}.}}
\end{remark}

\subsection{H\"older continuity of the paths of $X$}

In this part, we study the $\gamma-$H\"older  ($\gamma \in (0,H)$) regularity of the trajectories of the simulated process $\overline X^n$. To this end, we test the stability of the ratio 
\begin{equation}\label{eq:HolderLastRatio}
     \frac{\mathbb{E}[|\overline X_{t_n}^n-\overline X_{t_{n-1}}^n|^2]}{\left(\frac{T}{n}\right)^{2 H}} = \frac{\mathbb{E}[|\overline X_{T}^n-\overline X_{t_{n-1}}^n|^2]}{\left(\frac{T}{n}\right)^{2 H}},
     \end{equation} 
for different values of $n \in \{250,...,10000\}$ and $H \in \{0.1,0.2,0.3,0.4\}$. 

We denote with $H_{num}$ the value of $H$ used to simulate the trajectories of the process $\overline X^n$ in the numerator in Equation \eqref{eq:HolderLastRatio}, and with $H_{den}$ the value of $H$ appearing in the denominator.

\begin{table}[h]
\centering
\begin{tabular}{|c|c|c||c|c|}
\hline
n & $H_{num}=H_{den}=0.1$ & $H_{num}=H_{den}=0.4$ & $H_{num}=0.1$, $H_{den}=0.4$ & $H_{num}=0.4$, $H_{den}=0.1$ \\
\hline
250 & 4.221 & 0.790 & 126.060 & 0.029 \\
\hline
500 & 4.675 & 0.779 & 194.897 & 0.017 \\
\hline
1000 & 4.633 & 0.732 & 274.605 & 0.013 \\
\hline
5000 & 4.551 & 0.685 & 711.009 & 0.005 \\
\hline
10000 & 4.518 & 0.794 & 1036.610 & 0.003 \\
\hline
\end{tabular}
\caption{Values of the average ratio in \eqref{eq:HolderLastRatio} vs $n$ according to $H_{num}=H_{den}$ or $H_{num}\ne H_{den}$.}
\label{tab:meanValuesLastTimeStepEul}
\end{table}

{{In Table \ref{tab:meanValuesLastTimeStepEul} we see that the ratio in the second and in the third columns is stable, if compared to the fluctuations in the last two columns, i.e., the ratio is stable provided that $H_{num}=H_{den}$.}}

\begin{table}[h]
\centering
\begin{tabular}{|c|c|c||c|c|}
\hline
n & $H_{num}=H_{den}=0.1$ & $H_{num}=H_{den}=0.4$ & $H_{num}=0.1$, $H_{den}=0.4$ & $H_{num}=0.4$, $H_{den}=0.1$ \\
\hline
250 & 0.020 & 0.426 & 0.519 & 0.014 \\
\hline
500 & 0.011 & 0.349 & 0.446 & 0.008 \\
\hline
1000 & 0.006 & 0.328 & 0.399 & 0.005 \\
\hline
5000 & 0.002 & 0.220 & 0.246 & 0.001 \\
\hline
10000 & 0.001 & 0.208 & 0.214 & 0.001 \\
\hline
\end{tabular}
\caption{Values of the average ratio in \eqref{eq:HolderLastRatio} vs $n$ for the Projected-scheme according to $H_{num}=H_{den}$ or $H_{num}\ne H_{den}$.}
\label{tab:meanValuesLastTimeStepHybrid}
\end{table}

{{Finally, Table~\ref{tab:meanValuesLastTimeStepHybrid} shows a different behaviour for the Projected-Euler scheme: even when $H_{num}=H_{den}$, the normalized ratio decreases systematically as $n$ increases. This indicates that, at the grid scale tested here, the Projected-Euler paths are smoother than the target $H$-rough scaling.  This observation concerns the regularity of the discrete approximation and does not by itself contradict convergence of the scheme to the rough limiting process as the mesh is refined.}}

\section{Conclusions}

We have proposed a new family of stochastic models, based on Volterra-type SDEs of convolution type.
The key contribution of our approach is the two-sided explicit link that we are able to establish between an SDE of convolution type, with memory (the process $X$) and an associated standard SDE, without memory (the process $\xi$).

On the simulation side, for the Smart-Euler approximation scheme, we prove, in the fractional-kernel case, the strong upper bound $O(h^{\gamma\wedge1/2})$, independently of the Hurst parameter $H$. Thus the order is $1/2$ when the coefficients are at least $1/2$-H\"older continuous in time. This behavior is enabled by the specific Markovian-memory structure of our equation: the coefficients in the dynamics of $X$ depend on the de-convolved Markovian state $\xi=I^{1-\alpha}X$. Accordingly, comparisons with $H$-dependent rates for other stochastic Volterra equations must be understood as comparisons across different model classes.

\newpage

\appendix 


\section{Basics on fractional integrals  and Laplace transforms}\label{app:frac_Laplace}

\subsection{Fractional integral and derivative}\label{sec:fractional_op}

    As a useful recap, we recall here the definition of Riemann-Liouville fractional integral and derivative, following \cite[Chapter 2, Definition 2.1]{Samko}. 
    \begin{definition}[\textbf{Fractional integral}]\label{Def_st_int}
    	For $\beta >0$ and $f:(0,+\infty)\to \R$ in $L^1([0,T])$, the {\em Riemann-Liouville fractional integral of order $\beta$} is defined as
    	\begin{equation}\label{eq:Integral}
    		I^{\beta}f(t)= \frac{1}{\Gamma(\beta)}\int_0^t (t-s)^{\beta-1}f(s)ds.
    	\end{equation}
    \end{definition}
    For simplicity, we skip $0$ in the above notation, so that to avoid writing $I_{0+}^\beta$.
    
    \begin{remark}
    Exploiting the fractional kernel $K_\beta$ introduced in Table~\ref{Table_K} for $c=1$, we clearly have
    $$
    I^{\beta}f(t)= (K_\beta \star f)(t). 
    $$
    \end{remark}
    
    \begin{definition}[\textbf{Fractional derivative}]
    	For $\beta \!\in(0,1)$, the {\em Riemann-Liouville fractional derivative of order $\beta$} of $f$ reads
    	\begin{align}
    		D^{\beta} f(t) = \frac{1}{\Gamma (1 - \beta)}\frac{d}{dt}\int_0^t (t-s)^{-\beta}f(s) ds.
    	\end{align}
    	{{If $f \in \textrm{AC}([0,T])$, then the fractional derivative $D^{\beta} f$ exists almost everywhere on $[0,T]$.}} 
    \end{definition}
    
    \begin{remark}
    	 $(i)$ In this paper we deal with $\beta \in (0,1)$. Nevertheless, the fractional derivative is also defined for a general $\beta \ge 1$ as follows
    	\begin{equation}\label{eq:Dalphagene}
    		D^{\beta} f(t) = \frac{1}{\Gamma (n - \beta)}\frac{d^n}{dt^n}\int_0^t (t-s)^{n-1-\beta}f(s) ds,\quad\mbox{ where }\quad n=\lfloor \beta \rfloor+1.
    	\end{equation} 
    	A  sufficient condition for this to exist is $f \in \textrm{AC}^{\lfloor \beta \rfloor }([0,T])$.
    \smallskip
    
    	\noindent $(ii)$  When $\beta =1$, $D^{\beta}$ coincides with the regular differentiation operator. 
    \end{remark}
    
    The following result, which corresponds to \cite[Thm 2.4]{Samko}, might be useful as well.
    \begin{theorem}
    	Let $\beta >0$. Then
    	\begin{equation}\label{eq:DcompI}
    		D^{\beta}\circ I^{\beta} f = f
    	\end{equation}
    	is true for any $f \in L^1([0,T])$. On the other hand, the equality
    	\begin{equation}\label{eq:IcompD}
    		I^{\beta}\circ D^{\beta} f = f
    	\end{equation}
    	is valid for $f \in \mathcal I^\beta(L^1)$, where $\mathcal I^\beta(L^1)$ denotes the space of functions $f$ that can be represented as the fractional integral of order $\beta$ of an integrable function, namely $f = I^\beta \varphi$ for some $ \varphi \in L^1([0,T])$.
    \end{theorem}
    
    Direct computations allow to prove the next lemma \cite[Section 2.3]{Samko}.
    \begin{lemma}\label{lem:I_Leb}
    	Under Assumption~\ref{ass:Fubini},  for every $\alpha, \beta >0$, we have the well-known composition formula:
    	\begin{equation}\label{eq:I_prop}
    		I^\beta \circ I^\alpha f =I^{\alpha + \beta}f.
    	\end{equation}
    \end{lemma}

\subsection{The Laplace transform as a useful tool}\label{sec:laplaceTr}
We briefly provide here some background on the Laplace transform, since it is a very efficient tool to deal with the key Equation ~\eqref{eq:KKtilde}. 
Let us recall that the Laplace transform associated to (a kernel) $K$ always exists and reads, for $\zeta >0$ 
\begin{align}
	L_K(\zeta) := \int_0^{+\infty} e^{- \zeta u}K(u)du.
\end{align}		
Exploiting the convolution Theorem \cite[Theorem 2.8 $(iii)$]{Grip90},
\begin{align*}
	L_K(\zeta)\cdot L_{\widetilde{K}}(\zeta) = L_{ K \star \widetilde{K}}(\zeta),
\end{align*}
and the injectivity of this transform, we obtain that Equation~\eqref{eq:KKtilde} reads,  with $\rho >0$,
\begin{align}\label{eq=KKtilde_laplace}
	L_K(\zeta)\cdot L_{\widetilde{K}}(\zeta) =    \int_0^\infty e^{-(\rho + \zeta)u}du = \frac{1}{\zeta +\rho}, \text{ for all } \zeta >0
\end{align}
and this gives an easy way of finding $\widetilde K$, given $K$, as we are going to see below.

\begin{example}\label{ex:gammaKtilde_Laplace}
	When $K$ is the Gamma kernel $K_{c,\alpha, \rho}$ in Table~\ref{Table_K}, for $c>0, \alpha \in (0,1), \rho >0$, then, introducing $v =u(\zeta+\rho)$, we have
	\begin{align*}
		L_{K_{c, \alpha, \rho}}(\zeta)
		= \int_0^\infty c e^{-(\zeta+\rho)u} \frac{ u^{\alpha-1}}{\Gamma(\alpha)} du
		= \frac{ c (\zeta+\rho)^{-\alpha}}{\Gamma(\alpha)} \int_0^\infty e^{-v} v^{\alpha-1}dv
		=  c (\zeta+\rho)^{-\alpha}.
	\end{align*}
	It is then easy to see that $\widetilde{K}_{c, \alpha,\rho} =  K_{\frac1c, 1-\alpha,\rho}$, namely
	$$
	\widetilde{K}_{c,\alpha,\rho}(u) 
	= K_{\frac 1c, 1-\alpha,\rho}(u)
	= e^{-\rho u} K_{\frac 1c, 1-\alpha,0}(u)
	= \frac{e^{-\rho u}}{c} \frac{u^{-\alpha}}{\Gamma(1-\alpha)}.
	$$ 
	\end{example}

In the next final example we focus on the simpler case when $\rho=0$, i.e., when $(K\star \widetilde K) = 1$.
	
\begin{example}[\bf The case $(K\star \widetilde K) = 1$]\label{ex:Ktilde_Laplace}
    We now deal with Equation~\eqref{eq:KKtilde} in the case $\rho =0$.
	\begin{itemize}
    	\item[a)]  For the constant kernel, $K(u) = c>0$ for all $u \in [0,T]$, we get $L_K(\zeta)=\frac{c}{\zeta}$ and so $ L_{\widetilde{K}}(\zeta) \equiv 1/c$.  In this case $\widetilde K$ only exists as a measure, namely $\widetilde K = \frac 1c \delta_0$ (Dirac mass at $0$).
        Indeed (recall Remark \ref{rem:resolvent}) in this case $\widetilde K$ is the \emph{resolvent of the first kind of $K$}.
    	
    	\smallskip
    	\item[b)] When $K$ is the fractional kernel $K_{c,\alpha,0}$ in Table~\ref{Table_K}, for $\alpha \in (0,1)$, this corresponds to $\alpha$-fractional integration since (see, e.g., \cite[Eq. (7.5)]{Samko})
    	$$
    	L_{K_{c,\alpha,0}}(\zeta)= c\int_0^\infty e^{-\zeta u} \frac{u^{\alpha-1}}{\Gamma(\alpha)} d u =  c\, \zeta^{-\alpha}.
    	$$
    	Then, $L_{\widetilde{K}_{c, \alpha,0}}(\zeta)=\frac 1c \zeta^{-(1-\alpha)}$ and so, for $\alpha \in (0,1)$
    	\begin{align}\label{kappatildepower}
    		\widetilde{K}_{c,\alpha,0}= K_{\frac 1c,1-\alpha, 0}.
    	\end{align}
    	
	\end{itemize}
\end{example}


\section{Proofs\label{AppendixB}}\label{app:proofs}

\subsection{Proof of Lemma~\ref{lem:regularityX}}\label{sec_proof_lem:regularityX}

The proof exploits Kolmogorov continuity criterion (see e.g. \cite[Theorem 3.23]{Kal02})\footnote{Let $ X$ be a stochastic process with values in the Polish metric space $(S, \rho)$. If there exist $a,b, c >0$ such that 
	$$
	\mathbb E \{ \rho( X_s, X_t)\}^a \le  c | s - t| ^{b+d}, \quad s,t \in \mathbb R^d
	$$
	then $X$ admits a continuous modification and there exists a modification whose paths are H\"{o}lder continuous of order $\gamma$, for every $\gamma \in (0, \frac{b}{a})$.}
 
{
First, assumptions i)--ii) imply $K\in L^2(0,T)$. Hence, by It\^o's isometry,
\[
\|X_t\|_2
\le \|\xi^0\|_2
+\left(\int_0^tK(t-s)^2\|Y_s\|_2^2ds\right)^{1/2}
\le \|\xi^0\|_2+\|K\|_{L^2(0,T)}\sup_{s\in[0,T]}\|Y_s\|_2<\infty.
\]
Thus the stochastic convolution is well defined in $L^2(\mathbb P)$. Under the additional $L^p$ assumptions in iii), the Burkholder--Davis--Gundy (BDG) and generalized Minkowski inequalities\footnote{For every $p, r \in (0, \infty)$, with $r \le p$, we have:
	$$
	\Big|\Big|  \left( \int | X_t|^r  d\mu(t) \right)^\frac{1}{r} \Big|\Big|_p  		\le  \left( \int \big\|   X_t \big\|_p^r  d\mu(t) \right)^\frac{1}{r}. 	$$
} similarly give
\[
\|X_t\|_p
\le \|\xi^0\|_p+C_p^{\rm BDG}\|K\|_{L^2(0,T)}\sup_{s\in[0,T]}\|Y_s\|_p<\infty.
\]
}

Moreover, for $0 \le s \le t$, exploiting again BDG inequality and the generalized Minkowski inequality  with $r=2$ and $p \ge 2$, we obtain
\begin{align*}
    \Big|\Big| \int_0^t K(t - u ) & Y_u dW_u - \int_0^s K (s - u ) Y_u dW_u  \Big|\Big|_p \\
    & \le  \Big|\Big| \int_0^s \left( K(t - u ) -  K(s - u ) \right) Y_u dW_u  \Big|\Big|_p +  \Big|\Big|  \int_s^t K(t - u ) Y_u dW_u \Big|\Big|_p \\
	& \le  C_p^{\textrm{BDG}} \Big\{ \Big|\Big| \left( \int_0^s \left( K(t - u ) -  K(s - u ) \right)^2 Y_u^2 du \right)^\frac12 \Big|\Big|_p   +  \Big|\Big|  \left( \int_s^t K^2(t - u ) Y_u^2 du\right)^\frac12 \Big|\Big|_p  \Big\} \\
	&  \stackrel{p \ge 2}{\le}  C_p^{\textrm{BDG}}   \Big\{  \left( \int_0^s \left( K(t - u ) -  K(s - u ) \right)^2 \big|\big| Y_u \big|\big|_p^2 du \right)^\frac12 +    \left( \int_s^t K^2(t - u ) \big|\big| Y_u\big|\big|_p ^2 du\right)^\frac12  \Big\}
	\\
	& \le  C_p^{\textrm{BDG}} \sup_{u \in [0,T]} \big|\big| Y_u\big|\big|_p  \Big\{  \left( \int_0^s \left( K(t - u ) - K(s - u )\right)^2du \right)^\frac12 + \left( \int_s^t K^2(t - u ) du\right)^\frac12  \Big\} .
\end{align*}
Hence, by two elementary changes of variable we obtain
    \begin{align}
	  \nonumber\Big|\Big| \int_0^t &K(t - u ) Y_u dW_u - \int_0^s K (s - u ) Y_u dW_u  \Big|\Big|_p\\
    \label{eq:finalbound}  &\le   C_p^{\textrm{BDG}} \sup_{u \in [0,T]} \big|\big| Y_u\big|\big|_p  \Big\{  \left( \int_0^s \left( K(t-s+v  ) - K(v)\right)^2 {{dv}} \right)^\frac12 + \left( \int_0^{t-s} K^2(v ) dv\right)^\frac12  \Big\} \\
       \nonumber     &\le C_p^{\textrm{BDG}} C_K {{\left[ (t-s)^{\theta} + (t-s)^{\frac{\beta-1}{2 \beta}}\right] }}.
	\end{align}
where, in the last step, we have used  Assumption $iv)$ {{and H\"older inequality.}}

\subsection {Proof of Theorem~\ref{thm:existence}.}\label{Sec:proof_thm:existence}

We split the proof into three steps: we first deal with the Markovian stochastic process $\xi$, we then deal with $X$ and finally we link them.

{\sc Step~1}. {\em The diffusion SDE in Equation~\eqref{eq:Wdiff}}. This SDE can be rewritten as follows. We set $\widetilde \kappa(t) :=e^{\rho t} (\widetilde K\star\mbox{\bf 1})(t) =  e^{\rho t} \widetilde \varphi(t)$, with $\widetilde{\varphi}(t)=(\widetilde K\star\mbox{\bf 1})(t)$, and we define $(\mathbb Y_t)_{t\in [0,T]}$ as 
\[
    \mathbb Y_t := (Y^1_t,Y^2_t ):= (\xi^0,\xi_t- \widetilde \kappa(t) \xi^0), \quad t\!\in [0,T],
\] 
so that 
$$
    \xi_t 
= Y_t^2 +  \widetilde \kappa(t) Y_t^1
= Y_t^2 +  \widetilde \kappa(t) \xi^0.
$$
Then, for $t \in [0,T]$, $(\mathbb{Y}_t)_{t\in [0,T]}$ satisfies a SDE that component-wise reads as
\begin{align}
    \label{eq:VersionY1} Y^1_t  &= \xi^0+\int_0^t 0\, ds +\int_0^t 0\; dW_s, \\
    \label{eq:VersionY2} Y^2_t &= \int_0^t \widetilde b(s, Y^2_s+ \widetilde \kappa (s) Y_s^1 )ds +\int_0^t  \widetilde \sigma (s, Y^2_s+\widetilde \kappa (s) Y^1_s ) dW_s, 
\end{align}
or, equivalently,
\begin{equation}\label{eq:Y}
    d \mathbb Y_t = \overline b(t,\mathbb Y_t) dt +\overline\sigma(t,\mathbb Y_t) d W_t, \quad \mathbb Y_0 = \begin{pmatrix} \xi^0\\ 0\end{pmatrix},
\end{equation}
with 
\begin{equation}\label{eq:bar_b_sigma}
\overline b\big(t,(y^1,y^2)\big)=  \begin{pmatrix}  0\\ \widetilde b (t, y^2+\widetilde \kappa(t) y^1) \end{pmatrix}
 \quad \mbox{ and }\quad \overline \sigma(t,(y^1,y^2)):= \begin{pmatrix} 0\\ \widetilde \sigma (t, y^2+\widetilde \kappa(t) y^1) \end{pmatrix}.
\end{equation}
 
Now,  by condition~\eqref{eq:LipxUt}, $b$ and $\sigma$ are Lipschitz continuous in $x$ uniformly in $t\!\in [0,T]$, so that 
$\widetilde b(t,x)$ and $ \widetilde \sigma(t,x)$, defined by~\eqref{Eq:def_over_b_sigma} clearly satisfy the same assumption with 
\[
    [\widetilde b]_{{\rm Lip},x} \le  e^{\rho T}[b]_{{\rm Lip},x} 
    \quad \mbox{ and }\quad 
    [\widetilde \sigma]_{{\rm Lip},x} \le  e^{\rho T}[\sigma]_{{\rm Lip},x} ,
\]
and 
$$
\sup_{t\in [0,T]}( |\widetilde \sigma(t,0)|+  |\widetilde b(t,0)| )
{{\le}} e^{\rho T} \sup_{t\in [0,T]} ( |\sigma(t,0)|+|b(t,0)| ) < + \infty.
$$

So, as a third sub-step, {one easily checks that $\overline b$ and $\overline\sigma$, introduced in Equation~\eqref{eq:bar_b_sigma}, are Borel measurable on $[0,T]\times\mathbb R^2$.}
Furthermore, they satisfy a $2$-dimensional version of the Lipschitz assumption in Equation ~\eqref{eq:LipxUt} 
in $\mathbf y=(y^1,y^2)$, uniformly in $t\!\in [0,T]$
since $\widetilde \kappa$ is non-negative and non-decreasing with $\widetilde \kappa (T)<+\infty$, by definition. Indeed, by introducing $\mathbf y := (y^1,y^2)$  and $ \mathbf y' := ({(y^1)}', {(y^2)}')$ we have 
\begin{align*}
    \|  \overline b\big(t, \mathbf y \big)- \overline b\big(t,\mathbf y' \big) \| & = \big|   \widetilde b \left (t, y^2+\widetilde \kappa(t) y^1 \right) - \widetilde b \left( t, {(y^2)}' + \widetilde \kappa(t) {(y^1)}' \right) \big| \\
    & \le [\widetilde b]_{{\rm Lip},x} \left( \big|  y^2 -  {(y^2)}' \big| + \widetilde  \kappa(T) \big|  y^1  - {(y^1)}' \big| \right) \\
    & \le [\widetilde b]_{{\rm Lip},x} (1 \vee \widetilde  \kappa(T))  \left( \big|  y^2 -  {(y^2)}' \big| + \big|  y^1  - {(y^1)}' \big| \right) \\
    & =  [\widetilde b]_{{\rm Lip},x} (1 \vee \widetilde  \kappa(T))  \|  \mathbf y - \mathbf  {y}'\|_1   \le \sqrt 2 [\widetilde b]_{{\rm Lip},x} (1 \vee \widetilde  \kappa(T))  \|  \mathbf y - \mathbf  {y}'\|_2,
\end{align*} 
where $|| \cdot ||_1$ and $|| \cdot ||_2$ denote, respectively, the $L^1$ and the $L^2$ norms of a real vector and the same holds for $\overline \sigma$. Moreover,
\begin{align*}
    \sup_{t\in [0,T]} \| \overline b(t,0) \| + \| \overline \sigma (t,0) \| & \le  \sup_{t\in [0,T]} | \widetilde b(t,0) | + | \widetilde \sigma (t,0) | < + \infty,
\end{align*} 
and so the above two-dimensional SDE~\eqref{eq:Y} has a unique pathwise continuous, $\mathbb F^{\xi^0,W}$-adapted solution starting from $\mathbb Y_0$ (see \cite[Thm.~IX.2.1]{RevuzYor99} and \cite[Theorem~A3.3, Chapter~5]{BouLep}). {For a merely almost-surely finite initial variable $\xi^0$, this conclusion follows by solving with the bounded truncations $\xi^0\mathbf 1_{\{|\xi^0|<m\}}$ and pasting the solutions by pathwise uniqueness.}

Moreover, the regularity of $\overline b$ and $\overline \sigma$ above implies that, for some positive constant $C >0$,
$$
 \sup_{t\in [0,T]} \| \overline b(t,\mathbf y) \| + \| \overline \sigma (t, \mathbf y) \| \le C (1 + \| \mathbf y \|_2)
$$
so that, by \cite[Prop.~7.6(a)]{GilPag}, {whenever $\xi^0\in L^p(\mathbb P)$, the following moment estimate holds for $p\in(0,+\infty)$,}
\begin{equation}\label{eq:sup_Y}
  \Big\| \sup_{t \in [0,T]} \| \mathbb Y_t \| \Big\|_p \le C_{\widetilde b, \widetilde \sigma, T, p} (1 + \| \mathbb Y_0 \|_p),
\end{equation}
where $ C_{\widetilde b, \widetilde \sigma, T, p} >0$. {Hence, using $\xi_t=Y_t^2+\widetilde\kappa(t)\xi^0$, we immediately derive}
$$
\sup_{t \in [0,T]}| \xi_t | \le \sup_{t \in [0,T]} |Y_t^2| + \widetilde \kappa (T) | \xi^0|
$$
which finally implies
\begin{equation}\label{eq:LpboundX}
\Big\| \sup_{t \in [0,T]} | \xi_t| \Big\|_p \le C'_{ b, \sigma, T, p} (1 + \| \xi^0 \|_p).
\end{equation}

{Moreover, after localization, the drift and diffusion integrands of $\mathbb Y$ are bounded. Standard BDG increment estimates and Kolmogorov's criterion therefore show that $\mathbb Y$ has pathwise $a$-H\"older trajectories for every $a\in(0,1/2)$, up to $\mathbb P$-indistinguishability.}
As consequence {{$\xi_t = Y^2_t + \widetilde \kappa(t) \xi^0$}} acquires the lowest pathwise regularity between the one of $Y^2$ and $\widetilde \varphi$, so that,  if $\widetilde \varphi$ is $\eta$-H\"older continuous  for some $\eta\!\in (0, \frac 12)$, {then $\xi$ is locally $(a\wedge\eta)$-H\"older continuous.}

{
\medskip
\noindent {\sc Step~2}. {\em Construction of a candidate solution to Equation~\eqref{eq:pathdepVolt}.}
Let $\xi$ be the unique strong solution of Equation~\eqref{eq:Wdiff} constructed in Step~1, and define
\begin{equation}\label{eq:4.2bis}
X_t=\xi^0+\int_0^tK(t-s)b(s,e^{-\rho s}\xi_s)ds
+\int_0^tK(t-s)\sigma(s,e^{-\rho s}\xi_s)dW_s.
\end{equation}
Because $\xi$ is adapted and pathwise continuous, the two coefficient processes in Equation~\eqref{eq:4.2bis} are progressively measurable. The linear-growth property of $b$ and $\sigma$ and the pathwise boundedness of $\xi$ on $[0,T]$ show that they are pathwise bounded. Moreover, condition a) implies $K\in L^1(0,T)\cap L^2(0,T)$. Therefore, both integrals in Equation~\eqref{eq:4.2bis} are well defined.

Assume first that $\xi^0\in L^p(\mathbb P)$ for some
\[
p>\max\left\{2,\frac{1}{\theta\wedge\frac{\beta-1}{2\beta}}\right\}.
\]
By Equation~\eqref{eq:LpboundX} and linear growth,
\[
\sup_{t\in[0,T]}\|\sigma(t,e^{-\rho t}\xi_t)\|_p<\infty.
\]
Lemma~\ref{lem:regularityX} then yields a pathwise continuous modification of the stochastic-convolution term in Equation~\eqref{eq:4.2bis}. The deterministic-convolution term is continuous by translation continuity of $K$ in $L^1(0,T)$ and pathwise boundedness of its integrand. Thus $X$ admits a pathwise continuous modification.

We now remove the moment assumption on $\xi^0$. Let
\[
A_k:=\{k\le|\xi^0|<k+1\},\qquad k\ge0,
\]
and set $(\xi^0)^{(k)}:=\xi^0\mbox{\bf 1}_{A_k}$. Let $\xi^{(k)}$ and $X^{(k)}$ denote the processes obtained from Equations~\eqref{eq:Wdiff} and~\eqref{eq:4.2bis}, respectively, with initial variable $(\xi^0)^{(k)}$. Since $(\xi^0)^{(k)}$ is bounded, the preceding argument provides a pathwise continuous modification $\overline X^{(k)}$ of $X^{(k)}$. By pathwise uniqueness and the locality of ordinary and stochastic integration, $\xi^{(k)}=\xi$ and $X^{(k)}=X$ on $A_k$. Consequently,
\[
\overline X_t:=\sum_{k\ge0}\mbox{\bf 1}_{A_k}\overline X_t^{(k)},\qquad t\in[0,T],
\]
is a pathwise continuous modification of the process $X$ defined in Equation~\eqref{eq:4.2bis}.

\medskip
\noindent {\sc Step~3}. {\em Identification, uniqueness and representation.}
Convolving Equation~\eqref{eq:4.2bis} with $\widetilde K$, using associativity and the deterministic and stochastic Fubini arguments justified in Remark~\ref{rem:fubini_check}, gives
\begin{align*}
\widetilde K\star X
&=\xi^0(\widetilde K\star\mbox{\bf 1})
+e^{-\rho\cdot}\star b(\cdot,e^{-\rho\cdot}\xi_{\cdot})
+e^{-\rho\cdot}\stackrel{W}{\star}\sigma(\cdot,e^{-\rho\cdot}\xi_{\cdot}).
\end{align*}
Multiplication by $e^{\rho t}$ and Equation~\eqref{Eq:def_over_b_sigma} yield
\begin{equation}\label{eq:X_xi}
e^{\rho\cdot}(\widetilde K\star X)
=\xi^0e^{\rho\cdot}(\widetilde K\star\mbox{\bf 1})
+\int_0^{\cdot}\widetilde b(s,\xi_s)ds
+\int_0^{\cdot}\widetilde\sigma(s,\xi_s)dW_s
=\xi_{\cdot}.
\end{equation}
Substituting Equation~\eqref{eq:X_xi} into Equation~\eqref{eq:4.2bis} proves that $X$ solves Equation~\eqref{eq:pathdepVolt}.

For uniqueness, let $X'$ be another pathwise continuous strong solution of Equation~\eqref{eq:pathdepVolt} and define
\[
\xi'_t:=e^{\rho t}(\widetilde K\star X')_t.
\]
The same convolution argument shows that $\xi'$ solves Equation~\eqref{eq:Wdiff}. By pathwise uniqueness in Step~1, $\xi'=\xi$. Substitution into Equation~\eqref{eq:pathdepVolt} then gives, for every $t\in[0,T]$,
\begin{align*}
X'_t
&=\xi^0+\int_0^tK(t-s)b(s,e^{-\rho s}\xi'_s)ds
+\int_0^tK(t-s)\sigma(s,e^{-\rho s}\xi'_s)dW_s\\
&=\xi^0+\int_0^tK(t-s)b(s,e^{-\rho s}\xi_s)ds
+\int_0^tK(t-s)\sigma(s,e^{-\rho s}\xi_s)dW_s=X_t.
\end{align*}
Finally, convolving Equation~\eqref{eq:X_xi} with $K$ gives
\[
K\star(e^{-\rho\cdot}\xi)
=(K\star\widetilde K)\star X
=e^{-\rho\cdot}\star X
=e^{-\rho\cdot}\int_0^{\cdot}e^{\rho s}X_sds.
\]
After multiplication by $e^{\rho t}$,
\[
\int_0^t e^{\rho s}X_sds=\big((e^{\rho\cdot}K)\star\xi\big)_t.
\]
Since $X$ is pathwise continuous, differentiation yields the representation in Equation~\eqref{eq:Xrepresentation}.
}

\subsection{Proof of Theorem~\ref{thm:RateEulercont}}\label{sec_proof_thm:RateEulercont}
In the computations that follow, the constants $C_{b,\sigma, T,p}$ may vary from line to line but they do not depend neither on the step of the schemes, nor on the (pseudo-)starting value $\xi^0$.
The proof is divided into three steps: in the first one, in Equation \eqref{eq:Lp-boundtildeX}, we establish uniform bounds for the solutions of the SDE and of its Euler schemes. In the second step, for $p \ge 2$, the strong error rate for the Euler scheme is proved, with the difficulty of dealing with the non-null initial condition. In the third and final step we treat the case $p \in (0,2)$.

 \smallskip
\noindent {\sc Step~1}: Recall that $\mathbb{Y}$, defined in the proof of Theorem~\ref{thm:existence} as (here $\widetilde \kappa(t) =\widetilde \varphi(t)$ as $\rho=0$)
\[
\mathbb Y_t = (Y^1_t,Y^2_t ):= (\xi^0,\xi_t- \widetilde \varphi(t) \xi^0), \quad t\!\in [0,T],
\] 
 is solution to the regular SDE
 \begin{equation*}
d \mathbb Y_t = \overline b(t,\mathbb Y_t) dt +\overline\sigma(t,\mathbb Y_t) d W_t, \quad \mathbb Y_0 = \begin{pmatrix} \xi^0\\ 0\end{pmatrix},
\end{equation*}
with drift and diffusion coefficients which are Lipschitz in space, uniformly in time,
\begin{equation*}
\overline b\big(t,(y^1,y^2)\big)=  \begin{pmatrix}  0\\ \widetilde b (t, y^2+\widetilde \varphi (t) y^1) \end{pmatrix}
 \quad \mbox{ and }\quad \overline \sigma(t,(y^1,y^2)):=
 \begin{pmatrix} 0\\ \widetilde \sigma (t, y^2+\widetilde \varphi(t) y^1) \end{pmatrix}.
\end{equation*}
Exploiting  \cite[ Proposition~7.2]{GilPag}, we have, denoting by $\overline{\mathbb Y}^h= (\overline{Y}_t^{h,1},\overline{Y}_t^{h,2}) $ the continuous time Euler scheme of $\mathbb Y$  with time step $h= \frac Tn$,
\[
    \Big \| \sup_{t\in [0,T]}| \mathbb{Y}_t| \Big\|_p + \Big\| \sup_{t\in [0,T]} |\overline{\mathbb{Y}}^h_t | \Big\|_p  \le C_{b,\sigma,T,p}(1+\|\mathbb{Y}_0\|_p)
\]  

where    $C_{b,\sigma,T,p}$ is  a real constant not depending on $n$, i.e., on the step $h$.
We know that, by definition,   $\mathbb Y =  (\xi^0,\xi_t- \widetilde \varphi(t) \xi^0)$ and we straightforwardly check that for every step $h$
\[
\overline{\xi}_t^h = {{\overline{{Y}}_t^{h,2}}}+\widetilde \varphi(t) \xi^0, \quad t\!\in [0,T].
\]
Proceeding similarly as in the proof of Theorem~\ref{thm:existence}
and exploiting the properties of $\widetilde{\varphi}$, namely $\widetilde \varphi(0)=0$ and $\widetilde \varphi$ non-decreasing, we also have that, for every $p>0$, there exists a  a real constant $C_{b,\sigma,T,p}>0$ such that, for all $ n\ge 1$ and $h= \tfrac Tn$,

\begin{equation}\label{eq:Lp-boundtildeX}
    \Big\|\sup_{t\in [0,T]} |\xi_t| \Big\|_p+  \Big\|\sup_{t\in [0,T]} |\overline{\xi}^h_t| \Big\|_p\le C_{b,\sigma,T,p}(1+\|\xi^0\|_p).
\end{equation}
 
\noindent {\sc Step~2}. 
Now, let us assume $p\ge 2$. 
We first notice that, for a standard diffusion (e.g., if $\widetilde \varphi= {\bf 1}$), from Equations~\eqref{Xtilde_rho0} and~\eqref{eq:Euler_cont_xi}, one has 
\begin{equation}\label{eq:diff_xi_t}
    \xi_t - \overline{\xi}_{t}= \int_0^t \big( b(s,  \xi_s) - b(\underline s, \overline{\xi}_{\underline  s})\big)ds + \int_0^t \big( \sigma(s,  \xi_s)   -\sigma(\underline s, \overline{\xi}_{\underline  s})\big)dW_s.
\end{equation}
Then, we follow the lines of the proof of~\cite[Theorem~7.2, Section~7.8.4, p.331]{GilPag}, which is based on Burkholder-Davis-Gundy (BDG),  generalized Minkowski inequalities and a variant of Gronwall's lemma ~\cite[Lemma~7.3, p.327]{GilPag}, and we apply them on the decomposition
\[
\xi_s - \overline  \xi_{\underline s} =  \xi_s-\xi_{\underline s} + \xi_{\underline s} - \overline \xi_{\underline s} , \quad s\!\in [0,T],
\] 
exploiting the obvious fact that $\sup_{s\in[0,T]}\|\xi_{\underline s} - \overline \xi_{\underline s}\|_p\le \sup_{s\in[0,T]}\|\xi_{s} - \overline \xi_{ s}\|_p$. Then we derive, using the (uniform in time) Lipschitz condition from Equation~\eqref{eq:bsigHolLip}, 
\begin{equation}\label{eq:Lp-rate_I}
\forall\, t\!\in [0,T], \quad     \Big\|\sup_{s\in [0,t]} |\xi_s -\overline{\xi}_s| \Big\|_p\le C_{b,\sigma,T,p}\Bigg( h^{\gamma} (1+\|\xi^0\|_p)+ \int_0^t \| \xi_s -\xi_{\underline s}\|_p ds + \Big( \int_0^t \| \xi_s -   \xi_{\underline s} \|_p^2ds\Big)^{1/2}\Bigg) .
\end{equation}

The process $(\Lambda_t)_{t\in [0,T]}$ defined by 
\begin{align*}
    \Lambda_t := \int_0^t b(s,  \xi_s) ds + \int_0^t  \sigma(s,  \xi_s) dW_s, \quad t\!\in [0,T],
\end{align*}
is an It\^o  process  such that $\big\|\sup_{t\in [0,T]}| b(t,\xi_t)|  \big\|_p+\big\|\sup_{t\in [0,T]} | \sigma(t,\xi_t) | \big\|_p\le C_{b,\sigma, T,p} (1+\|\xi^0\|)_p$, owing to Equations~\eqref{eq:bsigHolLip} and~\eqref{eq:Lp-boundtildeX}. Consequently,  from~\cite[Lemma~7.4, Section 7.8.3, p.329]{GilPag}, it follows
\begin{align*}
    \Big\| \Lambda_t-\Lambda_s\Big\|_p\le C_{b,\sigma,T,p} |t-s|^{1/2} (1+\|  \xi^0\|_p).
\end{align*}
Then, noticing that
\begin{align*}
    \xi_t -  \xi_{\underline t} = \xi^0(\widetilde \varphi(t)- \widetilde \varphi(\underline t) \big)+ \Lambda_t-\Lambda_{\underline t}
\end{align*}
and since $\widetilde \varphi$ is non-decreasing, we find
\begin{equation}\label{eq:LpregulX}
    \| \xi_t - {\xi}_{\underline t} \|_p \le \|\xi^0\|_p \big( \widetilde \varphi(t) - \widetilde \varphi(\underline t)\big) + C_{b,\sigma,T,p} \cdot h^{1/2}\big(1+\|\xi^0\|_p\big).
\end{equation}

Plugging this inequality into~\eqref{eq:Lp-rate_I} 
yields  
\begin{align}
    \Big\|\sup_{t\in [0,T]} |\xi_t -\overline{\xi}_t| \Big\|_p & \le C_{b,\sigma,T,p}\Bigg( h^{\frac 12\wedge\gamma} (1+\|\xi^0\|_p)+ {{\|\xi^0\|_p}} \int_0^T  \big( \widetilde \varphi(t) - \widetilde \varphi(\underline t)\big)dt + {{\|\xi^0\|_p}} \Big( \int_0^T \big( \widetilde \varphi(t) - \widetilde \varphi(\underline t)\big)^2dt\Big)^{1/2}\Bigg) \nonumber \\
    & \le C_{b,\sigma,T,p}\Bigg( h^{\frac 12\wedge\gamma} (1+\|\xi^0\|_p)+  {{\|\xi^0\|_p}} (1+\sqrt{T})  \Big( \int_0^T \big( \widetilde \varphi(t) - \widetilde \varphi(\underline t)\big)^2dt\Big)^{1/2}\Bigg), \label{eq:sup_tildeX}
\end{align}
where we used the elementary inequality $\sqrt{a+b} \le \sqrt{a} +\sqrt{b}$, for $a$, $b\ge 0$ in the first line and {{Cauchy-Schwarz}}  inequality in the second one.
As $\widetilde \varphi$ is non-decreasing and concave, for every $t\!\in (0,T]$, we have
 \[
  0\le \widetilde \varphi(t) - \widetilde \varphi(\underline t)\le \widetilde \varphi'_r(\underline t) (t-\underline t) 
 \]
where $\widetilde \varphi'_r$ is the right derivative of $\widetilde \varphi$ on $(0,T]$. Let us notice that $\widetilde \varphi'_r(0)$ exists but it can be equal to $+\infty$. Consequently, as $\underline t = 0$ if $t \in [0, \frac{T}{n})$ and $\widetilde \varphi (0)=0$, we obtain
\begin{align*}
    \int_0^T \big( \widetilde \varphi(t) - \widetilde \varphi(\underline t)\big)^2dt
    & \le \int_0^{\tfrac Tn} \widetilde \varphi^2(t)dt +\int_{\frac Tn}^T( \varphi_r'(\underline t))^2(t-\underline t)^2dt
    \le \tfrac Tn \widetilde \varphi^2\big(\tfrac Tn\big) + \sum_{k=1}^{n-1}\int_{\frac{kT}{n}}^{\frac{(k+1)T}{n}}(\widetilde \varphi_r'\big(\tfrac{kT}{n}\big))^2\big(t-\tfrac{kT}{n}\big)^2dt\\
    & \le \tfrac Tn\Big( \widetilde \varphi^2\big(\tfrac Tn\big) + \big( \tfrac Tn \big)^2\sum_{k=1}^{n-1}(\widetilde \varphi_r'\big(\tfrac{kT}{n}\big))^2 \Big).
\end{align*}
Hence, the fact that $(\sum_i a_i^2)^{1/2}\le \sum_i a_i$, for $a_i\ge 0$, together with the fact that $\widetilde \varphi_r'$ is non-increasing by concavity of $\widetilde \varphi$ yield
\begin{align}
    \left(\int_0^T \big( \widetilde \varphi(t) - \widetilde \varphi(\underline t)\big)^2ds \right)^{1/2} &\le \sqrt{\tfrac Tn}\Big( \widetilde \varphi\big(\tfrac Tn \big) +\tfrac Tn \sum_{k=1}^{n-1} \widetilde \varphi_r '\big(\tfrac{kT}{n} \big)\Big) \le  \sqrt{\tfrac Tn} \Big( \widetilde \varphi\big(\tfrac Tn \big)  +   \int_0^{\frac{(n-1)T}{n}}\widetilde \varphi'_r(t)dt\Big) \nonumber \\
    &  \le  2\, \widetilde \varphi\big(T \big)  \sqrt{\tfrac Tn}. \label{eq:ineq_varphi}
\end{align}
One concludes that, for every step $h$, 
\begin{align*}
    \Big\|\sup_{t\in [0,T]} |\xi_t -\overline{\xi}_t| \Big\|_p \le C_{b,\sigma,T,p} \,h^{\frac 12\wedge\gamma} (1+\|\xi^0\|_p).
\end{align*}
  
\noindent {\sc Step~3}.  Now, let us consider $0<p<2$ {{and denote by $\mathcal W \in C([0,T], \R)$ the sample path of the Brownian motion $W$.}}  
There exist functionals $F$ and $\overline F: \R\times  C([0,T], \R)\to C([0,T], \R)$ such that 
\begin{align*}
    \xi 
    = F(\xi^0,\mathcal W) \quad \mbox{ and }\quad  \overline{\xi} = \overline F(\xi^0, \mathcal W). 
\end{align*}
This is a straightforward consequence of Blagoven$\check{{\rm s}}\check{{\rm c}}$enkii-Freidlin Theorem \cite[Theorem V.13.1]{RogersWilliamsII} applied to $\mathbb{Y}$ and the fact that $\xi= Y^2 + \widetilde \varphi(\cdot) \xi^0$ and $\overline{\xi}= \overline Y^2 -\widetilde \varphi(\cdot) \xi^0$. Then, temporarily denoting by $\xi^{(x_0)}$ and $\overline{\xi}^{(x_0)}$ the process $\xi$ and its Euler scheme, respectively, starting at $\xi^0= x_0$, we obtain
\begin{align*}
    \int \sup_{t\in [0,T]} |\xi_t -\overline{\xi}_t|^p d\mathbb{P}
    &= \int \mathbb E \left(   \sup_{t\in [0,T]} |\xi^{(x_0)}_t -\overline{\xi}^{(x_0)}_t|^p \right) d\mathbb{P}_{\xi^0}(x_0)
    \le  \int  \left[  \mathbb E \left(  \sup_{t\in [0,T]} |\xi^{(x_0)}_t -\overline{\xi}^{(x_0)}_t|^2 \right) \right]^{\frac p2} d\mathbb{P}_{\xi^0}(x_0)\\
    & \le (C_{b,\sigma, T, 2})^{p} {{h^{(\frac 12\wedge\gamma)p}}}\int  (1+|\xi^0|)^p  d\mathbb{P}_{\xi^0}(x_0) 
    \le 2^{p} (C_{b,\sigma, T, 2})^{p} (1+ \mathbb E|\xi^0|^p),
\end{align*}
so that
\begin{align*}
    \Big\|\sup_{t\in [0,T]} |\xi_t -\overline{\xi}_t| \Big\|_p  \le 2\,C_{b,\sigma, T, 2}\, h^{\frac 12\wedge\gamma}(1+ \|\xi^0\|_p). 
\end{align*}
This completes the proof. \hfill

\subsection{Proof of Corollary~\ref{cor:L1RateEulercstpm}}\label{sec_proof_cor:L1RateEulercstpm}

    Let us temporarily assume $p\ge 1$, so that $\|\cdot\|_p$ is a norm. 
    One has, for every $t\!\in [0, T]$,
    \begin{align*}
        \|\xi_t -\overline{\xi}_{\underline t}\|_p & \le \|\xi_t -  \xi_{\underline t}\|_p+  \|\xi_{\underline t} -  \overline{\xi}_{\underline t}\|_p
        \le  \|\xi_t -  \xi_{\underline t}\|_p+  \Big \|\sup_{t\in [0,T]} \big| \xi_{\underline t} -  \overline{\xi}_{\underline t}\big| \Big\|_p.
    \end{align*}
    Hence, for every $r>0$, we get 
    \begin{align*}
        \left( \int_0^T \|\xi_t -\overline{\xi}_{\underline t}\|^r_pdt\right)^{1/r} \le 2^{(r-1)^+}\left(\Big( \int_0^T  \|\xi_t -  \xi_{\underline t}\|^r_pdt\Big)^{1/r} +T^{1/r}  \Big \|\sup_{t\in [0,T]} \big| \xi_t -  \overline{\xi}_t\big| \Big\|_p\right). 
    \end{align*}
    Finally, one relies on Equation~\eqref{eq:LpregulX} and Theorem~\ref{thm:RateEulercont} to provide upper-bounds for the first and the second term on the right hand side of the above inequality, respectively, to get
    \begin{align*}
        \left( \int_0^T \|\xi_t -\overline{\xi}_{\underline t}\|^r_pdt\right)^{1/r} \le C_{b,\sigma, T,p,r}(1+\|\xi^0\|_p)\left( \Big(\frac Tn\Big)^{\gamma\wedge \frac 12}+\Big( \int_0^T(\widetilde \varphi(t)-\widetilde \varphi(\underline t))^rdt \Big)^{1/r} \Big)\right).
    \end{align*}
    If $0<p<1$, one gets analogous results (with different constants) using the sub-additivity of $\|\cdot\|_p$.

\subsection{Proof of Theorem~\ref{thm:est_eu_sch_X}}\label{sec_proof_thm:est_eu_sch_X} We may assume without loss of generality that $\xi_0\in L^p(\mathbb{P})$. 
The proof is split into two steps: the first dealing with the case $p\ge 2$ and the second with the case $p \in (0,2)$.
    
    {\sc Step~1} ({\em Case $p\ge 2$}).
    We start by considering the following equation
    \[
        X_t = \xi^0 + \int_0^t K(t-s)  \big(b(s,\xi_s) ds +\sigma(s,\xi_s)dW_s \big)
    \]
    and its Euler like (continuous time) discretization  with time step $h= \frac Tn$, defined  by  
    \[
      \overline X_t = \xi^0 + \int_0^t K(t-s) \big(b(\underline s,\overline\xi_{\underline s}) ds +\sigma(\underline s,\overline \xi_{\underline s})dW_s \big).
    \]
    Then, by a  standard application of BDG inequality and generalized Minkowski inequality, we have, for $p\ge 2$,
   \begin{align*}
        \| X_t -\overline X_t\|_p & \le \int_0^t K(t-s)\| b(s,\xi_s) -b(\underline s,\overline{\xi}_{\underline s})\|_pds + C^{BDG}_p \left(\int_0^t K(t-s)^2\| \sigma(s,\xi_s) -\sigma (\underline s,\overline{\xi}_{\underline s})\|^2_pds  \right)^{1/2}\\
        & \le [b]_{HolLip} \int_0^t K(t-s) \big((s-\underline s)^{\gamma} + \big\| \xi_s -  \overline{\xi}_{\underline s}\big\|_p\big )ds  \\
        &\qquad +C^{BDG}_p [\sigma]_{HolLip}\left( \int_0^t K(t-s)^2 \big((s-\underline s)^{2\gamma} + \big\| \xi_s- \overline{\xi}_{\underline s}\big)\big\|^2_p\big)ds\big)\right)^{1/2}.
    \end{align*}
    Using the elementary inequality $\sqrt{a+b} \le \sqrt{a} +\sqrt{b}$, for $a,\,b\ge 0$, we obtain
    \begin{align}\label{eq:X_barX}
        \| X_t -\overline X_t\|_p & \le \left([b]_{HolLip} \int_0^TK(s)ds+C^{BDG}_p [\sigma]_{HolLip}\left( \int_0^T K(s)^2ds\right)^{1/2}\right)\Big(\frac Tn\Big)^{\gamma} \\ \nonumber
        & \quad  + [b]_{HolLip} \int_0^t K(t-s)\big\| \xi_s- \overline{\xi}_{\underline s} \big\|_pds+ C^{BDG}_p [\sigma]_{HolLip}\left( \int_0^t K(t-s)^2  \big\| \xi_s-  \overline{\xi}_{\underline s}\big\|^2_p ds \right)^{1/2}.
    \end{align}
    As for the first term on the right-hand side of the above inequality, one first notices that, using Cauchy-Schwarz inequality and Equation \eqref{eq:Eulercstpm}  with $r=2$ in  Corollary~\ref{cor:L1RateEulercstpm}, 
    \begin{align}\label{eq:term2_X}
        \int_0^t K(t-s)\big\| \xi_s-  \overline{\xi}_{\underline s} \big\|_p ds & \le \left(\int_0^T K^2(s)ds\right)^{1/2}
        \left(\int_0^t \big\| \xi_s - \overline{\xi}_{\underline s} \big\|^2_p ds \right)^{1/2}\\ \nonumber
        &\le C_{K,b,\sigma,T,p}\left( \Big(\frac Tn\Big)^{\gamma\wedge \frac 12}+\Big(\int_0^T(\widetilde \varphi(t)-\widetilde \varphi(\underline t))^2 dt \Big)^{1/2} \right)(1+\|\xi^0\|_p)\\ \nonumber
        &\le  C_{K,b,\sigma,T,p}\Big(\frac Tn\Big)^{\gamma\wedge \frac 12}(1+\|\xi^0\|_p),
    \end{align} 
    as we have $\Big(\int_0^T(\widetilde \varphi(t)-\widetilde \varphi(\underline t))^2 dt \Big)^{1/2} = O\big(\sqrt{\frac Tn}\big)$ from Step~2 of Theorem~\ref{thm:RateEulercont}.  
   
    The second term is more demanding and its discussion relies on assumption {{$i)$ in the statement}}. 
    By H\"older inequality with conjugate exponents $\beta$ and $\frac{\beta}{\beta-1}$, one has, for every $t\!\in [0,T]$, 
    \begin{align*}
        \left(\int_0^t K(t-s)^2  \big\| \xi_s-  \overline{\xi}_{\underline s} \big\|^2_p ds\right)^{1/2}&
        \le \left( \int_0^T K^{2\beta}(s)ds \right)^{\frac{1}{2\beta}} \Big( \int_0^T\big\| \xi_s-  \overline{\xi}_{\underline s}\big\|^{\frac{2\beta}{\beta-1}}_p dt \Big)^{\frac{\beta-1}{2\beta}}.
    \end{align*}
   
    It follows from the assumption made on the kernel $K$ that $\widetilde \varphi =( \widetilde K \star \mbox{\bf 1})$ is null at $0$, {{non-decreasing}} and concave. 
    Thus an application of Corollary~\ref{cor:L1RateEulercstpm} with $r= \frac{2\beta}{\beta-1}$ yields
    \[ 
        \left (\int_0^T \big\| \xi_s-  \overline{\xi}_{\underline s}\big\|^{\frac{2\beta}{\beta-1}}_pds\right) ^{\frac{\beta-1}{2\beta}}\le T^{\frac{\beta-1}{2\beta}} C_{K,b,\sigma,T,p,\beta}(1+\|\xi^0\|_p)\left( \Big(\frac Tn\Big)^{\gamma\wedge \frac 12}+\Big( \int_0^T(\widetilde \varphi(s)-\widetilde \varphi(\underline s))^{\frac{2\beta}{\beta-1}}ds \Big)^{\frac{\beta-1}{2\beta}} \right).
    \]
     
    \noindent {\sc Step~2}. The case $1\le p<2$ 
    can be formally handled following the lines of Step~3 in the proof of Theorem~\ref{thm:RateEulercont}.

\subsection{Proof of Corollary~\ref{lem:continuous_version}}\label{sec_proof_lem:continuous_version}

{
By It\^o's isometry and $\alpha>1/2$,
\[
\|X_t\|_2
\le \|\xi^0\|_2
+\frac{1}{\Gamma(\alpha)}
\left(\int_0^t(t-s)^{2(\alpha-1)}\|Y_s\|_2^2ds\right)^{1/2}
\le \|\xi^0\|_2
+\frac{t^{\alpha-1/2}}{\Gamma(\alpha)\sqrt{2\alpha-1}}
\sup_{s\in[0,T]}\|Y_s\|_2<\infty.
\]
Thus the stochastic convolution in Corollary~\ref{lem:continuous_version} is well defined in $L^2(\mathbb P)$.
}

An application of Equation~\eqref{eq:finalbound}, for $0 \le s \le t $,  paired with BDG inequality and the generalized Minkowski inequality, for $r=2$ and $p \ge 2$, yields
\begin{align*}
    &  \Big|\Big| \int_0^t (t - u )^{\alpha - 1} Y_u dW_u - \int_0^s (s - u )^{\alpha - 1} Y_u dW_u  \Big|\Big|_p \le  \Big|\Big| \int_0^s \left( (t - u )^{\alpha - 1} - (s - u )^{\alpha - 1} \right) Y_u dW_u  \Big|\Big|_p \\
		& \le  C_{\textrm{BDG}} \sup_{u \in [0,T]} \big|\big| Y_u\big|\big|_p  \Big\{  \left( \int_0^s \left( (t - u )^{\alpha - 1} - (s - u )^{\alpha - 1} \right)^2du \right)^\frac12 + \left( \int_0^{t-s} u^{2(\alpha - 1)} du\right)^\frac12  \Big\} .
\end{align*}
	
We now notice that the second term in the last line can be rewritten as
\begin{align*}
    \left(  \int_0^{t-s} u^{2(\alpha - 1)} du \right)^\frac12 = \frac{(t-s)^{\alpha - \frac12}}{\sqrt{2 \alpha - 1}}
\end{align*}
and so it remains to deal with $\int_0^s \left( (t - u )^{\alpha - 1} - (s - u )^{\alpha - 1} \right)^2du$. Via the change of variable: $v := \frac{s-u}{t-s}$ we find
\begin{align*}
	\int_0^s \left( (t - u )^{\alpha - 1} - (s - u )^{\alpha - 1} \right)^2du = (t-s)^{2 \alpha -1} \int_0^{\frac{s}{t-s}} \left( (1 + v )^{\alpha - 1} - v^{\alpha - 1} \right)^2dv.
\end{align*}
Now, we claim that the integral $\int_0^{\frac{s}{t-s}} \left( (1+ v )^{\alpha - 1} - v^{\alpha - 1} \right)^2dv$ is uniformly bounded by a finite positive constant $\kappa_{\alpha}$ which does not depend on $t$, $s$ and $T$ as soon as $\alpha > \frac12$.
{{To prove this, fix $\tau>0$ and define the integral (and use strict positivity of the integrand)
$$
\Psi_\tau := \int_0^\tau \left\{ (x+1)^{\alpha-1} - x^{\alpha-1} \right\}^2 dx \le \int_0^\infty \left\{ (x+1)^{\alpha-1} - x^{\alpha-1} \right\}^2 dx.
$$
We split the integral into two parts: over $[0,1]$ and $[1,\infty)$.

On $[0,1]$: since $\alpha < 1$, we have $0 < (x+1)^{\alpha-1} < x^{\alpha-1}$, which implies $\vert{}(x+1)^{\alpha-1} - x^{\alpha-1}\vert{} \le x^{\alpha-1}$. Thus:$$\int_0^1 \left\{ (x+1)^{\alpha-1} - x^{\alpha-1} \right\}^2 dx \le \int_0^1 x^{2(\alpha-1)} dx = \frac{1}{2\alpha - 1}.
$$
On $[1,\infty)$: by the Mean Value Theorem, for each $x \ge 1$ there exists $\xi_x \in (x, x+1)$ such that$$(x+1)^{\alpha-1} - x^{\alpha-1} = (\alpha - 1) \xi_x^{\alpha - 2}.
$$
Since $\alpha < 1$, the function $y \mapsto y^{\alpha - 2}$ is decreasing, so $\xi_x^{\alpha - 2} \le x^{\alpha - 2}$. Therefore:
$$\int_1^\infty \left\{ (x+1)^{\alpha-1} - x^{\alpha-1} \right\}^2 dx \le (\alpha - 1)^2 \int_1^\infty x^{2\alpha - 4} dx = \frac{(\alpha - 1)^2}{3 - 2\alpha}.
$$
Combining both bounds, we conclude that for any $\tau \ge 0$:$$\Psi_\tau \le \frac{1}{2\alpha - 1} + \frac{(\alpha - 1)^2}{3 - 2\alpha} =: \kappa_\alpha,
$$
}}
proving the claim.
So we finally have
\begin{align*}
    \Big|\Big| \int_0^t (t - u )^{\alpha - 1} Y_u dW_u & - \int_0^s (s - u )^{\alpha -1} Y_u dW_u  \Big|\Big|_p \le  C_{\textrm{BDG}} \sup_{u \in [0,T]} \big|\big| Y_u\big|\big|_p \Big(  \sqrt{\kappa_{\alpha}} +  \frac{1}{\sqrt{2 \alpha - 1}}  \Big)  |t-s|^{\alpha -\frac12}
\end{align*}
and we apply Kolomogorov continuity criterion with $d=1$ and $b + d = b+1 = p (\alpha - \frac12)$, as soon as $b >0$, which here corresponds to 
$
p \left( \alpha - \frac12 \right) - 1 >0 \Longleftrightarrow p > \frac{1}{ \alpha - \frac12 }.
$
Hence, the lemma is proved and $ X$ admits a modification which is $\gamma$-H\"{o}lder continuous with
$
\gamma \in \left(0, \alpha-\frac12-\frac1p  \right).
$

\subsection{Proof of Lemma~\ref{lem:I_Ito}}\label{app:Lem_Ito_frac}
{
The stochastic Fubini condition follows directly from the assumptions. Indeed, using the Beta-integral identity and $\sup_{u\in[0,T]}\|Y_u\|_2<\infty$,
\begin{align*}
&\mathbb E\int_0^t\left(\int_u^t\frac{(t-s)^{\beta-1}(s-u)^{\alpha-1}}{\Gamma(\beta)\Gamma(\alpha)}\,ds\right)^2|Y_u|^2du\\
&\qquad\le C\sup_{u\in[0,T]}\|Y_u\|_2^2\int_0^t(t-u)^{2(\alpha+\beta-1)}du<\infty,
\end{align*}
because $\alpha+\beta>1/2$. Hence stochastic Fubini applies, and the definition of the fractional integral in Equation~\eqref{eq:Integral} yields
}

\begin{align*}
    I^\beta \left( \int_0^\cdot \frac{ (\cdot - u )^{\alpha - 1}}{\Gamma(\alpha)} Y_u dW_u  \right)(t) 
    &= \frac{1}{\Gamma(\beta)}\int_0^t (t-s)^{\beta-1} \left( \int_0^s \frac{(s - u )^{\alpha - 1}}{\Gamma(\alpha)} Y_u dW_u\right)  ds \\
	& = \frac{1}{\Gamma(\beta) \Gamma(\alpha)}\int_0^t   Y_u  \left(  \int_u^t  (t-s)^{\beta-1} (s - u )^{\alpha - 1} ds \right) dW_u \\
	& = \frac{1}{\Gamma(\beta) \Gamma(\alpha)}\int_0^t   Y_u  \left(  \int_0^1  (t- u)^{\alpha +\beta-1} \tau^{\alpha - 1} (1- \tau)^{\beta -1} d\tau \right) dW_u\\
	& = \frac{ B(\alpha, \beta)}{\Gamma(\beta) \Gamma(\alpha)}\int_0^t   (t-u)^{\alpha +\beta-1} Y_u  dW_u =  \frac{1}{\Gamma(\alpha + \beta)}\int_0^t   (t-u)^{\alpha +\beta-1} Y_u  dW_u.     
\end{align*}
where in the next-to-last  passage we used $\tau:=\frac{s-u}{t-u}$ and $B(\alpha, \beta)=\frac{ \Gamma(\alpha) \Gamma(\beta)}{\Gamma(\alpha +\beta)}$.

\subsection{Proof of Theorem~\ref{thm:partK}}\label{sec_proof_thm:partK}

Exploiting the proof of Theorem \ref{thm:est_eu_sch_X}, we are left to prove 
\begin{align}\label{eq:stat}
    \sup_{t\in [0,T]} \int_0^t K_{1,\alpha}(t-s)^2 \big(\widetilde \varphi(s)-  \widetilde \varphi(\underline s) \big)^2ds = O\Bigg(\frac Tn\Bigg). 
\end{align}
Indeed, exploiting the over-bound in Equation \eqref{eq:X_barX} and the fact that the first two terms on the right-hand side are respectively of order $O\left(\left(\frac{T}{n}\right)^{\gamma}\right)$ and  $O\left(\left(\frac{T}{n}\right)^{\gamma\land\frac12}\right)$, where the order of the second term comes from the estimation in Equation \eqref{eq:term2_X}, we should only address the third term, namely
\begin{align}\label{eq:int}
	\left( \int_0^t K_{1, \alpha}(t-s)^2  \big\| \xi_s-  \overline{\xi}_{\underline s}\big\|^2_p ds \right)^{1/2}.
\end{align}
To study this term, we first of all exploit the triangular inequality $ \big\| \xi_s-  \overline{\xi}_{\underline s}\big\|_p\leq  \big\| \xi_s-  {\xi}_{\underline s}\big\|_p +  \big\| \xi_{\underline s}-  \overline{\xi}_{\underline s}\big\|_p \leq  \big\| \xi_s-  {\xi}_{\underline s}\big\|_p +  \big\| \sup_{s\in[0,T]}|\xi_{s}-  \overline{\xi}_{s}|\big\|_p$ to say that we have actually to handle two terms for which we already have bounds in terms of $\widetilde \varphi$.
In particular, the first term can be handled as in Equation \eqref{eq:LpregulX}, leading to
\begin{equation}
	\| \xi_s - {\xi}_{\underline s} \|_p \le \|\xi^0\|_p \big( \widetilde \varphi(s) - \widetilde \varphi(\underline s)\big) + C_{b,\sigma,T,p} \cdot \left(\frac{T}{n}\right)^{1/2}\big(1+\|\xi^0\|_p\big),
\end{equation}
while the second can be over-bounded as in Equation \eqref{eq:Eulercont} in Theorem \ref{thm:RateEulercont}
\begin{equation}
     \Big\|\sup_{t\in [0,T]} |\xi_t -\overline{\xi}_t| \Big\|_p  \le \,C_{b,\sigma, T, p}\, \left(\frac{T}{n}\right)^{\frac 12\wedge\gamma}(1+ \|\xi^0\|_p),
\end{equation}
so that
\begin{align}
	 \big\| \xi_s-  \overline{\xi}_{\underline s}\big\|_p^2 
	 \leq 2 \|\xi^0\|_p^2 \big( \widetilde \varphi(s) - \widetilde \varphi(\underline s)\big)^2 + C_{b,\sigma, T, p}\, \left(\frac{T}{n}\right)^{1\wedge2\gamma}(1+ \|\xi^0\|_p)^2,
\end{align}
which yields that the term in Equation \eqref{eq:int} reduces to \eqref{eq:stat}. To deal this with it, we first get rid of the constants so that it remains to consider the quantity
\begin{equation}\label{eq:fract_calc}
    \sup_{t\in [0,T]} \int_0^t (t-s)^{2(\alpha-1)}  \big(s^{1-\alpha}-\underline s^{1-\alpha} \big)^2ds . 
\end{equation}
As $\alpha \!\in (1/2,1)$, the map $u\mapsto u^{1-\alpha}$ is $(1-\alpha)$-H\"older, non-decreasing and concave, so that, for every $u \in [0,T]$,
\begin{align*}
  0 \le   u^{1-\alpha}-\underline  u^{1-\alpha} \le (u-\underline  u)^{1-\alpha}\quad\mbox{ and }\quad 0 \le u^{1-\alpha}-\underline  u^{1-\alpha} \le(1-\alpha) \underline u^{-\alpha}(u-\underline  u).
\end{align*}
We now work on the integral in Equation \eqref{eq:fract_calc} by distinguishing two cases: in step one we treat the case $t=t_k$ for some $k=1,\ldots,n$ and in step two we deal with the more general $t\!\in [0,T]$, hence $t_k = \underline t$. 

\noindent {\sc Step~1}. ({\em $t=t_k= \frac{kT}{n}$, $k=1,\ldots,n$}). 
We decompose the integral in Equation \eqref{eq:fract_calc} into the sum of two terms  
\begin{align}\label{eq_two_terms}
    \int_0^{t_k} (t_k-s)^{2(\alpha-1)} \big(s^{1-\alpha}-\underline s^{1-\alpha} \big)^2ds  = \underbrace{\int_0^{t_1} \cdots\; ds}_{(a)} +\underbrace{\int_{t_1}^{t_k}\cdots \;ds}_{(b)} .
\end{align}
First we focus on $(a)$ and we observe that $\underline s=0$ and so, using that $2(\alpha - 1) + 1=2 \alpha - 1 \in (0,1)$, we find \begin{align}\label{eq_first_term}
    (a) \le \Big( \frac Tn \Big)^{2(1-\alpha)}\frac{t_k^{2(\alpha-1)+1}-(t_k-t_1)^{2(\alpha-1)+1}}{2(\alpha-1)+1}\le \frac{1}{2(\alpha-1)+1} \Big( \frac Tn \Big)^{2(1-\alpha)+1+2(\alpha-1)}= \frac{1}{2(\alpha-1)+1} \frac Tn.
\end{align}
On the other hand, setting $\varpi= 1-\frac{1}{2\alpha}\!\in (0,1)$, we get 
\begin{align*}
    (b) & \le \sum_{\ell=1}^{k-1} \int_{t_{\ell}}^{t_{\ell+1}}(t_k-s)^{2(\alpha-1)}  \big(s^{1-\alpha}-\underline s^{1-\alpha} \big)^{2(1-\varpi)} \big(s^{1-\alpha}-\underline s^{(1-\alpha)} \big)^{2\varpi}ds\\
    & \le  \sum_{\ell=1}^{k-1} \int_{t_{\ell}}^{t_{\ell+1}}(t_k-s)^{2(\alpha-1)}  (s -\underline s)^{2(1-\varpi)(1-\alpha)} {{{(1-\alpha)}^{2\varpi}}} \underline s ^{-2\alpha\varpi} (s -\underline s)^{2\varpi}ds\\
    &\le \Big( \frac Tn\Big)^{2\varpi +2(1-\varpi)(1-\alpha)}  {{{(1-\alpha)}^{2\varpi}}} \sum_{\ell=1}^{k-1} \int_{t_{\ell}}^{t_{\ell+1}}(t_k-s)^{2(\alpha-1)}\Big( \frac{\ell T}{n}\Big)^{-2\alpha \varpi}ds.
\end{align*}
Now,  {{${(1-\alpha)}^{2\varpi} < 1$ }} and one checks that 
\begin{align*}
    \Big(\frac{\ell T}{n}\Big)^{-2\alpha\varpi} =  \Big(1+\frac{1}{\ell}\Big)^{2\alpha \varpi} \Big(\frac{(\ell+1)T}{n}\Big)^{-2\alpha\varpi}\le   (1+2\alpha \varpi) \big(\frac{(\ell+1)T}{n}\big)^{-2\alpha\varpi},
\end{align*}
since $2\alpha \varpi  = 2\alpha-1 <1$ and where we used that $(1+u)^a \le 1+ au$ for all $u \ge 0$ and $a \in (0,1]$.
 
Hence, exploiting the fact that $ \big(\frac{(\ell+1)T}{n}\big)^{-2\alpha\varpi}\le s^{-2\alpha\varpi} $, for every $s \in [t_{\ell}, t_{\ell+1}]$,
\begin{align*}
    (b) & \le 2\alpha \Big( \frac Tn\Big)^{2\varpi +2(1-\varpi)(1-\alpha)}  \int_{t_1}^{t_k}(t_k-s)^{2(\alpha-1)} s^{-2\alpha \varpi} ds \\
    & \le 2\alpha \ \frac Tn \int_0^{t_k}(t_k-s)^{2(\alpha-1)} s^{-2\alpha \varpi} ds \\
    & = 2\alpha  \ \frac Tn \ t_k^{2(\alpha-1)-2\alpha \varpi +1} B\big(2(\alpha-1)+1, 1-2\alpha\varpi\big), 
\end{align*}
where $B(a,b) = \int_0^1 (1-u)^{a-1}u^{b-1}du$.
At this stage one checks that $2(\alpha-1)-2\alpha \varpi +1= 0$, which yields
\begin{align}\label{eq_second_term}
    (b) \le 2\alpha  \frac Tn  B\left( 2(\alpha-1)+1, 2(1-\alpha) \right).
\end{align}
This bound does not depend on $k$ and  consequently it is valid for all $k=1,\ldots,n$.
So the inequalities in Equation \eqref{eq_first_term} and \eqref{eq_second_term} yield the desired result.
 
\noindent {\sc Step~2}. ({\em $t\!\in [0,T]$}). 
Let $t_k = \underline t$. 
We observe, back to \eqref{eq:fract_calc}, 
\begin{align*}
    \int_0^{t} (t-s)^{2(\alpha-1)} \big(s^{1-\alpha}-\underline s^{1-\alpha} \big)^2ds 
    &\le  \int_0^{t_k}(t_k-s)^{2(\alpha-1)} \big(s^{1-\alpha}- \underline s^{1-\alpha} \big)^2ds \\
    &\quad  + \int_{t_k}^t  (t-s)^{2(\alpha-1)} \big(s^{1-\alpha}-t_k^{1-\alpha} \big)^2ds.
\end{align*}
Since the first term on the right is $O\Big(\frac Tn\Big)$ by Step~1, one just needs to evaluate the second term on the right hand side. 
We thus have
\begin{align*}
    \int_{t_k}^t (t-s)^{2(\alpha-1)} \big(s^{1-\alpha} - t_k^{1-\alpha} \big)^2ds& \le (t^{1-\alpha}-t_k^{1-\alpha})^2\int_0^{t-t_k}u^{2(\alpha-1)} du \\
    & \le \frac{ (t- t_k)^{2(1-\alpha) +1+2(\alpha-1)}}{2 \alpha -1} =\frac{t-t_k}{2 \alpha -1} \le \frac {1}{(2 \alpha - 1)} \frac Tn,
\end{align*}
since $u^{1-\alpha}$ is $(1-\alpha)$-H\"older. 
This bound does not depend on $t$ and so the proof is complete.

\subsection{Proof of Lemma~\ref{lem:asympt_average}}\label{sec_proof_lem:asympt_average}

In what follows, for simplicity, we denote by $K$ and $\widetilde K$ the fractional kernel $K_{1,\alpha,0}$ and its co-kernel. We first focus on the process $\xi$, which is defined as in Equation \eqref{eq:def_Y} (recall that the initial condition for the Markovian process, see Equation \eqref{eq:Wdiff}, is $\xi^0\,e^{\rho t} (1\star\widetilde K) (t)$ and here $\rho=0$):
\begin{align*}
    \xi_t = \xi^0  (\widetilde K\star\mbox{\bf 1})(t) +\int_0^t  (\mu - \lambda \xi_s) ds +  \int_0^t   \sigma(\xi_s)\, dW_s.
\end{align*}
Its expectation at time $t$, $m_t := \mathbb E (\xi_t), t \in [0,T]$, reads
\begin{align}
    m_t = \mathbb E(\xi^0) \int_0^t \widetilde K(u) du +\int_0^t  (\mu - \lambda m_s) ds,
\end{align}
thus satisfying the following ODE: $m'_t = - \lambda m_t + \mu+ \mathbb E(\xi^0) \widetilde K(t)$. 
This equation has an explicit solution:
\begin{equation}\label{eq:expectation_Y}
    m_t = \mathbb E(\xi^0) (e^{- \lambda \cdot} \star \widetilde K)(t) + \frac{\mu}{\lambda} (1 - e^{- \lambda t}).
\end{equation}
In order to compute the limit when $t$ goes to infinity for $m_t$, we need to study first the behaviour of $(e^{- \lambda \cdot} \star \widetilde K)(t)$ at infinity. Since $\lim_{t \rightarrow +\infty} \widetilde K(t) = 0$, then it is possible to use Lemma \ref{lem:asympt_tilde_K} below to prove that also 
$$
\lim_{t \rightarrow +\infty} (e^{- \lambda \cdot} \star \widetilde K)(t) = 0.
$$
It is now straightforward that $\lim_{t \rightarrow +\infty} m_t = \lim_{t \rightarrow +\infty} \mathbb E[\xi_t] = \frac{\mu}{\lambda}$.
\medskip
We now pass to the process $X$, which is defined as in Equation \eqref{eq:def_Z}:
\begin{align*}
    X_t= \xi^0 + \int_0^t K(t-s)  (\mu - \lambda \xi_s)\, ds +  \int_0^t K(t-s) \sigma\big(\xi_s\big) \,dW_s,
\end{align*}
and whose expectation is 
\begin{equation}
\mathbb E(X_t) = \mathbb E(\xi^0) + \int_0^t K(t-s)  (\mu - \lambda m_s)\, ds.
\end{equation}
So, exploiting Equation \eqref{eq:expectation_Y} we find
\begin{align*}
    \mathbb E(X_t) & = \mathbb E(\xi^0) + \int_0^t K(t-s)  \left[ \mu - \lambda \left( \mathbb E(\xi^0) (e^{- \lambda \cdot} \star \widetilde K)(s) + \frac{\mu}{\lambda} (1 - e^{- \lambda s}) \right) \right] \, ds \\
    & = \mathbb E(\xi^0) + \int_0^t \mu K(t-s) e^{- \lambda s} ds - \int_0^t  \lambda \mathbb E(\xi^0) K(t-s)   (e^{- \lambda \cdot} \star \widetilde K)(s) ds \\
    & = \mathbb E(\xi^0) + \mu (K \star e^{- \lambda \cdot})(t) - \lambda \mathbb E(\xi^0) \left( K \star ( \widetilde K \star e^{- \lambda \cdot} ) \right)(t) \\
    & = \mathbb E(\xi^0) + \mu (K \star e^{- \lambda \cdot})(t) - \lambda \mathbb E(\xi^0) \left( \mbox{\bf 1} \star e^{- \lambda \cdot}  \right)(t) \\
    & = \mu (K \star e^{- \lambda \cdot})(t) {{+}} \mathbb E(\xi^0) e^{-\lambda t}
\end{align*}
where we have exploited commutativity and associativity of convolution and Definition \ref{def:coker}. We now exploit once more Lemma \ref{lem:asympt_tilde_K} and we find that $\lim_{t \rightarrow +\infty} \mathbb E(X_t)  = 0$. 

\begin{lemma}\label{lem:asympt_tilde_K}
    If  $\lim_{t \rightarrow +\infty}  K(t) = 0$, for a given kernel $K$, then also holds
    \begin{align*}
        \lim_{t \rightarrow +\infty} (e^{- \lambda \cdot} \star K)(t) = 0.
    \end{align*}
\end{lemma}
\begin{proof}
    First of all, we notice that the limit $\lim_{t \rightarrow +\infty} (e^{- \lambda \cdot} \star K)(t)$ has to be greater or equal to zero and that if $\lim_{t \rightarrow +\infty}  K(t) = 0$, then for every $\varepsilon >0$ it is possible to find $t_\varepsilon >0$ such that for every $u \ge t_\varepsilon$, $0 \le  K (u) \le \varepsilon$ and so we have
\begin{align*}
    (e^{- \lambda \cdot} \star  K)(t) &= \int_0^t e^{- \lambda (t-s)} K(s) ds = e^{- \lambda t} \left[ \int_0^{t_\varepsilon}  e^{\lambda s} K(s) ds + \int_{t_\varepsilon}^t e^{\lambda s} K(s) ds\right] \\
    & \le  e^{- \lambda t} \left[ \int_0^{t_\varepsilon}  e^{\lambda s} K(s) ds + \varepsilon \frac{e^{\lambda t} - e^{\lambda t_{\varepsilon}}}{\lambda} \right] \\
    &= e^{- \lambda t}\int_0^{t_\varepsilon}  e^{\lambda s} K(s) ds + \varepsilon \frac{1 - e^{-\lambda (t-t_\varepsilon)}}{\lambda}.
\end{align*}
Since the term $\int_0^{t_\varepsilon}  e^{\lambda s} K(s) ds$ is finite, we pass now to the limit for $t\rightarrow +\infty$ and we find 
\begin{align*}
    \lim_{t \rightarrow +\infty} (e^{- \lambda \cdot} \star  K)(t) \le \frac{\varepsilon}{\lambda},
\end{align*}
which holds for every $\varepsilon >0$, hence the conclusion follows.
\end{proof}




\section{On the simulation of the continuous time, semi-integrated Smart-Euler scheme~\eqref{eq:EulerXdisc1} with ``rough'' kernels}\label{app:simu_det}

To simulate the Smart scheme defined by Equation~\eqref{eq:EulerXdisc1} on the time grid $t_k =t^n_k =\frac{kT}{n}, k=0, \dots, n$ and in the fractional kernel case, namely   $\overline X^n_{0}=\xi^0$ and for every $k=1,\ldots,n$,  
\begin{equation*} 
\overline X^n_{t_{k}} 
= \xi^0 +  \sum_{\ell=1}^{k} \left( b(t_{\ell-1}, \overline{\xi}_{t_{\ell-1}}^n) \int_{t_{\ell-1}}^{t_{\ell}} K(t_{k} -s)ds   + \sigma(t_{\ell-1}, \overline{\xi}_{t_{\ell-1}}^n)  \int_{t_{\ell-1}}^{t_{\ell}} K(t_{k} -s) \, dW_s \right),
\end{equation*}
one has to deal with both the deterministic and stochastic integrals in the discretization.  Let us denote by $C$ the $n\times n$ lower triangular matrix involving the deterministic integrals, namely 

$$ C=(C_{k\ell})_{k,\ell=1:n}= \left(
\int_{t_{\ell-1}}^{t_\ell} K(t_k-s)ds\mbox{\bf 1}_{\{1\le \ell\le k\le n\}}
\right).
$$
The deterministic terms can be easily computed using high performance  numerical integration methods, see e.g. the library \url{https://docs.scipy.org/doc/scipy/reference/generated/scipy.integrate.quad.html}. The main problem, of course, concerns the simulation of the random terms $\int_{t_{\ell-1}}^{t_{\ell}}K(t_{k}-s)dW_s$, namely the  $(n+1)\times n$ matrix 
$$
G= (G_{k\ell})_{ k=1:n+1, \ell=1: n}= \left( \begin{array}{l}
\int_{t_{\ell-1}}^{t_\ell}K(t_k-s)dW_s\mbox{\bf 1}_{\{1\le \ell \le k\le n\}}\\
\Delta W_{t_\ell},\,k=n+1,\, \ell=1:n
\end{array} \right),
$$
where the last line has been introduced in order to be able to perform a consistent joint simulation of $\bar X^n$ and  the Euler scheme of the Markovian process $\xi$ ``impulsed'' by $\xi^0$ independent of $W$ given by
\begin{equation}\label{eq:xi_bar^N}
\overline{\xi}^n_{t_{k}}=\overline{\xi}^n_{t_{k-1}}+\xi^0(\widetilde \varphi(t_{k})-\widetilde \varphi(t_{k-1}))+b(t_{k-1},\overline{\xi}^n_{k-1}) \frac{T}{n} +\sigma(t_{k-1},\overline{\xi}^n_{k-1}) \Delta W_{k}.
\end{equation}

Then the following relation holds: $\overline X_{0}^n=\xi^0$ and for every $k=1,\dots,n$
$$ 
\big( \overline X^n_{t_{k}} \big)_{k=1:n}= \xi^0\mbox{\bf 1}+C_{k,\cdot}\big( b(t_{j-1},\overline{\xi}^n_{t_{j-1}}))_{j=1:n} +G_{k,\cdot}\big( \sigma(t_{j-1},\overline{\xi}^n_{t_{j-1}}))_{j=1:n}
$$
and every $\ell= 1,\ldots,n$,
$$
\overline{\xi}^n_{t_{\ell}}=\overline{\xi}^n_{t_{\ell-1}}+\xi^0\big(\widetilde \varphi(t_{\ell})-\widetilde \varphi(t_{\ell-1})\big)+b(t_{\ell-1},\overline{\xi}^n_{t_{\ell-1}}) \frac{T}{n} +\sigma(t_{\ell-1},\overline{\xi}^n_{t_{\ell-1}}) G_{n+1,\ell}\; \mbox{ with }\;\xi_0=0. 
$$

Let us denote by $ G^\ell $, $\ell \in \{1, ..., n\}$ the $\ell$-th column of the matrix $ G $ without the zero-elements, that is the Gaussian (column) vector of size $n - \ell + 2 $:

\begin{equation}\label{eq:Gell}
G^\ell = \big( G_{\ell-1+k,k}\big)_{k=1:n-\ell+2}.
\end{equation}
The $(n-\ell +2)\times (n-\ell +2)$ symmetric covariance matrix of $ G^\ell $, denoted by $\Sigma^\ell$, is given by:
$$
\Sigma^\ell=\begin{pmatrix}
B^\ell  & V_\ell\\
V_\ell^* & T/n
\end{pmatrix},
$$
where $B^\ell$ is an $(n-\ell+1)\times (n-\ell+1)$ positive definite symmetric matrix reading
\begin{equation}\label{eq:Bell}
B^\ell=(B^\ell_{ij})_{i,j=1:n-\ell+1} \in \mathbb{R}^{(n-\ell+1)\times (n-\ell+1)}\;\mbox{ such that}\; B^\ell_{ij}= \int_{t_{\ell-1}}^{t_\ell} K(t_{\ell+i-1}-s)K(t_{\ell+j-1}-s)  \, ds 
\end{equation}
and $V_\ell$ is the column  vector reading $V_\ell=\left(\displaystyle  \int_{t_{\ell - 1}}^{t_{\ell}} K(t_{t_k} - s) \, ds \right)_{k=\ell:n}\!\in \R^{n\times 1}$ with nonnegative components.

Note that $\Sigma^\ell$ being a  positive definite matrix has a (unique) Cholesky decomposition $\Sigma ^\ell = L^{\ell} (L^\ell)^*$ where $L^\ell$ is lower triangular with positive diagonal terms. However the  regular Cholesky procedure often crashes when some eigenvalues are too close to zero due to the propagation of  rounding errors and the presence of square roots in the procedure.
We then apply the extended version of the Cholesky decomposition method on each $B^\ell$, that turns out to be much more stable. 

The \textit{extended Cholesky decomposition} or $\Lambda D\Lambda^*$ \textit{decomposition} {{or \textit{root-free Choleski} was initially introduced in \cite{MPW1965}}} of a positive definite matrix $A = (A_{ij})_{1 \le i,j \le n}$ is as follows: there exists a unique lower triangle matrix $\Lambda$ with $\Lambda_{ii}=1$, $ i=1:n$ and a unique diagonal matrix $D$ with positive entries such that $A =\Lambda^*D\Lambda$. This decomposition can be computed as follows: for $k = 1,\dots,n$
\begin{align}\label{eq:extCholesky}
D_{kk} &= A_{kk} - \sum_{\ell = 1}^{k-1} \Lambda_{k\ell}^{2} \, D_{\ell\ell}
\text{ and for } i > k,\quad
\Lambda_{ik} = \frac{1}{D_{kk}}
\Bigl(
A_{ik} - \sum_{\ell=1}^{k-1} \Lambda_{i\ell}\,\Lambda_{k\ell}\,D_{\ell}
\Bigr),
\end{align}
and $\Lambda_{ik}=0$ for $i<k$. Once this decomposition is computed, we recover the classical Cholesky decomposition $A=LL^*$ taking the  (lower triangular) matrix $L=\Lambda\sqrt{D}$. This algorithm turns out to be more stable than the regular Cholesky procedure.

The   Cholesky decomposition of $\Sigma^\ell$ can then be calculated as follows: we 
first compute the decomposition of the matrix $B^\ell$ through the lower triangular matrix $L^{B^\ell}$. Then 
\begin{equation}\label{eq:Lell}
L^\ell := \begin{pmatrix}
L^{B^\ell} & 0 \\
\delta_{\ell}^* & \kappa
\end{pmatrix}
\end{equation}
where  $\delta_\ell = (L^{B^\ell})^{-1} V_\ell$ and $\kappa = \sqrt{\frac{T}{n} - \delta{_\ell}^* \delta_\ell}$ since 
$$
L^\ell (L^{\ell})^* = \begin{pmatrix}
L^{B^\ell} & 0 \\
\delta_{\ell}^* & \kappa
\end{pmatrix} \begin{pmatrix}
(L^{B^\ell})^* & \delta_{\ell} \\
0 & \kappa
\end{pmatrix} = \begin{pmatrix}
L^{B^\ell} (L^{B^\ell})^* & L^{B^\ell} \delta_{\ell} \\
\delta_{\ell}^* (L^{B^\ell})^* & \delta_{\ell}^* \delta_{\ell} + \kappa^2
\end{pmatrix} = \begin{pmatrix}
B^\ell & V_\ell \\
V_{\ell}^* & \frac{T}{n}
\end{pmatrix} = \Sigma ^{\ell}.
$$

Finally,  we can simulate $G^\ell$ via the equality in distribution
\begin{equation}\label{eq:SimuGell} 
G^\ell \stackrel{\mathcal L}{=} L^\ell Z^\ell, \text{ with } Z^\ell \sim \mathcal{N}(0,\mathcal{I}_{n-\ell+1}). 
\end{equation}
From an algorithmic point of view, the following proposition is crucial, since it implies that we only have to compute one Cholesky decomposition, namely that of the matrix $B$ defined below.
\begin{proposition} Let the step size $\frac Tn$ be fixed.
	
	\noindent  $(a)$ Let $B\!\in \R^{n\times n}$ be the symmetric matrix with entries
	\begin{equation}\label{eq:Bij}
	B_{ij}= \tfrac Tn\int_{0}^{1} K\big(\tfrac Tn(i-v)\big) K\big(\tfrac Tn(j-v)\big)dv, \quad i,j \in \{1, \dots, n\}.
	\end{equation}
	Then all matrices $B^\ell$, $\ell \in \{1,\ldots,n\}$,  from~\eqref{eq:Bell} are sub-matrices of $B$ in the sense that $B^\ell= [B_{ij}]_{i,j= 1:n-\ell+1}$ (and $B=B^1$). Moreover, with obvious notation, $L^{B^\ell}= [L^B_{ij}]_{i,j=1:n-\ell+1}$, $\ell=1,\ldots,n$.
	
	\noindent $(b)$ When $K$ is a (singular) fractional kernel $K (t)=K_{1,\alpha,0}(t) = \frac{t^{\alpha-1}}{\Gamma(\alpha)}, u \in [0,T]$, where $\alpha \!\in (1/2,1)$, one derives from~\eqref{eq:Bij} that 
	$$
	B=  \big(\tfrac Tn\big)^{2\alpha-1}\frac{1}{\Gamma(\alpha)^2}\big(\mathbf{B}_{ij}\big)_{i,j=1:n} \;\mbox{ and }\;\; B^\ell=  \big(\tfrac Tn\big)^{2\alpha-1}\frac{1}{\Gamma(\alpha)^2}\big(\mathbf{B}_{ij}\big)_{i,j=1:n-\ell+1},\quad\ell=1,\ldots,n,
	$$ 
	where $\mathbf{B}$ is defined by 
	\[
	\mathbf{B}_{ij}=\int_0^1 (i-v)^{\alpha-1}(j-v)^{\alpha-1}dv= \int_0^1 (i-1+v)^{\alpha-1}(j-1+v)^{\alpha-1}dv,\quad i,j\ge 1.
	\]
	So, if the Cholesky decomposition of  $[\mathbf{B}_{ij}]_{i,j=1:n}$ reads $[\mathbf{B}_{ij}]_{i,j=1:n}=\mathbf{L}^{\mathbf{B}}(\mathbf{L}^{\mathbf{B}})^*$ then  
	$$
	L^{B^\ell} = \big(\tfrac Tn\big)^{\alpha-\frac 12}\frac{1}{\Gamma(\alpha)}[\mathbf{L}^{\mathbf{B}}]_{i,j=1:n-\ell+1},\quad\ell=1,\ldots,n.
	$$
	\end{proposition}
\begin{proof} $(a)$ 
	Let $\ell \in \{1,...,n\}$ and and $i,j=1, \dots,n-\ell+1$. It follows from~\eqref{eq:Bell} that         
	\begin{align*}
	B_{ij}^\ell &=\int_{t_{\ell-1}}^{t_\ell} K(t_{\ell+i-1}-s) K(t_{\ell+j-1}-s)ds \stackrel{(s=u+t_{\ell-1})}{=} \int_{0}^{h} K(t_{\ell+i-1}-t_{\ell-1}-u) K(t_{\ell+j-1}-t_{\ell-1}-u)du\\
	&=\int_{0}^{h} K(t_{i}-u) K(t_{j}-u)du = h\int_{0}^{1} K\big(h(i-u)\big) K\big(h(j-u)\big)du =  B_{ij}.
	\end{align*}
	Moreover, one easily checks the procedure of the Cholesky decomposition (or, here, on  its extended version~\eqref{eq:extCholesky}): if $B=L^B(L^B)^*$ is the Cholesky decomposition of $B$  then  $B^\ell = L^{B^\ell}(L^{B^\ell})^*$ with $L^{B^\ell}= [L^B_{ij}]_{i,j= 1:n-\ell+1}$.
\hspace{1cm} $(b)$ This is immediate once the kernel $K$ is specified.\end{proof}

\noindent {\bf Remark}. Moreover note that the diagonal terms of $\mathbf{B}$ have a closed form
\[
\mathbf{B}_{ii}= \frac{1}{2\alpha-1}\big(i^{2\alpha-1}-(i-1)^{2\alpha-1}\big), \, i\ge 1,
\]
(still with $h= \frac Tn$) and  the non-diagonal entries of the  first line  can be rewritten without singular integrals as follows
\[
\mathbf{B}_{1j} = \frac{j^{\alpha-1}}{\alpha}-\frac{\alpha-1}{\alpha}\int_0^1u^\alpha(j-1+u)^{\alpha-2}du, \; j\ge 2.
\]

\noindent{\em Practitioner's corner}. $\bullet$ The  (non-diagonal) coefficients of $\mathbf{B}$ can be computed  by a cubature formula (trapezoid, midpoint, Simpson, higher order Newton-Cote integration formulas or Gauss-Legendre (weighted) points, \dots). 

\noindent $\bullet$ Then, the step size $h= \frac Tn$ being fixed, one compute the entries of matrix $B$. We perform an extended Cholesky transform to $B$ to derive the lower triangular matrix $L^B$ and by successive extractions the matrices $L^{B^\ell}$. This in turn allows to compute the $L^\ell$ matrices~\eqref{eq:Lell} to simulate the Gaussian vector $G^\ell$, defined for every $\ell \!\in \{1,\ldots,n\}$  by~\eqref{eq:Gell}, by
\[
G^\ell = L^\ell Z^{(n-\ell+2)}, \quad Z^{(n-\ell+2)}\stackrel{d}{=} {\mathcal N}(0, I_{n-\ell+2}),
\]
where the random normal vectors $ Z^{(n-\ell+2)}$  are independent so that  the vectors $G^\ell $ are in turn independent.

\noindent $\bullet$ The computation of $L^\ell$ requires inverting $L^{B^\ell}$, but as a lower triangular sub-matrix of the   lower triangular  matrix  $L^B$, it is clear that $(L^{B^\ell})^{-1} = \big[((L^B)^{-1})_{ij}]_{i,j=1: n-\ell+1}$.

\noindent $\bullet$ One checks that for small values of $H= \alpha -\frac 12$, such as $H=0.1$, the columns of $B$ are all numerically zero (say lower than$10^{-4}$) from the fourth column onward. Taking this into account allows a crucial speed-up of the simulations of the ``Smart Euler scheme''. 
$\bullet$ To sum up, prior to the Euler simulation, one only needs to perform, for a given $n$:

-- the (extended) Cholesky decomposition of $[\mathbf{B}]_{i,j=1:n}= L^{\mathbf{B}} {L^{\mathbf{B}}}^*$ 

-- the inversion of the lower triangular matrix $L^{\mathbf{B}}$.

\newpage

\section*{Declarations}

\textbf{Acknowledgements:} 

The authors thank Edgar Neboit for his precious assistance in the numerical part. The authors would like to thank Fabienne Comte and Andrea Pallavicini for fruitful discussions on the model and they thank all the participants to the Workshop ``Volatility is rough: now what?'', and in particular E. Abi Jaber, C. Cuchiero, J. Guyon, A. Jacquier, M. Rosenbaum, for their precious comments and remarks. 

\vspace{0.5cm}

\textbf{Funding and/or Conflicts of interests/Competing interests:}

The first author acknowledges financial support from the EPSRC grant EP/T032146/1. 
The fourth author acknowledges financial support from the chaire ``Risques financiers'' from the {\em Fondation du Risque}. Financial support has also been provided by the University of Padova grant BIRD227115 - "Equilibrium approaches in financial and energy markets".

The authors declare that they have no conflicts of interest or competing interests to disclose in relation to this work.

\vspace{0.5cm}

{\textbf{Data access statement}

There are no data associated with this article.}

\vspace{0.5cm}

{
\textbf{Declaration of Generative AI and AI-Assisted Technologies in the Writing Process}

The use of LLMs in this paper has been limited to minimal proofreading.}

\newpage
\small
\bibliographystyle{siam}

\bibliography{references}

 \end{document}